\documentclass[traditabstract]{aa}

\pdfoutput=1
\usepackage{graphicx} 
\usepackage{txfonts}
\usepackage{amssymb}

\usepackage{natbib}
\bibpunct{(}{)}{;}{a}{}{,}

\usepackage{color}

\newcommand{\hi}{\ion{H}{i}}
\newcommand{\vs}{$v_{\rm S}$}
\newcommand{\cc}{\mbox{cm$^{-3}$}}
\newcommand{\kms}{\mbox{km\,s$^{-1}$}}
\newcommand{\fcnm}{\mbox{$f_{\rm CNM}$}}
\newcommand{\sigturb}{$\sigma_{\rm turb}$}
\newcommand{\sigtherm}{$\sigma_{\rm therm}$}
\newcommand{\sigtot}{$\sigma_{\rm tot}$}
\newcommand{\machtheo}{$\mathcal{M}_{\rm theo}$}
\newcommand{\machobs}{$\mathcal{M}_{\rm obs}$}

\begin{document}

\title{The structure of the thermally bistable and turbulent atomic gas in the local interstellar medium}
\author{E. Saury \inst{\ref{inst1},\ref{inst2}}
\and M.-A. Miville-Desch\^enes \inst{\ref{inst1},\ref{inst2}}
\and P. Hennebelle \inst{\ref{inst3}}
\and E. Audit\inst{\ref{inst4}}
\and W. Schmidt\inst{\ref{inst5}} }

\institute{Institut d'Astrophysique Spatiale, CNRS UMR 8617, Universit\'e Paris-Sud 11, B\^atiment 121, 91405, Orsay, France\label{inst1}\\ email : eleonore.saury@ias.u-psud.fr
\and Canadian Institute for Theoretical Astrophysics, University of Toronto, 60 St. George St., Toronto, ON M5S 3H8, Canada\label{inst2}
\and Laboratoire de Radioastronomie, CNRS UMR 8112, \'Ecole Normale Sup\'erieure, 24 rue Lhomond, 75231 Paris, France\label{inst3}
\and Laboratoire AIM, CEA/DSM - CNRS - Universit\'e Paris Diderot, IRFU/SAp, 91191, Gif-sur-Yvette, France\label{inst4}
\and Institut f\"ur Astrophysik, Universit\"at G\"ottingen, Friedrich-Hund Platz 1, D-37077 G\"ottingen, Germany\label{inst5}}

\date{Received January 15, 2013 / Accepted ??}

\titlerunning{The structure of the thermally bistable and turbulent atomic gas in the local interstellar medium}

\authorrunning{Saury et al.}

\abstract{This paper is a numerical study of the condensation of the warm neutral medium (WNM) into cold neutral medium (CNM) structures under the effect of turbulence and thermal instability. It addresses the specific question of the CNM formation in the physical condition of the local interstellar medium (ISM). Using low resolution simulations we explored the impact of the WNM initial density and properties of the turbulence (stirring in Fourier with a varying mix of solenoidal and compressive modes) on the cold gas formation to identify the parameter space that is compatible with well established observational constraints of the \hi\ in the local ISM. Two set of initial conditions which match the observations were selected to produce high resolution simulations (1024$^3$) allowing to study in detail the properties of the produced dense structures.\\
For typical values of the density, pressure and velocity dispersion of the WNM in the solar neighborhood, the turbulent motions of the \hi\ can not provoque the phase transition from WNM to CNM, whatever their amplitude and their distribution in solenoidal and compressive modes. On the other hand we show that a quasi-isothermal increase in WNM density of a factor of 2 to 4 is enough to induce the phase transition, leading to the transition of about 40\,percent of the gas to the cold phase within 1\,Myr.
Given the observed properties of the \hi\ in the local ISM, the WNM and individual CNM structures in the local ISM are sub or transsonic and their dynamics are tightly interwoven. The velocity field bears the evidence of subsonic turbulence with a 2D power spectrum following the Kolmogorov law as $P(k) \propto k^{-8/3}$ while the density is highly contrasted with a singificantly shallower power spectrum, reminiscent of what is observed in the cold ISM. Supra-thermal line width observed for CNM might be the result of relative velocity between cold structures. Finally, the cold structures denser than 5\,\cc\ reproduce well the laws $M\propto L^{2.25-2.28}$ and $\sigma(|\mathbf{v}|) \propto 0.5-0.8\,L^{1/3}$ generally observed in molecular clouds.}

\keywords{turbulence - instabilities - hydrodynamics - ISM:clouds - ISM:structure}

\maketitle

\section{Introduction}

Understanding the star formation process remains one of the main area of research of modern astrophysics.
It is directly linked to the way interstellar gas is organized and how dense and cold structures form.
Turbulence plays a fundamental role here. 
It creates local increases of density that are amplified by gravity and that, in some cases, might lead to gravitational contraction.
At the same time the kinetic energy of turbulent motions needs to be dissipated in order for the gravitational collapse to happen.
The overall star formation process is therefore directly linked to the way turbulence shapes the density structure of the ISM
and how kinetic energy is transferred from large to small scales.

Molecular clouds, where star formation occurs, exhibit supra-thermal line widths and large contrast of 
density revealed by a log-normal PDF of column density \citep{vazquez2001, goodman2009, federrath2010, brunt2010a} and by power spectra of column density shallower \citep[-2.5,][]{bensch2001} than predicted for 
subsonic or transonic turbulence (-3.66), and than what is observed in the diffuse \hi\ \citep{mamd2003b}. 
For these reasons they are often modeled as isothermal and supersonic turbulent flows.
On the other hand, because turbulence should dissipate in one dynamical time if it is not maintained at large scale it is thought 
that star formation might occur within a dynamical (or free fall) time.
In this scenario the gravitational collapse occurs along the cloud formation.
Therefore, the initial conditions of the star formation process depend on the structure 
of the cold clouds in the atomic phase (\hi).

Most of the mass of the Galactic interstellar medium (ISM) is \hi\ and 
there are clear observational evidences that it is thermally bi-stable \citep{dickey2003}; it consists of two 
thermally stable states at significantly different temperatures (the Cold Neutral Medium - CNM - and the Warm Neutral Medium - WNM). 
This is a natural result of the density and temperature dependance of the cooling processes at play in the diffuse ISM.
Like molecular clouds the structure of the \hi\ is highly complex with large density contrasts
(the CNM fills only about 1 percent of the volume but has a density 100 times larger than the WNM)
but here the combination of turbulence and thermal instability \citep{field1965,field1969} is the main driver of the structure.
It is important to note that the nature of the density fluctuations in an isothermal and 
supersonic flow is conceptually different than in a thermally bi-stable fluid.
An isothermal turbulent flow with supersonic velocity fluctuations will naturally develop strong density 
fluctuations \citep[$\Delta \rho / \rho \sim M_s^2$,][]{vazquez2012}.
In such a flow the density fluctuations are transient, associated with the compression and depression of the turbulent motions. 
In a two-phase medium, once a high density structure is created by a local compression it remains at high density
even when the gas re-expand because it is in pressure equilibrium with the inter-cloud medium. 
A two-phase medium will therefore produce long-lived high density structures
contrary to supersonic flows in which the fluctuations appear and disappear on dynamical timescales. 
Therefore the thermally bistable \hi\ offers a natural way of producing highly structured initial conditions for the star formation
in molecular clouds.

Because of the complex dynamical processes involved, 
dedicated numerical simulations are useful to develop insights on the exact scenario that leads to the formation of cold 
structures in the \hi.
Several studies based on numerical simulations have explored the thermally bi-stable physics of the \hi\ 
since the early one-dimensional work of \citet{hennebelle1999} who showed the effect of compression in a WNM
converging flow on the triggering of the phase transition and the production of CNM structures. 
Similarly \citet{koyama2000,koyama2002} studied the impact of the propagation of a shock wave into WNM. They showed how CNM structures
form in the post-shock region through the thermal instability which operates faster than the duration of the shock (galactic spiral shock
or supernova). In both cases the resulting density contrast between the CNM and WNM is $\sim 100$. 
To obtain a similar contrast in isothermal supersonic turbulence, Mach numbers of the order of 10 are required. 
These studies have been extended to two and three dimensions simulations of thermally bistable flows with turbulence 
induced by a converging flow \citep{audit2005,hennebelle2007a,hennebelle2007b,audit2010,heitsch2005,heitsch2006}, by the magnetorotational instability \citep{piontek2004,piontek2005,piontek2007,piontek2009} or by a driving in Fourier space \citep{gazol2005,gazol2010,seifried2011}.

It was established that compression of the WNM in transonic converging flows triggers the thermal instability;
the increase in density can be such that the gas cools rapidly to temperatures of a few tens of degrees.
The cold gas then fragments into clumps, in particular through the Kelvin-Helmholtz instability \citep{heitsch2006}.
The thermal pressure of the cold gas is in approximate equilibrium with the sum of the ram and thermal pressure of the
surrounding warm gas. Once pushed at high enough densities, the cold structures can have densities and temperature
close to values typical of molecular clouds \citep{vazquez-semadeni2006a,banerjee2009}. 

In general previous works showed that the efficiency of the formation of the CNM 
depends directly on the nature and amplitude of the injected energy.
One question we wanted to explore in this study is related to the driving of the compression. Most studies
have explored the case of converging flows where the increase of the WNM density needed to trigger the 
thermal instability is obtained by a sustained ram pressure. The origin of such large scale
convergent flows is unclear so here we want to see if compression in 
turbulent motions alone, driven at large scales, could lead to the production of cold gas. 
In more details, we wanted to see if turbulent motions, at the level they are observed in the WNM,
can induce the phase transition and if not, what are the required physical conditions of the WNM to trigger the formation
of cold structures.
To do so, like \citet{gazol2005,gazol2010,seifried2011}, we used turbulent forcing in Fourier space.
In addition, similarly to the study of \citet{federrath2010} based on isothermal simulations of molecular clouds, we wanted to study the impact of compressive and solenoidal modes in the injected turbulent field.

In this paper we present a parameter study of WNM initial density and turbulent forcing properties 
on the formation of the CNM in forced turbulence and thermally bi-stable flows. 
In a similar study, \citet{seifried2011} chose initial densities ($1$ to $3\,$cm$^{-3}$) that are in the thermally unstable regime, 
to expect the production of a two-phase medium.
Our intent is to expand the parameter space to identify the dynamical conditions that lead to the formation of 
the cold phase. In addition, one fundamental aspect of the current study is that we guide our parameter search
based on all available observational constraints of the local ISM. We present 90 simulations performed on a $128^3$ grid allowing
to identify more precisely the physical conditions that favor the formation of cold structures 
while reproducing in detail the observed properties of the \hi\ in the local ISM.
For two specific cases that match the observations we performed two $1024^3$ simulations in order to study in greater details
the properties of the \hi: temperature, pressure and density histograms, power spectra, power laws of the cold structures.

The paper is organized as follow. HERACLES, the numerical code used in this study is briefly presented in \S~\ref{sec:num}.
The observational constrains used to lead the parameter study are detailed in \S~\ref{sec:obs}.
The methodology is presented in \S~\ref{sec:methodology}. The results of the parametric study and of the high resolution simulations are presented in \S~\ref{sec:parametric_study} and \S~\ref{sec:1024}, and then discussed in \S~\ref{sec:discussion}. The main conclusions of the study are given in \S~\ref{sec:conclusion}.

\section{Numerical methods}
\label{sec:num}

Despite their great importance in the interstellar medium, we won't consider magnetic field or gravity in this study to focus on the impact of vortical and compressive turbulence motions on the formation of CNM. All the following simulations are thus hydrodynamical. 
We performed our simulations with HERACLES \citep{gonzalez2007}. The numerical methods are similar to those described in paper from \cite{audit2005}.

\begin{figure}
  \resizebox{\hsize}{!}{\includegraphics{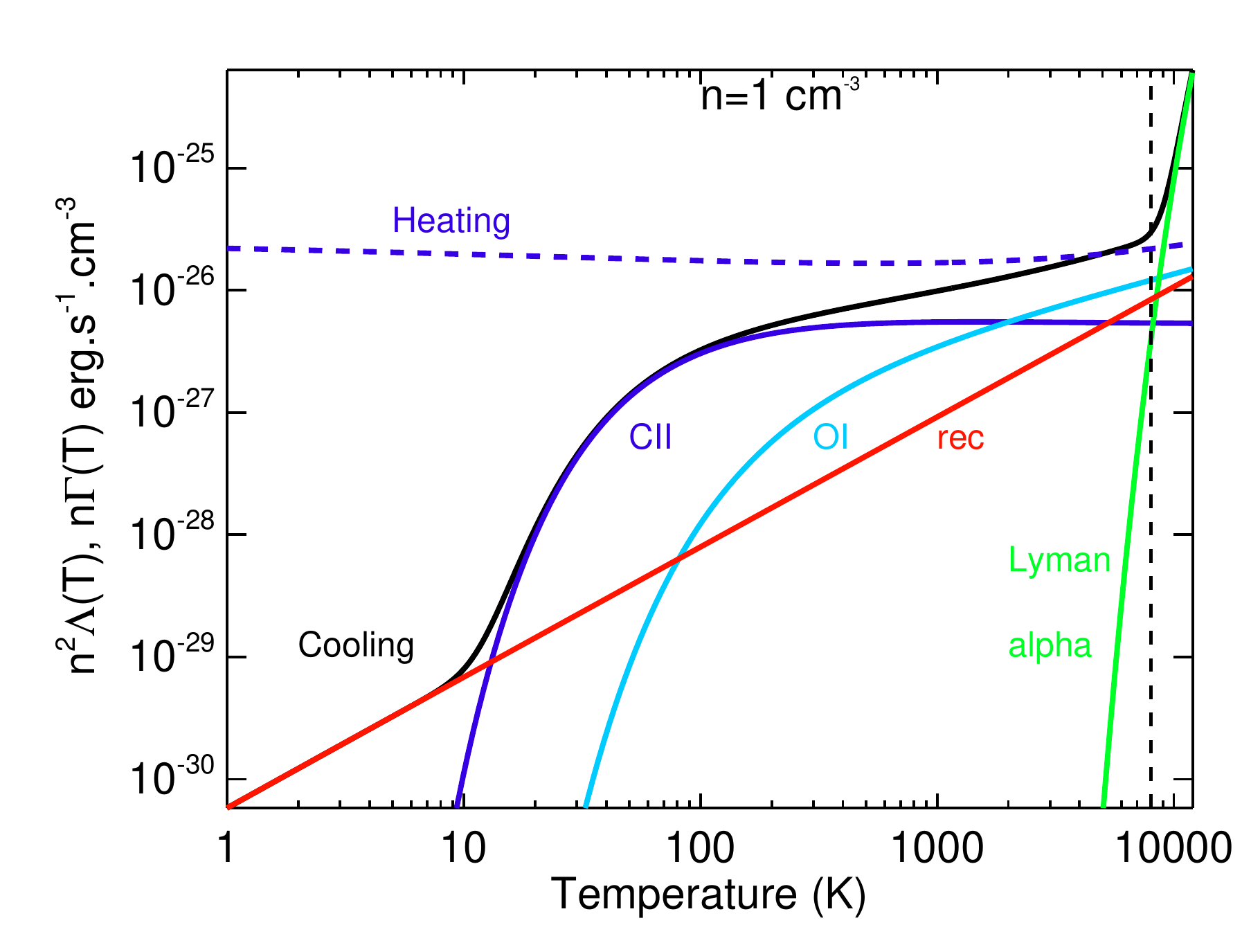}}
  \caption{Heating and cooling processes implemented in HERACLES based on \citet{wolfire2003}. We represent the respective energy densities versus the temperature at the fixed density \mbox{n=1 cm$^{-3}$}, typical of the WNM. The total cooling (black solid curve) is decomposed in its different components: CII (dark blue), OI (light blue), recombination on interstellar grains (red), Lyman $\alpha$ (green). The dashed blue line is the heating dominated by the photo-electric effect on small dust grains. We considered a spatially uniform radiation field with the spectrum and intensity of the Habing field ($G_0/1.7$ where $G_0$ is the Draine flux) \citep{habing1968,draine1978}.}
  \label{fig:coolheat}
\end{figure}

\subsection{Euler equations and Godunov scheme}

The Euler equations for a radiatively cooling gas are the classical equations of hydrodynamics in order to treat the flows at high Reynolds number, which is the case in the ISM. An external force $\mathbf{f}$ is adding to generate the turbulent motions. It has the dimension of an acceleration. We also include cooling and heatings terms based on the prescription of \citet{wolfire2003} (see Fig.~\ref{fig:coolheat}) in the net cooling rate $\mathcal{L}$.

\begin{eqnarray}
&& \partial_t\rho + \nabla\cdot[\rho \mathbf{u} ] = 0, \\
&& \partial_t[\rho \mathbf{u}] + \nabla\cdot[\rho \mathbf{u}\otimes \mathbf{u} +P] = \rho\mathbf{f}, \\
&& \partial_tE + \nabla\cdot[\mathbf{u}(E+P)] = \rho\mathbf{f}\cdot\mathbf{u}-\mathcal{L}(\rho,T)
\end{eqnarray}
where $\rho$ is the mass density, $\mathbf{u}$ the velocity, $P$ the pressure and $E$ the total energy. The considered atomic gas is perfect and ideal with an adiabatic index $\gamma = 5/3$ and a mean atomic weight $\mu = 1.4m_{\mathrm{H}}$ ($m_{\mathrm{H}}$ is the mass of the proton). The weight $\mu$ takes into account the cosmic abundance of other elements : helium and metals, and so the mass density $\rho = \mu\times n$. The net cooling function is defined as $\mathcal{L} = n^2\Lambda - n\Gamma$ (expressed in erg\,\cc\,s$^{-1}$), where $\Lambda$ is the cooling and $\Gamma$ the heating. 
They include the most relevant processes in the diffuse medium for our range of temperature [10K, 1000K] and are plotted and quickly described on the Figure \ref{fig:coolheat} \cite[for more details, see][]{audit2005}.
The equations solver in HERACLES is a second order Godunov scheme \cite[see][]{audit2010} for the conservative part. It is a MHD solver of type MUSCL-Handcook\citep{londrillo2000,fromang2006}. The cooling is applied after \cite[see][]{audit2005}.

\subsection{Turbulence forcing}
\label{subsec:forcing}

The force field $\mathbf{f}$ generates the large scales motions and is applied in Fourier space. We define the amplitude given to the field as $v_{\mathrm{S}}$. The magnitude of the applied force is thus \mbox{$F_{\mathrm{S}} = v_{s}^2/L_{\mathrm{S}}$} where \mbox{$L_{\mathrm{S}} = L_{\mathrm{box}}/k_0$} is the characteristic integration scale and $k_0 = 2$ the characteristic wave number of the stirring. The auto-correlation time of the process is given by \mbox{$T_{\mathrm{S}} = (L_{\mathrm{S}}/F_{\mathrm{S}})^{1/2} = L_{\mathrm{S}}/v_{\mathrm{S}}$}. The stirring is applied at large scale, and thus to the small wave numbers $k$ that fulfill \mbox{$0<|k|<2k_0$}. The modes of the turbulent forcing are computed with the stochastic process of Ornstein-Uhlenbeck \citep{eswaran1988a,schmidt2006}. The method is similar to \cite{schmidt2009} and \cite{federrath2010}. A Helmholtz decomposition is applied to the random field and project it on its compressive and solenoidal components:
\begin{equation}
\label{eq:helmoltz}
\mathcal{P}^{\zeta}_{ij} = \zeta\mathcal{P}^{\perp}_{ij} + (1-\zeta)\mathcal{P}^{\parallel}_{ij} = \zeta\delta_{ij} + (1-2\zeta)\frac{k_ik_j}{|k|^2}.
\end{equation}
Choosing a spectral weight $\zeta =$ 1 creates a physical force divergence free, i.e. purely solenoidal. Taking $\zeta <$ 1 generates dilatational components which compress or rarify the gas. Finally, $\zeta = 0$ creates a rotation-free field, i.e. purely compressive. In a natural state, meaning that the energy is naturally distributed between the compressive and solenoidal modes as the 2:1 mixing,  $\zeta$ takes the value of $0.5$. In this case, the Helmholtz operator is proportional to the identity operator and the projection does not alter the energy repartition between the different modes.


\section{Observational constraints}
\label{sec:obs}

In this section we present global properties of the Galactic \hi\ as deduced from the observations. 
These properties guided us in the choice of the initial conditions of the simulations and on the
qualification of the results obtained. 
The simulations presented in this study are dedicated to the \hi\ thermal instability and to the formation
of CNM out of pure WNM gas. What are reasonable properties of the WNM and what are exactly the observational constraints ?

\subsection{Pressure and thermal bi-stability of the \hi}
\label{sec:thermal}

\begin{figure}
  \resizebox{\hsize}{!}{\includegraphics{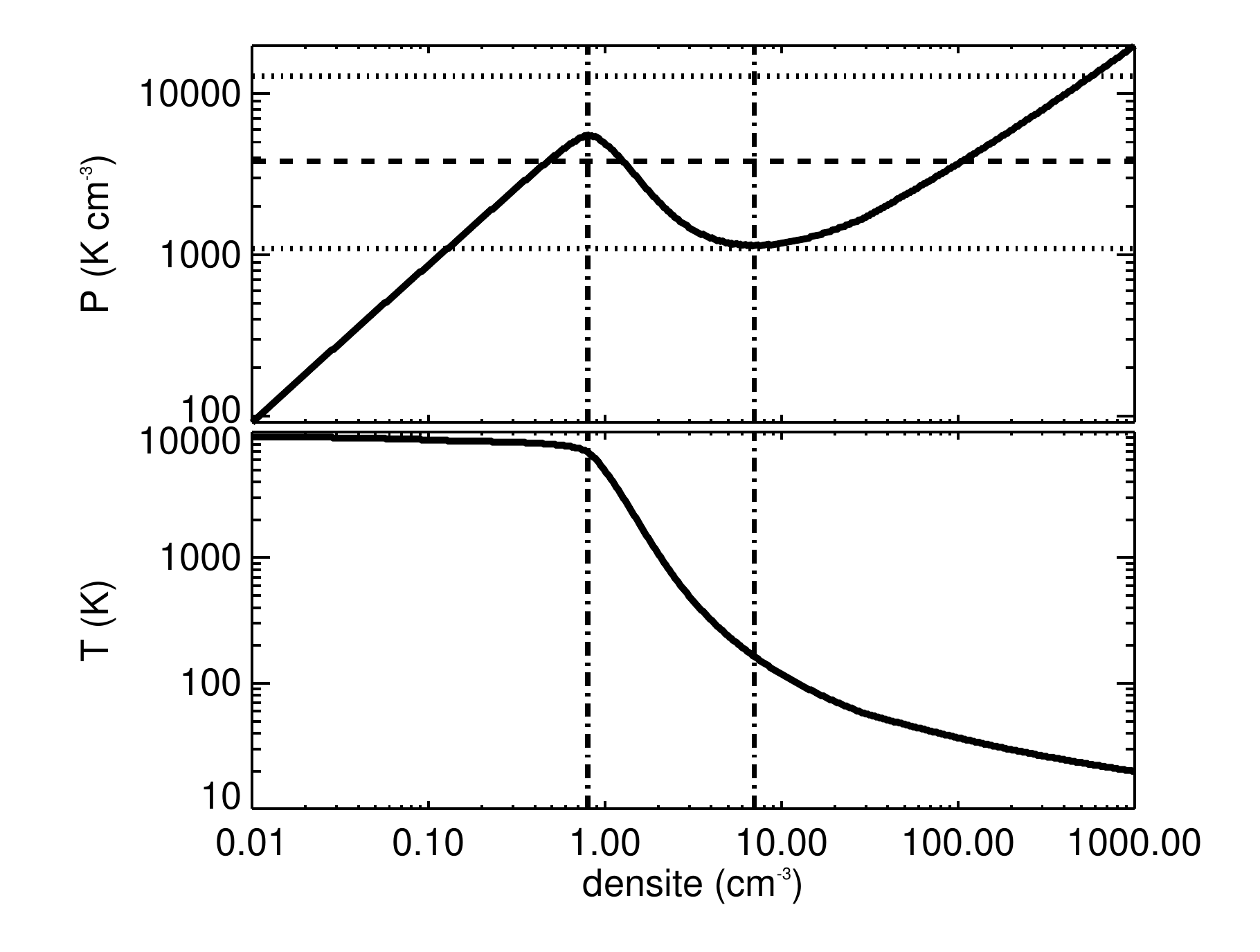}}
  \caption{Equilibrium temperature and pressure as a function of density. These equilibrium curves
    were computed equating the heating and cooling processes described in \citet{wolfire2003}.
  The dashed line corresponds to the average pressure of the cold diffuse ISM measured by \citet{jenkins2011}.
  The dotted lines show the maximum and minimum pressures that include 99 percent of the data of \citet{jenkins2011}. 
  The dash-dotted lines show the density range of the thermally unstable phase.}
  \label{fig:T_and_P_vs_n}
\end{figure}

Because of the density and temperature dependance of the heating and cooling processes in the diffuse ISM (see Fig.~\ref{fig:coolheat}) it has been shown early on by \citet{field1965,field1969} and updated by \cite{wolfire1995,wolfire2003} that there is a range of pressure where the neutral atomic gas can be in two thermally stable phases in pressure equilibrium.
This is generally displayed in diagrams like the ones shown in Fig.~\ref{fig:T_and_P_vs_n} where the solid curves in the two panels indicate the locus of thermal equilibrium in the $P-n$ and $T-n$ parameter spaces. 

The thermal stability is not preserved everywhere along these curve but only where $dP/dn > 0$. The two dash-dotted vertical lines mark the boundaries in density where the \hi\ is thermally unstable. Any perturbation of the gas sitting on the equilibrium curve in the thermally unstable range would move towards the cold or warm branches. From Fig.~\ref{fig:T_and_P_vs_n} one can appreciate the steepness of the equilibrium curve in $T-n$ space, highlighting the great difference in temperature and density of the two stable phases of the \hi.

It is important to point out that the bi-stability of the \hi\ is only present in a given range of pressure, determined by the maximum density of the warm branch ($n \sim 0.8 \,$cm$^{-3}$) and the minimum density of the cold branch 
($n \sim 7 \,$cm$^{-3}$). For $P<1000\,$K$\,$cm$^{-3}$ there is no CNM and for $P>6000\,$K$\,$cm$^{-3}$ all the \hi\ is cold. 

On the top panel the dashed curve gives the average pressure observed in the local ISM. 
According to \citet{jenkins2011} the pressure of the cold phase follows a log-normal distribution with an average of $\log_{10}(P/k)$ equals to 3.58
and minimum and maximum values (at the 99 percent level) equal to 3.04 and 4.11.
The minimum CNM pressure reported by \citet{jenkins2011} corresponds exactly to the minimum CNM pressure (3.06 dex) allowed by the thermal instability (see Fig.~\ref{fig:T_and_P_vs_n}) as predicted by the modeling of the heating and cooling processes of \citet{wolfire2003}.
On the other hand, the maximum pressure reported by \cite{jenkins2011} is significantly higher than the maximum pressure allowed for the thermally stable WNM (3.74).
In fact, the mass fraction of the CNM found at higher pressure by \citet{jenkins2011} is about 30 percent.
In those cases most of the WNM is in the thermally unstable regime or in a cooling phase; in any case the warm gas is an over-pressured state that is likely to be transient.

\subsection{WNM density and temperature}

Here we want to estimate what is the true density (as opposed to space-averaged) and temperature of the WNM in the local ISM.
Typical numbers cited are $n\sim0.2-0.5\,$cm$^{-3}$ and $T\sim6000-10^4\,$K \citep{ferriere2001},
but, as emphasized by \cite{kalberla2008}, 
the density of the WNM varies greatly with galacto-centric radius and galactic height $z$. It is also subject to local variations due to star formation activity 
\citep[e.g. supernovae, ][]{joung2009} and between the arm and inter-arm regions.

According to \citet{dickey1990} \citep[see also][]{celnik1979,malhotra1995,kalberla2007}, the distribution of the \hi\ density with $z$ close to the Sun can be decomposed in a narrow Gaussian of HWHM=$106\,$pc, a larger Gaussian of HWHM=$265\,$pc and an exponential with a scale height of $403\,$pc. Generally the narrow Gaussian is attributed to the CNM and the sum of the two others to the WNM. In this scheme, and still following \citet{dickey1990}, the space-average density of the CNM and WNM at $z=0$ are $0.395\,$cm$^{-3}$ and $0.172\,$cm$^{-3}$ respectively. These values are weighted by the volume filling factor $f$ and thus correspond to $n\,f$, where $n$ is the true volume density.  
The filling factor of the WNM is difficult to estimate; according to \citet{kalberla2009} it is $\sim 0.4$ which implies a true density of $n_{\mathrm{WNM}}=0.43\,$cm$^{-3}$. 

This value is close to the one that can be derived based on the knowledge of the heating and cooling processes.
For the average pressure of $P=3980\,$K$\,$cm$^{-3}$ found by \citep{jenkins2011}, using the heating and cooling curves of \citet{wolfire2003} and assuming a spatially uniform radiation field with the spectrum and intensity of the Habing field \citep[$G_0/1.7$ where $G_0$ is the Draine flux --][]{habing1968,draine1978}, the equilibrium density and temperature of the WNM
are $n_{\mathrm{wnm}}=0.5\,$cm$^{-3}$ and $T_{\mathrm{wnm}}=7960\,$K. 

\begin{figure}
  \resizebox{\hsize}{!}{\includegraphics{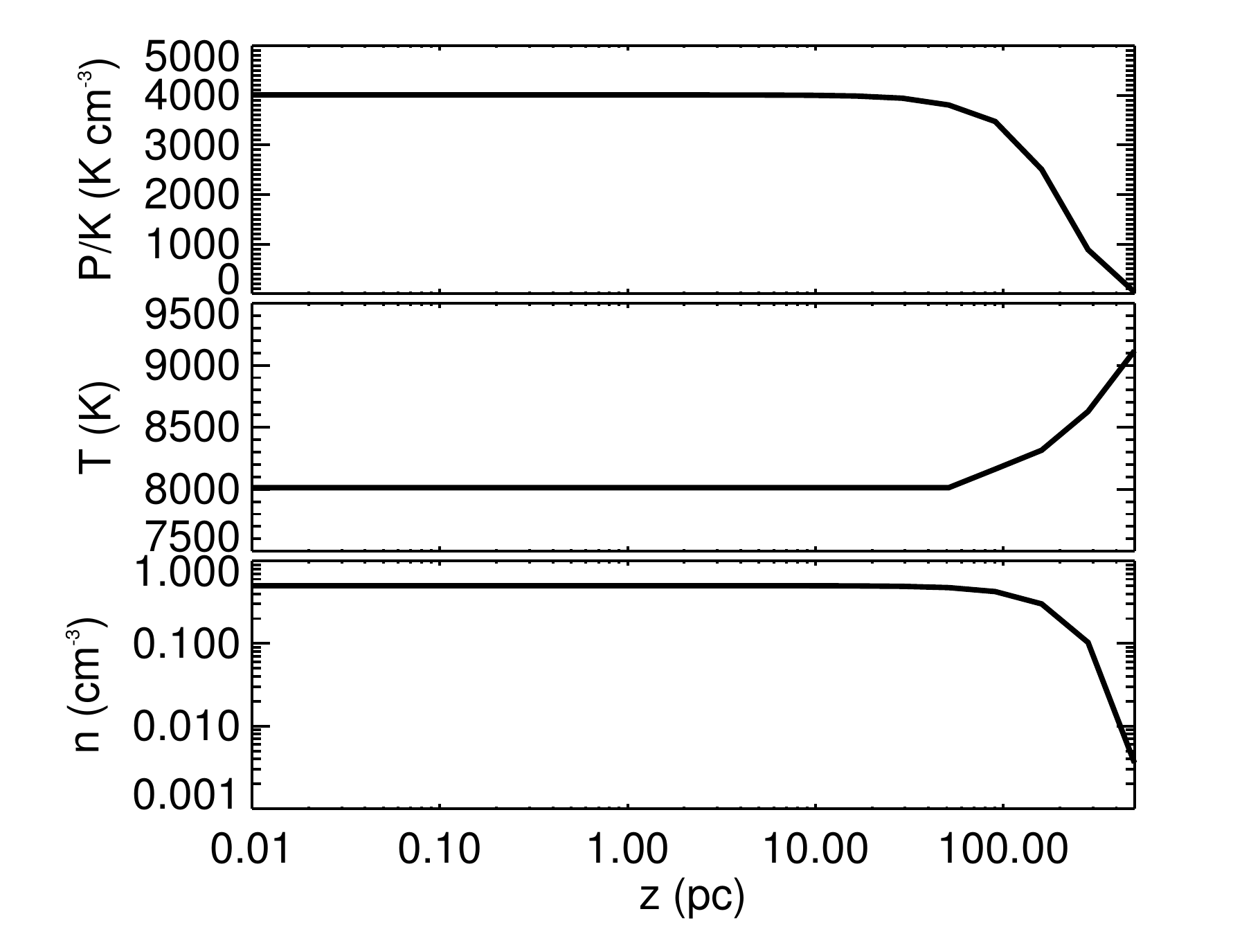}}
  \caption{Vertical distribution of the WNM density, temperature and pressure.
    The density profile assumes $n(z=0)=0.5\,$cm$^{-3}$, in accordance with the mean pressure
    of \citet{jenkins2011} and the heating and cooling processes described in \citet{wolfire2003}.
    The scale height of the WNM is the value of \citet{dickey1990} : HWHM = 265~pc.
    The temperature is computed at each position $z$ assuming thermal equilibrium at density $n(z)$. The pressure is simply $P/k=n\,T$.}
  \label{fig:n_t_vs_z}
\end{figure}

Fig.~\ref{fig:n_t_vs_z} shows how $n$, $T$ and $P$ vary with $z$ assuming a scale height of HWHM=265~pc for the WNM disk at $R=R_\odot$ \citep{dickey1990,malhotra1995,kalberla2007}. Assuming a true volume density $n_{\mathrm{WNM}}=0.5\,$cm$^{-3}$ at $z=0$, the corresponding temperature profile $T(z)$ of the WNM can be estimated using the thermal equilibrium curve based on the heating and cooling processes
described in \citet{wolfire2003}. 
It shows that the variations with $z$ of the WNM properties due to the hydrostatic equilibrium become important only at scales larger than 100~pc.

\begin{figure*}
  \centering\resizebox{0.4\hsize}{!}{\includegraphics{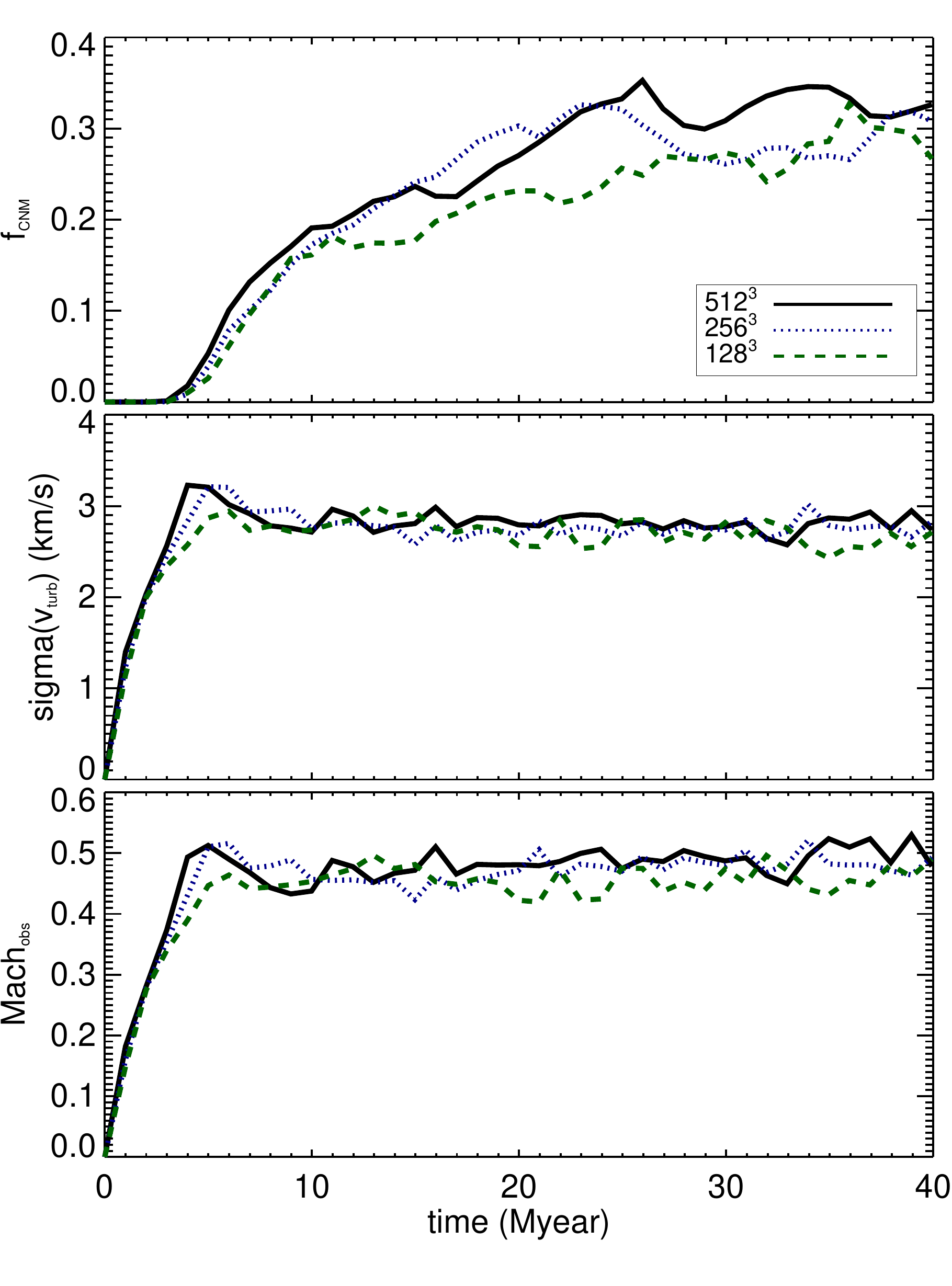}}
  \centering\resizebox{0.4\hsize}{!}{\includegraphics{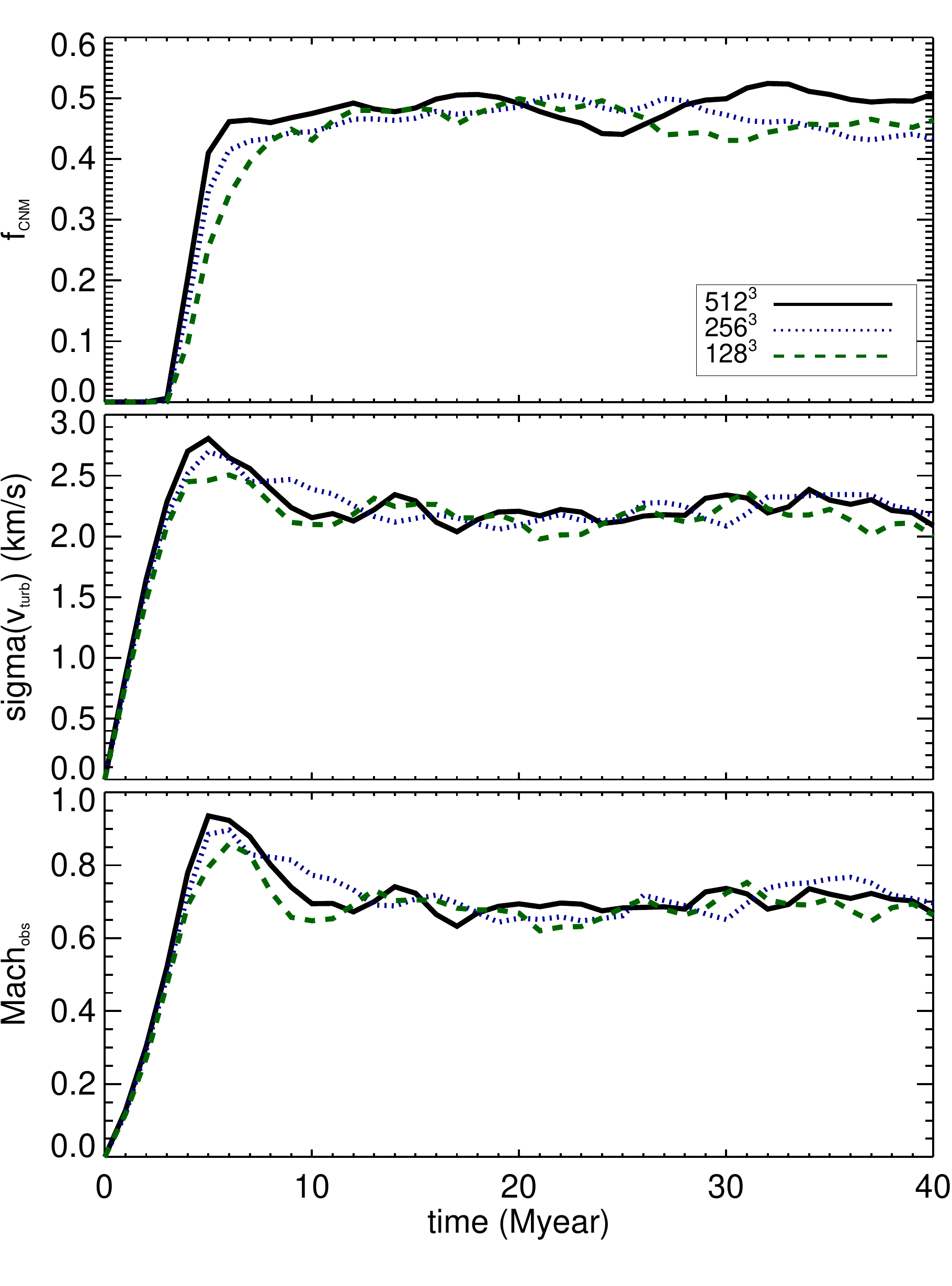}}
  \caption{Time evolution of the CNM mass fraction $f_{CNM}$, the mean velocity dispersion $\sigma_{turb}$ and the mean Mach number $\mathcal{M}_{\mathrm{obs}}$ for simulations with the following initial conditions: {\it on the left} $n_0=1 $cm$^{-3}$, $\zeta$=0.2, \mbox{$v_{\mathrm{S}} = 12.5$ km s$^{-1}$}, {\it on the right} n=2 cm$^{-3}$, $\zeta$=0.5, $v_{\mathrm{S}} = 7.5$ \mbox{km s$^{-1}$}, and different resolutions: 512$^3$ (solid black line), 256$^3$ (dotted blue line) and 128$^3$ cells (dashed green line).}
  \label{fig:resol0212}
\end{figure*}

\begin{figure}
  \centering\resizebox{0.8\hsize}{!}{\includegraphics{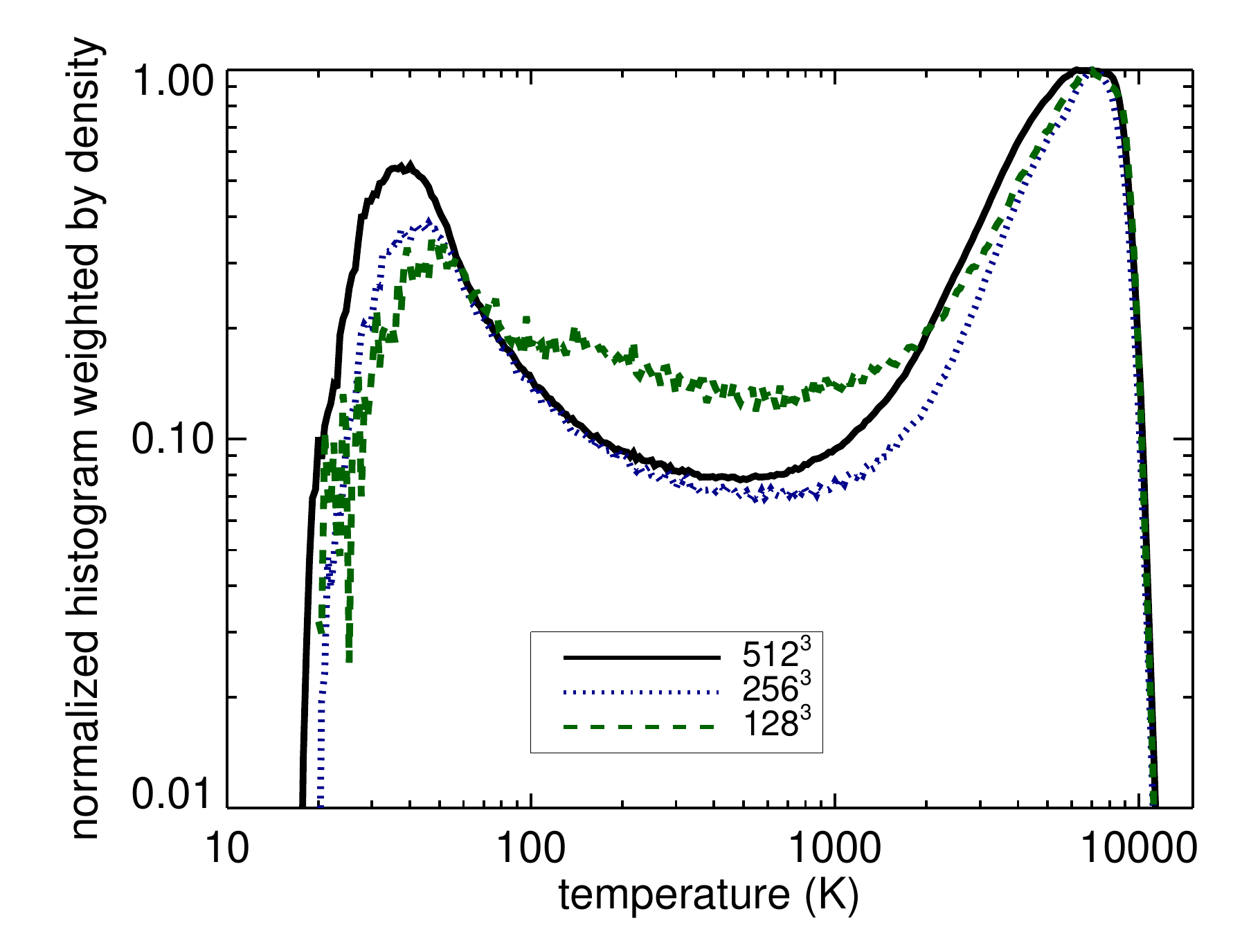}}
  \centering\resizebox{0.8\hsize}{!}{\includegraphics{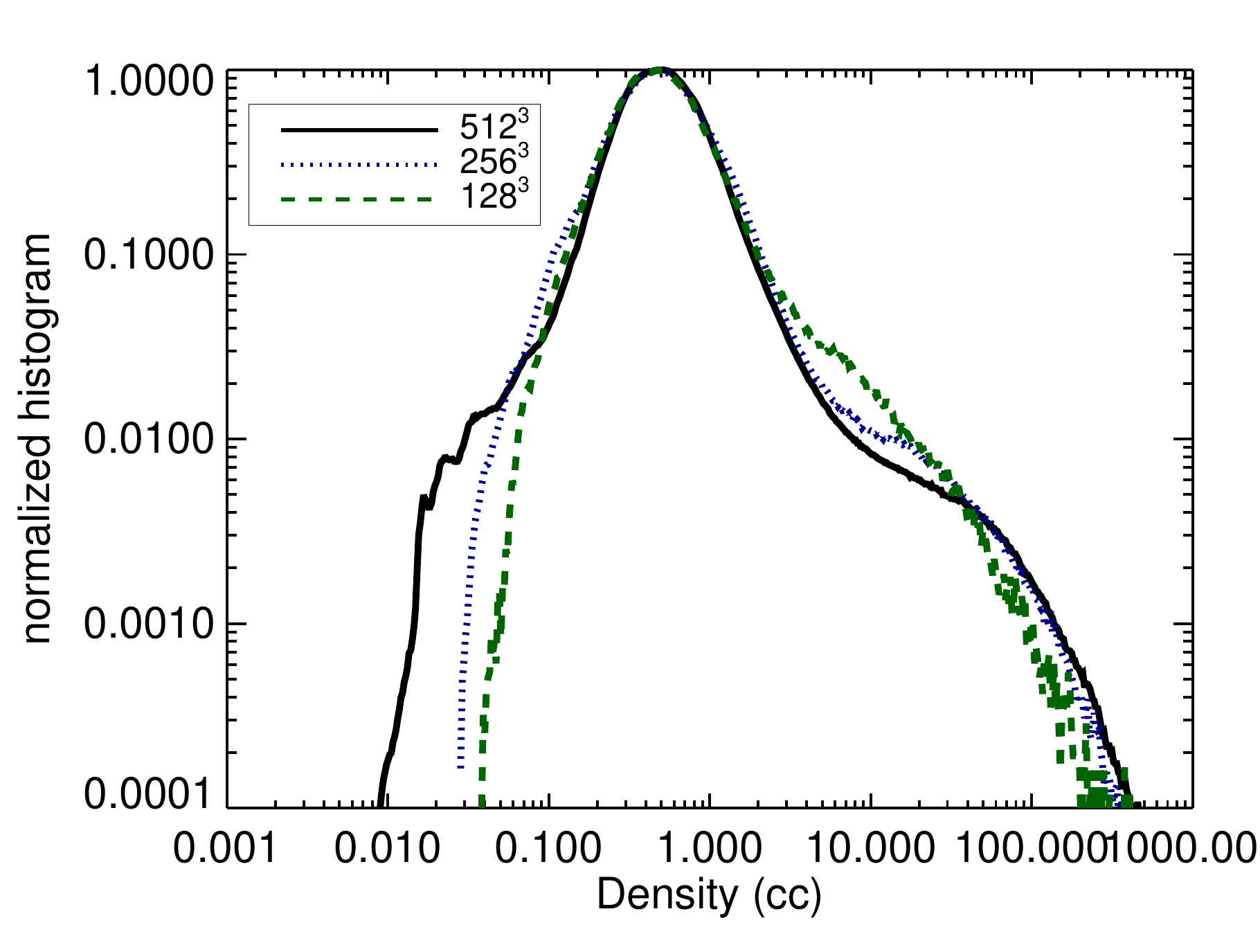}}
  \caption{Temperature histogram weighted by density ({\it on top}) and density pdf ({\it at the bottom}) for simulations with the following initial conditions: $n=1\,$cm$^{-3}$, $\zeta$=0.2, \mbox{$v_{\mathrm{S}} = 12.5$ km s$^{-1}$}, and different resolutions: 512$^3$ (solid black line), 256$^3$ (dotted blue line) and 128$^3$ cells (dashed green line) at 40~Myr.}
  \label{fig:historesol}
\end{figure}

\subsection{WNM turbulent velocity dispersion and Mach number} 
\label{sec:sigturb}

Apart from its density and temperature, the other relevant parameter for this study is the level of turbulence in the WNM, or equivalently its Mach number. The direct estimate of the Mach number of the WNM is difficult due to the fact that its temperature can not be easily measured. The traditional method to estimate the \hi\ temperature is based on a comparison of 21~cm absorption and emission measurements. Due to its low opacity, the warm phase produces almost undetectable absorption once observed in front of a strong radio-continuum source. Only a few detections were reported so far \citep[e.g.][]{carilli1998,kanekar2003,begum2010}, with deduced temperatures closer to the thermally unstable regime ($T\sim 500 - 6000\,$K) than to what is expected for the WNM ($T\sim 8000\,$K).

Nevertheless there are indirect evidences that the warm ISM is subsonic.
Recent findings on the warm ionized medium (WIM) by \citet{gaensler2011} confirmed the work of \citet{armstrong1995} and \citet{redfield2004} that showed that the WIM has the properties of a low Mach number and incompressible turbulent flow. 
Regarding the WNM, an indication of its low Mach number comes from the smoothness of the 21~cm profiles. \citet{haud2007} found that the simplest 21~cm spectra of the LAB data \citep{kalberla2005} at high Galactic latitudes are well modeled by a single Gaussian of $\sigma_{\mathrm{tot}}=10.2 \pm 0.3\,\mbox{km s}^{-1}$. Here we assume this value to be representative of the WNM total velocity dispersion in the Solar neighborhood.

The WNM velocity dispersion is the quadratic sum of the thermal ($\sigma_{\mathrm{therm}}$) and turbulent ($\sigma_{\mathrm{turb}}$) contribution to the gas motions, integrated along the line of sight.
Considering the specific line of sight at the Galactic pole, the thermal contribution to the line width of the WNM is an integral over Galactic height $z$: 
\begin{equation}
\label{eq:sigma_therm}
\sigma_{\mathrm{therm}} =  \left[ \frac{k}{m}\frac{\int_0^L n(z) T(z) dz}{\int_0^Ln(z)dz} \right]^{1/2}
\end{equation}
where $n(z)$ is assumed to be a Gaussian with HWHM=$265\,$pc \citep{dickey1990}. 
Using the corresponding $T(z)$ shown in Fig~\ref{fig:n_t_vs_z}, integrating Eq.~\ref{eq:sigma_therm} gives $\sigma_{\mathrm{therm}} = 8.3\,\mbox{km s}^{-1}$, a value essentially identical to the thermal width of an isothermal 8000~K gas ($\sigma_{\mathrm{therm}} = 8.1\,\mbox{km s}^{-1}$) which reflects the fact that the WNM temperature does not vary significantly with Galactic height.
Assuming $\sigma_{\mathrm{tot}} = 10.2\,\mbox{km s}^{-1}$ \citep{haud2007}, the contribution of turbulent motions 
to the observed line width is $\sigma_{\mathrm{turb}}=5.9\,\mbox{km s}^{-1}$.
The fact that $\sigma_{\mathrm{turb}} < \sigma_{\mathrm{therm}}$ is another indication that thermal motions dominates over turbulent motions at any scale in the Solar neighborhood.

Due to the nature of the turbulent cascade, the turbulent velocity dispersion increases with scale like 
$\sigma_{\mathrm{turb}} = \sigma(1)  l^{q}$, where $q\sim1/3$ for subsonic turbulence and, following the notation of \citet{wolfire2003},
$\sigma(1)$ is the velocity dispersion (in km$\,$s$^{-1}$) at a scale of $1\,$pc and $l$ is the scale (in pc).
Assuming that the turbulent contribution to the observed line width is representative of motions at the scale of the 
\hi\ scale height at the Sun location ($l=h_z=265\,$pc), one obtains $\sigma(1)=0.92\,\mbox{km s}^{-1}$.
This calculation is similar to the one of \citet{wolfire2003} who estimated $\sigma(1)=1.4\,\mbox{km s}^{-1}$ for the WNM.
The fact that we obtain a significantly lower value can be attributed mostly to the fact that we took into account the thermal 
contribution to the line width.

\subsection{CNM mass fraction and volume filling factor}

One important quantity we want to reproduce in the simulations is the amount of mass in the cold phase. 
The CNM mass fraction was estimated to 40~percent by \cite{heiles2003b}
by comparing 21~cm absorption and emission measurements on 79 random lines-of-sights. 
This is also compatible with the decomposition of the \hi\ density distribution with $z$ of \citet{dickey1990}; their narrow component ($h=106\,$pc) that can be associated to the CNM contributes to 44~percent of the total \hi\ surface density of the disk. Because the scale height of the CNM is smaller than that of the WNM, the CNM mass fraction varies with $z$. The CNM contributes to 60~percent of the \hi\ mass for $z<200\,$pc.

Like for the WNM one can compute the expected CNM volume density based on the heating and cooling processes \citep{wolfire2003}. For the average pressure of \citet{jenkins2011} ($P=3980\,$K$\,$cm$^{-3}$) the equilibrium density and temperature  are $n_{\mathrm{cnm}}=111\,$cm$^{-3}$, $T_{\mathrm{cnm}}=35.9\,$K. Following \cite{dickey1990}, the space-average density of the CNM is $\sim 0.4\,$cm$^{-3}$ at $z=0$, which implies a volume filling factor of the cold phase lower than 1\%.

\section{Methodology}
\label{sec:methodology}

Based on the observational constraints described in the previous section, we conclude that the WNM in the Solar neighborhood has a mean temperature $T\sim 8000\,$K, a mean density of $n \sim 0.2-0.5\,$cm$^{-3}$ and subsonic turbulent motions following $\sigma_{\mathrm{turb}} = 0.92 \, (l/\mathrm{1 pc})^{1/3}$ (km$\,$s$^{-1}$).
The first step of this study is to evaluate what are the plausible physical conditions that lead to the formation of CNM gas, of the order of 40 percent in mass, from WNM gas with such physical properties. 
We explored those conditions with 90 simulations at low resolution ($128^3$ cells). 
In this section we validate this choice by testing the use of $128^3$ simulations for the exploration of the parameter space by comparing with higher resolution ($256^3$ and $512^3$) simulations, after having described the characteristic scales involved in the simulations. 
Two sets of initial conditions that match the observations will then be used to produce high resolution simulations ($1024^3$).

\subsection{Characteristic scales}
\label{sec:scales}

\subsubsection{Dynamical time and cooling scale of the WNM}

If, during the perturbation of an initial WNM cloud, the cooling processes act slower than the compressive ones, the gas will tend to return to its initial state of temperature and density by decompression. The transition WNM-CNM becomes possible when the radiative time is lower than the dynamical time \citep{hennebelle1999}, allowing the condensed gas to keep its higher density even during the decompression:
\begin{equation}
t_{\mathrm{cool}} < t_{\mathrm{dyn}}.
\label{eq:tcooltdyn}
\end{equation}
The dynamical time is defined as the time needed by an atom to cross the cloud (of size $d$) at the sound speed velocity $c_{\mathrm{S}}$: 
\begin{equation}
t_{\mathrm{dyn}} = \frac{d}{c_{\mathrm{S}}}.
\end{equation}
The radiative time is the characteristic time needed by the medium to cool. It is thus equal to the ratio of the energy ($U$) and the energy transfer ($\epsilon$) applied to the medium: 
\begin{equation}
t_{\mathrm{cool}} = \frac{U}{\epsilon}. 
\label{eq:tcool}
\end{equation}

We thus can calculate the critical size below which the WNM will be unstable and will condense into the unstable regime or CNM structures. We consider typical conditions of the WNM (\mbox{$T=8000\,$K} and \mbox{$n=0.5\,$cm$^{-3}$}) and a static medium, assuming that the energy is fully thermal. Using equation \ref{eq:tcooltdyn} and the heat capacity per mass unity $C_{\mathrm{V}} = k_{\mathrm{B}}/m_{\mathrm{H}}(\gamma -1)$: 
\begin{eqnarray}
U & = & m_{\mathrm{H}}C_{\mathrm{V}}T \quad\textrm{the gas energy and}\\
\epsilon & = & n\Lambda \quad\textrm{the cooling applied to the WNM}
\end{eqnarray}
The cooling time thus becomes: 
\begin{equation}
t_{\mathrm{cool}} = \frac{k_{\mathrm{B}}T}{(\gamma -1)n\Lambda}.
\end{equation}
Considering the main cooling processes of the diffuse \hi\ described in Section \ref{sec:thermal} to evaluate $\Lambda$ and $c_{\mathrm{S}} = \sqrt{\gamma k_{\mathrm{B}}T/(1.4~m_{\mathrm{H}})}$, one obtains for the WNM cooling scale:
\begin{equation}
t_{\mathrm{cool}} < t_{\mathrm{dyn}} \Rightarrow d > \lambda_{\mathrm{cool}} = t_{\mathrm{cool}}\times c_{\mathrm{S,WNM}} \sim 22\,\mathrm{pc}.
\end{equation}
Thus, for the condition $t_{\mathrm{cool}} < t_{\mathrm{dyn}}$ to be observed and the thermal instability to develop, the initial scale $d$ of the perturbed zone must be greater than $\lambda_{\rm cool}\sim22$\,pc. Similarly, a perturbation of size lower than $\lambda_{\rm cool}$ do not efficiently trigger the phase transition.

\subsubsection{Sizes of the simulations}

To study the WNM-CNM transition it is important to evaluate the scales of cold structures to adjust the numerical resolution. 
To first order one can estimate the size of a cold structure resulting from the condensation of a WNM cloud with a typical size of 22\,pc. Because the CNM density is a hundred times higher than the WNM density, the CNM structures should be a hundred times smaller than the initial WNM cloud, leading to a rough size of 0.2\,pc.
In fact, as illustrated by \cite{hennebelle2007a}, the size of the CNM clumps follow a distribution with a maximal size of 0.6\,pc. Moreover, the Field length $\lambda_{\mathrm{F}}$ \citep{field1965,begelman1990}, which give the typical scale of the thermal front between the WNM and the CNM, and the front shock in the CNM are of the order of 10$^{-3}$\,pc. 
In principle, simulations of the thermally bistable \hi\ require a scale range from $10^{-3}$ to \mbox{$\sim20$ pc} but, as suggested by \cite{hennebelle2007a}, the important scale to resolve in simulations of thermally bistable and turbulent \hi\ is the 'sound crossing scale' which is the sound speed times the cooling time. As mentioned earlier, for typical values of the density, this scale is $\sim 22\,$pc and $\sim 0.2\,$pc in the WNM and CNM respectively. 

We chose to assure the phase transition with a box size equal to $40\,$pc and therefore larger than $\lambda_{\mathrm{cool}}$. 
In this study we used first a box size of $128^3$, leading to a resolution of 0.3 pc. Even though we barely resolve the
typical cold structure scale, using higher resolution simulations we will demonstrate that this set up can
be used to explore efficiently the parameter space for the formation of the CNM. The second part of the results presented in this paper is dedicated to the detailed study of $1024^3$ simulations (resolution of 0.04 pc) for specific initial conditions. 

\subsection{Effect of resolution}
\label{sec:resol}

\begin{figure*}[t]
  \centering\resizebox{0.40\hsize}{!}{\includegraphics{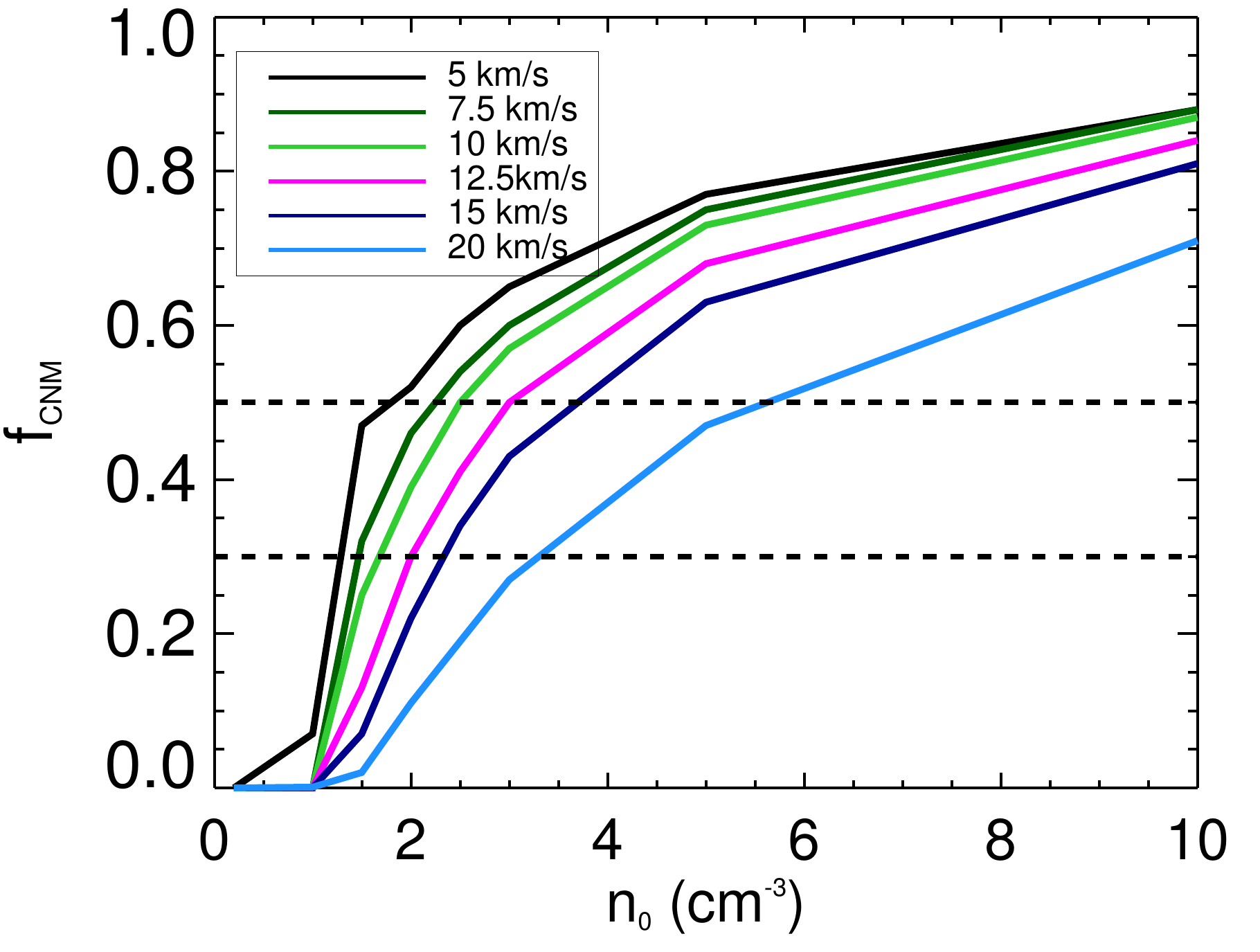}}
  \centering\resizebox{0.40\hsize}{!}{\includegraphics{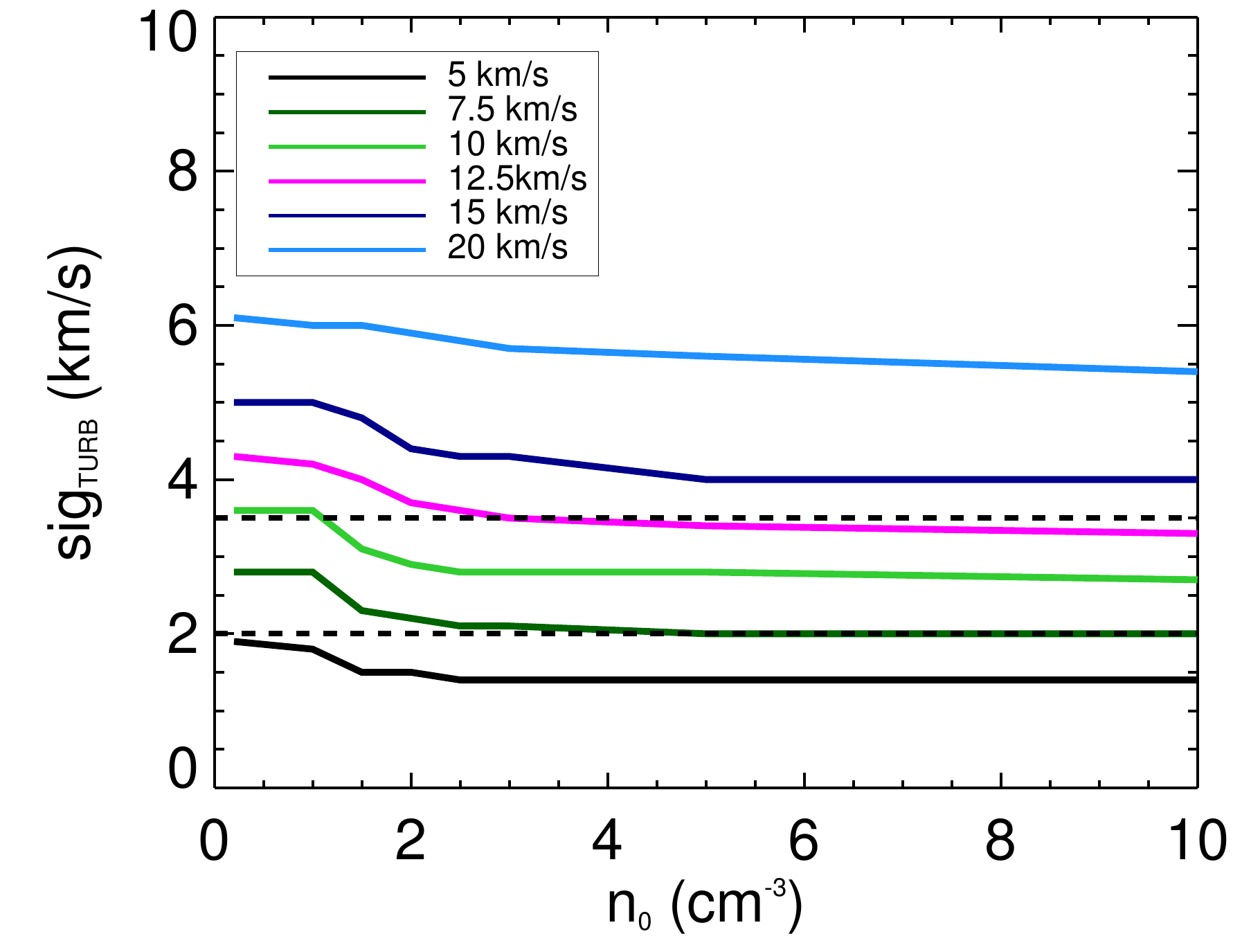}}
  \caption{\label{fig:setpw05} Behavior of \fcnm\ ({\it on the left}) and \sigturb\ ({\it on the right}) with the variation of the initial condition $n_0$, for the different values of the large scale velocity \vs\, (from 5 to 20\,\kms). The dotted lines frame the observational values of \fcnm\ and \sigturb\ that we want to reproduce.}
\end{figure*}

\begin{figure*}[t]
  \centering\resizebox{0.24\hsize}{!}{\includegraphics{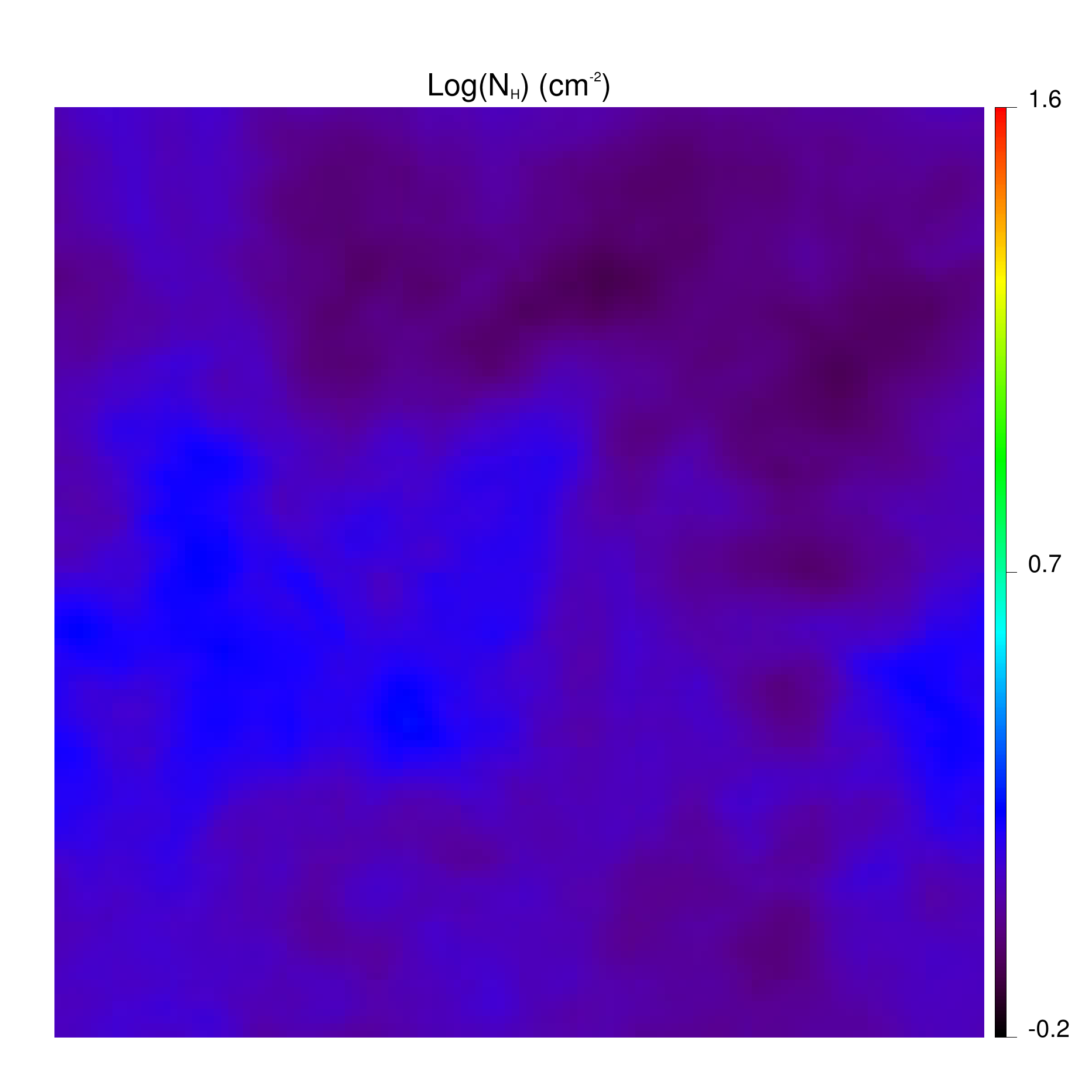}}
  \centering\resizebox{0.24\hsize}{!}{\includegraphics{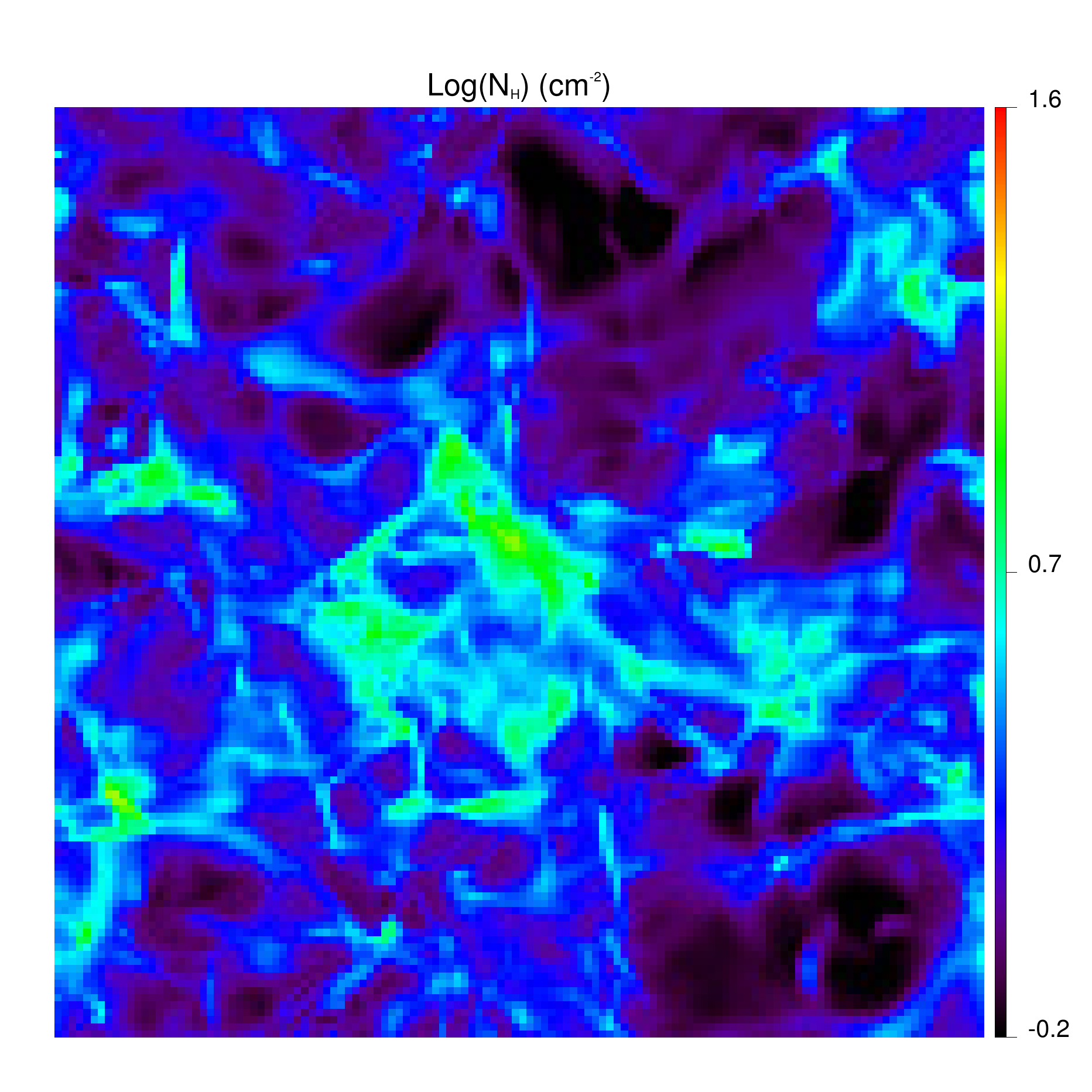}}
  \centering\resizebox{0.24\hsize}{!}{\includegraphics{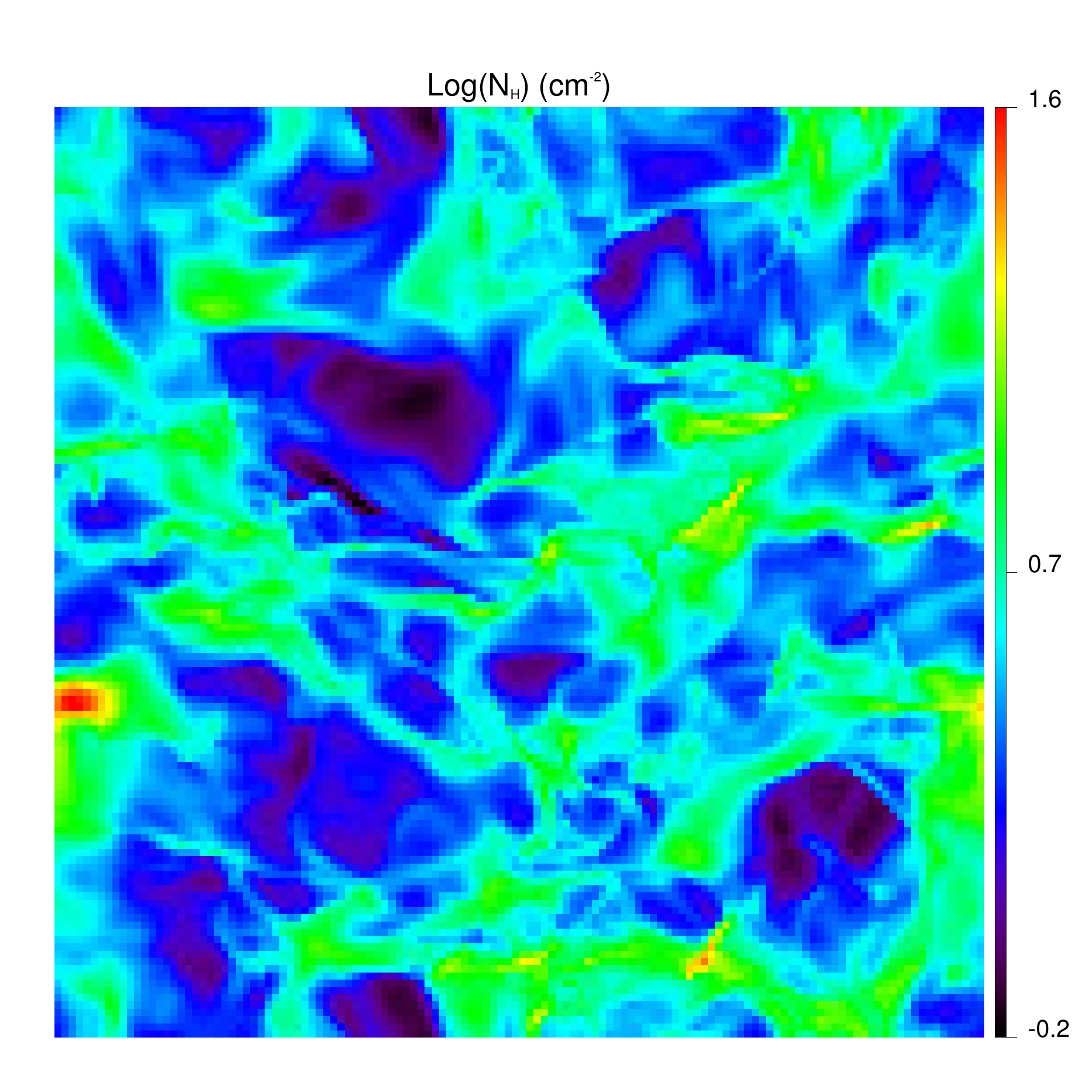}}
  \centering\resizebox{0.24\hsize}{!}{\includegraphics{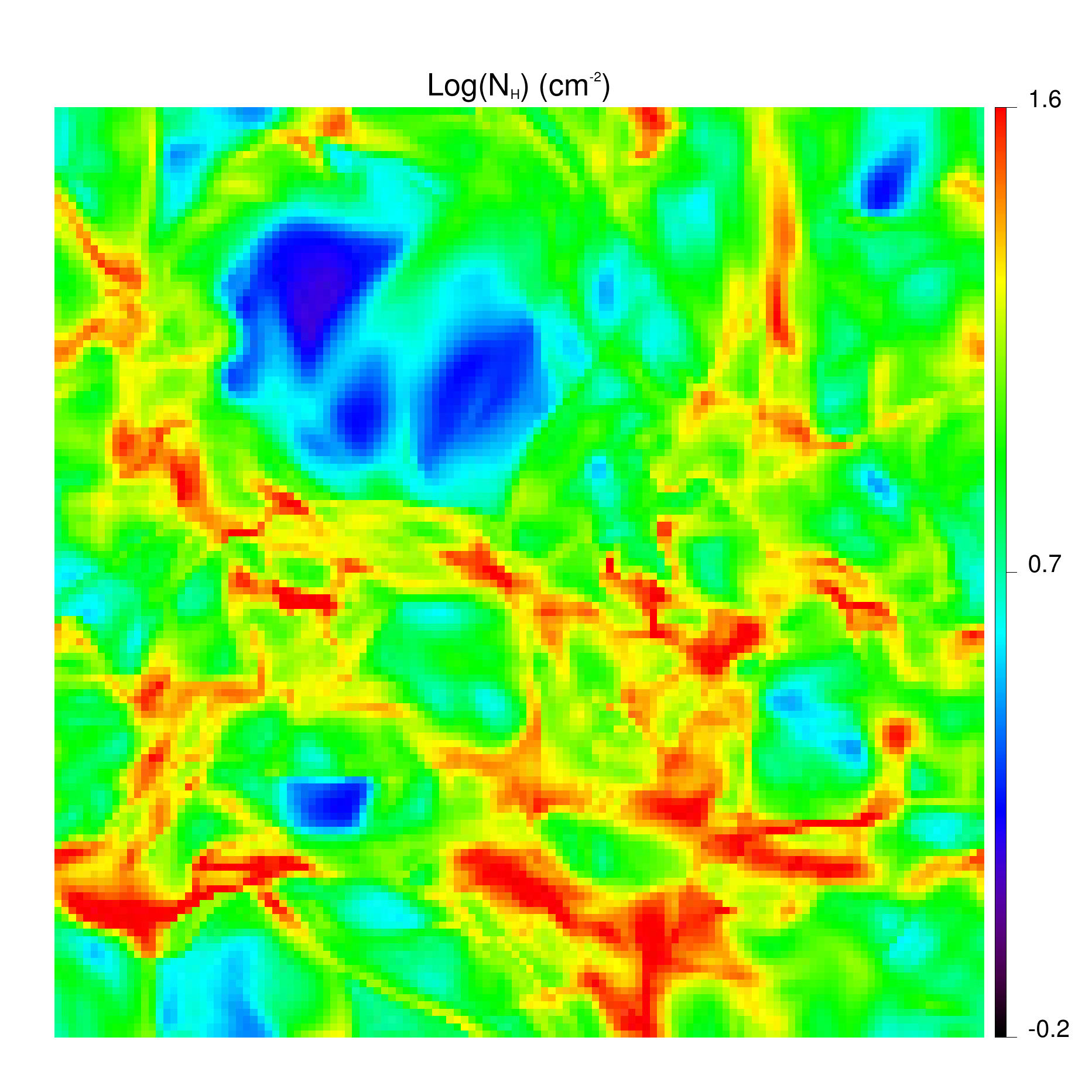}}
  \caption{\label{fig:cartespw05} Maps of the logarithm of the column density (density integrated along the direction $z$). The initial density increases from the {\it left} to the {\it right} and takes the following values: 1.0, 1.5, 3.0, 10.0\,\cc. The projection weight and the large scale velocity are fixed to $\zeta=0.5$ et \vs\,=\,7.5\,\kms. All maps are displayed with the same column density range [0.6\,10$^{20}$\,cm$^{-2}$, 40\,10$^{20}$\,cm$^{-2}$].} 
\end{figure*}

\begin{figure}[h]
  \resizebox{0.85\hsize}{!}{\includegraphics{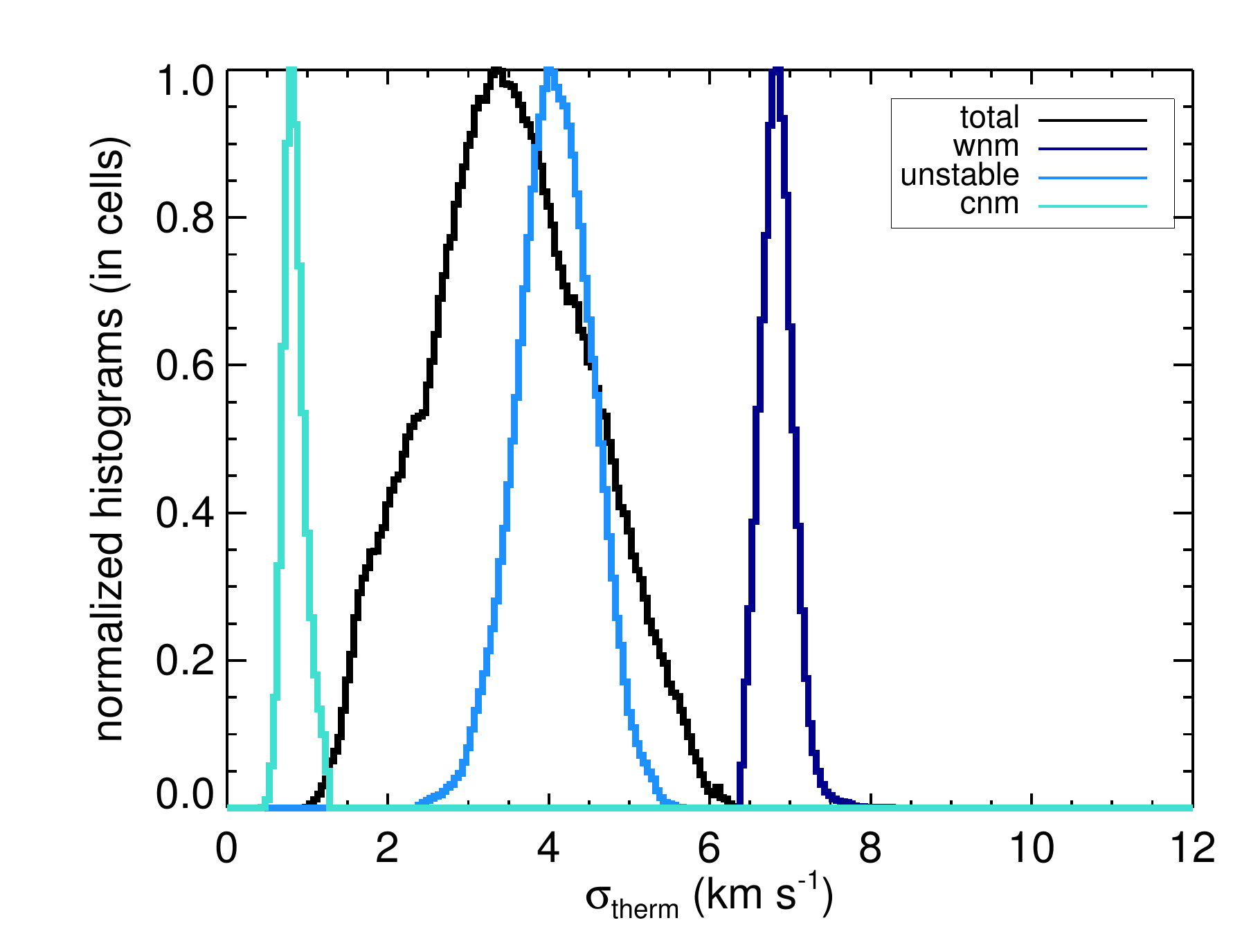}}
  \resizebox{0.85\hsize}{!}{\includegraphics{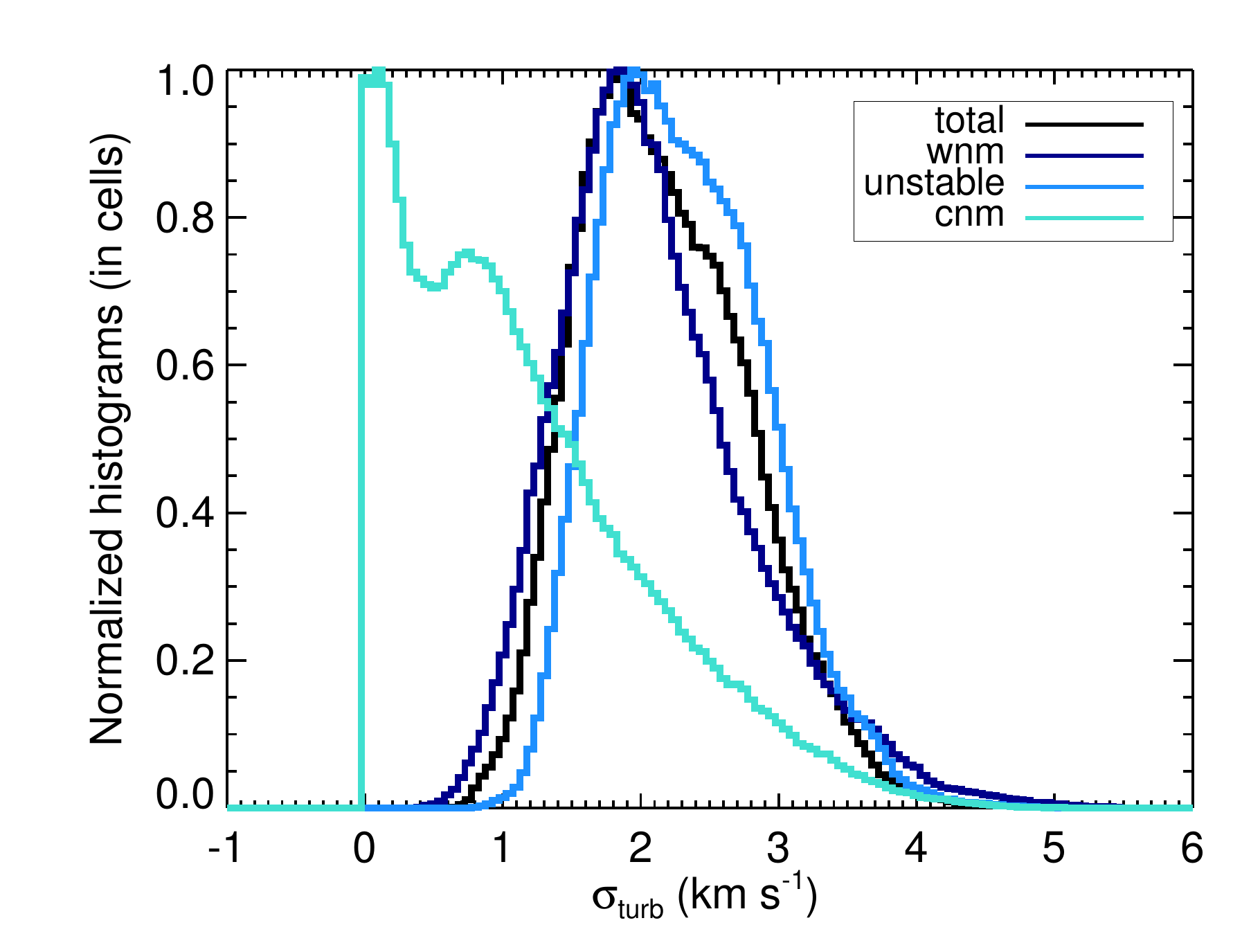}}
  \resizebox{0.85\hsize}{!}{\includegraphics{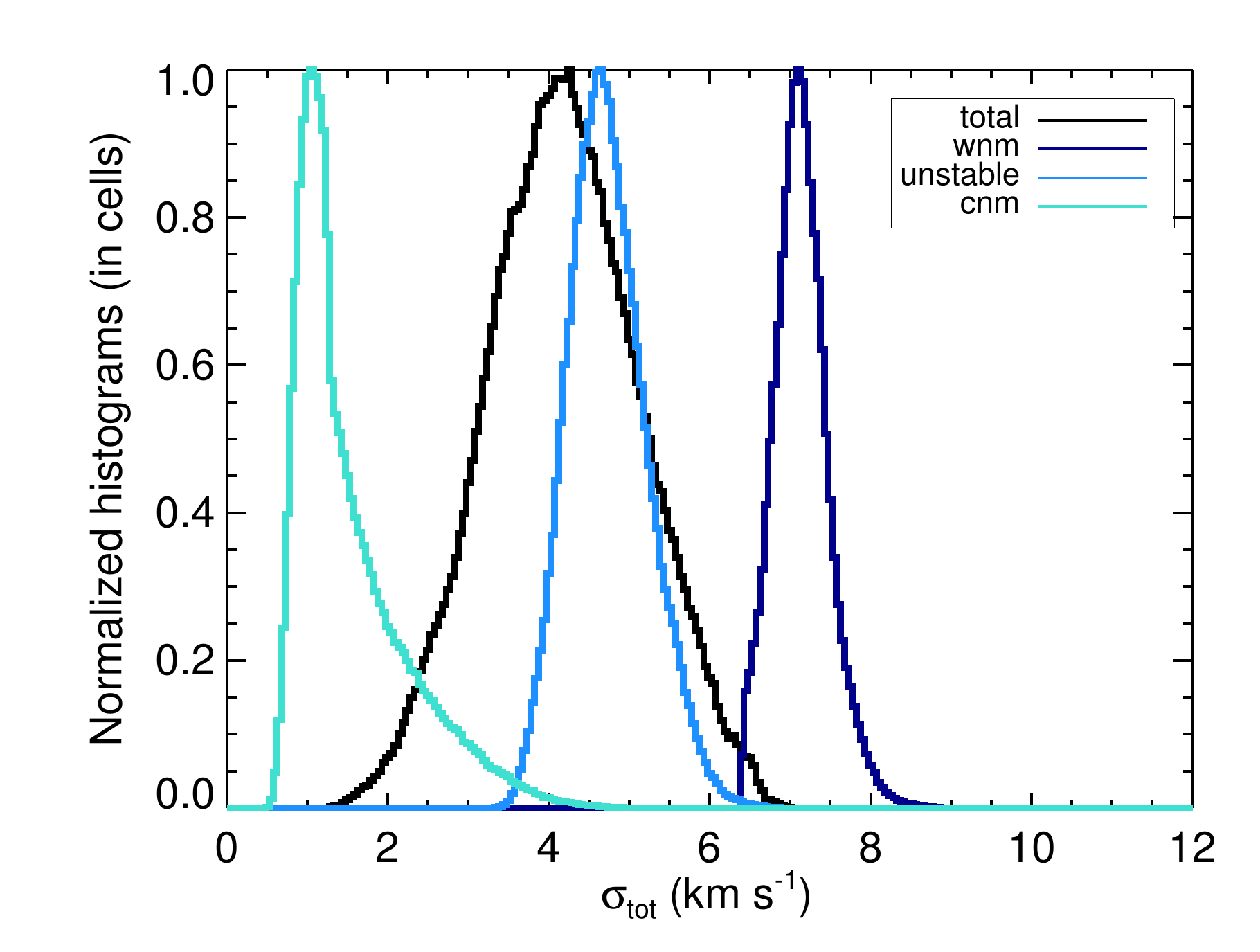}}
  \caption{\label{fig:histosigmas} Velocity dispersions histograms: {\it on top} \sigtherm, {\it in the middle} \sigturb\ and {\it at the bottom} \sigtot, for different temperature ranges: {\it in black} the all box, {\it in dark blue} the WNM ($T>5000$~K), {\it in blue} the unstable gas ($200\,K<T<5000$\,K) and {\it in light blue} the CNM ($T<200$\,K) for the following high resolution simulation: 1024$^3$, $n_0$=2.0~\cc, $\zeta=0.5$ et \vs=7.5~\kms.}
\end{figure}
\begin{figure}
    \resizebox{0.9\hsize}{!}{\includegraphics{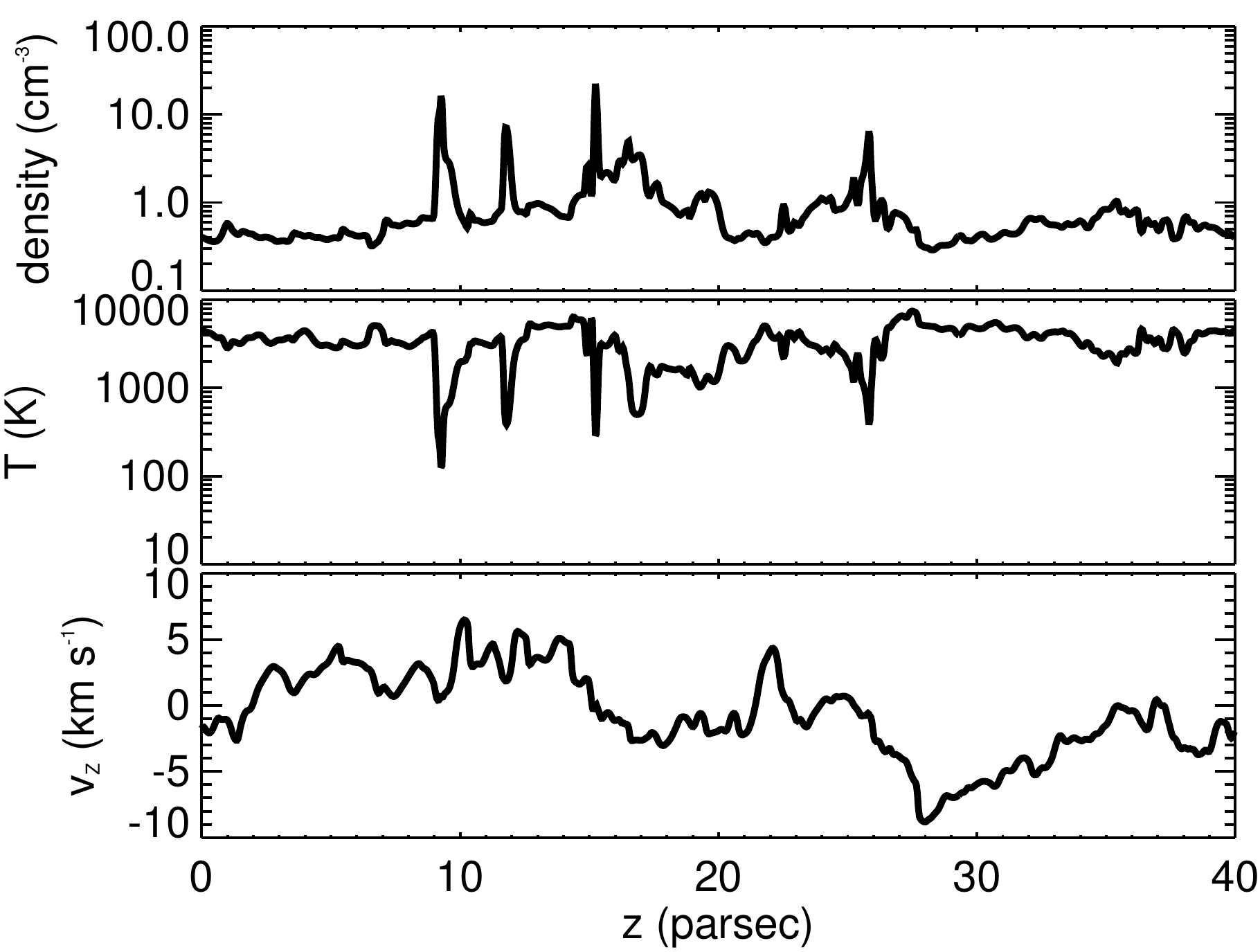}}
    \resizebox{0.9\hsize}{!}{\includegraphics{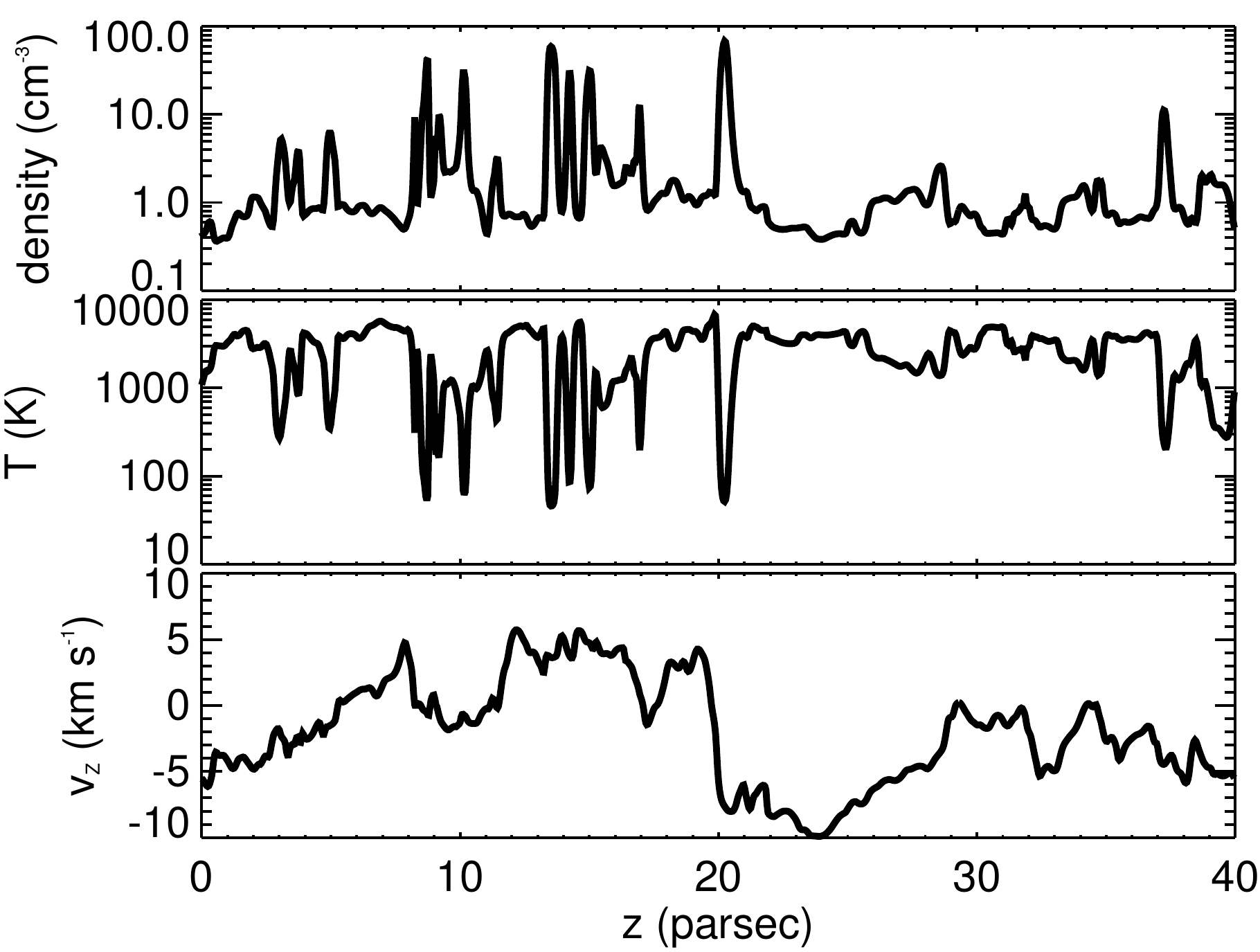}}  
    \caption{\label{fig:lofs128} Density, temperature and velocity ($z$-component) profiles for a given simulation (1024$^3$, $n_0$=2.0~\cc, $\zeta=0.5$ et \vs=7.5~\kms) and for two distinct lines of sight characterized by their value of \sigturb(CNM). {\it On top}: \sigturb(CNM)=0.14~\kms\ and {\it at the bottom}: \sigturb(CNM)=4.7~\kms.}
\end{figure}

Here we validate the use of low resolution simulations in the context of a parameter exploration study. 
Figure \ref{fig:resol0212} shows the evolution over time of the CNM massive
fraction, the dispersion of the line of sight turbulent velocity
\mbox{$\sigma_{\mathrm{turb}}$} and the observational Mach number for two representative simulations
([\mbox{n=1 cm$^{-3}$}, \mbox{$\zeta$=0.2}, \mbox{$v_{\mathrm{S}} = 12.5$ km s$^{-1}$}]
and [\mbox{n=2 cm$^{-3}$}, \mbox{$\zeta$=0.5}, \mbox{$v_{\mathrm{S}} = 7.5$ km s$^{-1}$}]) done on $128^3$, $256^3$ and $512^3$ grids.

We calculate the cold mass fraction $f_{\mathrm{CNM}}$ as the sum of the density in the pixels where \mbox{$T < 200$ K} over the total density in the box.
The turbulent velocity dispersion is computed the following way \citep[see][equation 8]{mamd2007}: 
\begin{equation}
\label{eq:sigturb}
\sigma_{\mathrm{turb}}^2(x,y) = \frac{\sum_{z'=0}^{Lbox} n(x,y,z')v_z^2(x,y,z')}{\sum_{z'=0}^{Lbox}n(x,y,z')} - C^2(x,y)
\end{equation}
where $C(x,y)$ is the velocity centroid map:
\begin{equation}
C(x,y) = \frac{\sum_{z'=0}^{Lbox}n(x,y,z')v_z(x,y,z')}{\sum_{z'=0}^{Lbox}n(x,y,z')}.
\end{equation}

In numerical simulations, the Mach number is often computed as the ratio of the velocity and the sound speed on each cell:
\begin{equation}
\mathcal{M}_{\mathrm{theo}} = <|\mathbf{u}|/c_{\mathrm{s}}>_{x,y,z}
\end{equation}
where $c_{\mathrm{S}} = \sqrt{(\gamma P(x,y,z))/\rho(x,y,z)}$.
Because this parameter can not be directly deduced from observations 
we consider an expression for the  Mach number calculated from integrated quantities along the line of sight: the turbulent velocity dispersion and the thermal velocity dispersion: 
\begin{equation}
\label{eq:machobs}
\mathcal{M}_{\mathrm{obs}} = <\frac{\sigma_{\mathrm{turb}}}{\sigma_{\mathrm{therm}}}>_{x,y}
\end{equation}
where the thermal velocity dispersion is defined as 
\begin{equation}
\label{eq:sigtherm}
\sigma_{\rm therm}^2 = \frac{k_{\rm B}}{m_{\rm H}} \frac{\sum_{z'=0}^{Lbox}n(x,y,z')T(x,y,z')}{\sum_{z'=0}^{Lbox}n(x,y,z')}.
\end{equation}
Obviously the total\footnote{or observed in the case of an optically thin line} velocity dispersion is simply
\begin{equation}
\label{eq:sigtot}
\sigma_{\rm tot}^2 = \sigma_{\rm{turb}}^2 + \sigma_{\rm therm}^2.
\end{equation}

All quantities shown in Fig~\ref{fig:resol0212} are spaced averaged over the box at different time steps. 
The mean values of $\sigma_{\mathrm{turb}}$ and $\mathcal{M}_{\mathrm{obs}}$ are in the same range, and their variations are of the same order, as illustrated in Figure \ref{fig:resol0212}. These two quantities are thus statistically well represented even in small resolution simulations.

The temperature and density pdfs (Fig.~\ref{fig:historesol}) and time evolution (Fig.~\ref{fig:resol0212}) show that the ratio of cold gas depends on the numerical resolution. Higher resolution runs lead to a slightly higher amount of cold gas. The cold peak of the temperature histogram (Fig.\,\ref{fig:historesol}) is lower in the $128^3$ and $256^3$ cases. The time evolution of $f_{\mathrm{CNM}}$ is increasing slower for the smallest resolution run and $f_{\mathrm{CNM}}$ is then lower. However the massive fractions of gas with \mbox{$T < 200$K} reached at \mbox{40 Myr} are similar (0.27, 0.31 and 0.33). We then deduce that the estimation of $f_{\mathrm{CNM}}$ on $128^3$ simulations is still worthwhile if we remember that it increases slightly with resolution.

The same resolution study on simulations with different initial conditions presents similar results. Like \citet{seifried2011}, we conclude that the resolution does not affect the time evolution of $\sigma_{\mathrm{turb}}$ and $\mathcal{M}_{\mathrm{obs}}$ and has a moderate effect on the formation of CNM. Therefore doing a statistical study on small simulations ($128^3$) is legitimate. Higher resolution simulations are needed to study in details the CNM properties, especially its structure.

\subsection{Convergence}

In order to evaluate the time needed for the $128^3$ simulations to reach a stationary state, we evaluated $\sigma_{\mathrm{turb}}$, $\mathcal{M}_{\mathrm{obs}}$ and $f_{\mathrm{CNM}}$ at each Myr. Most simulations converge in \mbox{10 Myr} but some needed a much longer time to transit to cold phase and never achieve a stationary state for the massive fraction of CNM. However we choose to stop all simulations at \mbox{40\,Myr} because of the typical time between two supernovae explosions estimated to be around \mbox{20 Myr} \citep[see][equation 17 at $z=0$ in a volume of \mbox{(40 pc)$^3$}]{ferriere2001}.

\section{Parametric study}

\label{sec:parametric_study}

According to Fig.~\ref{fig:n_t_vs_z}, we can consider constant $n$ and $T$ as adequate initial conditions for a box size of 40$\,$pc 
centered on $z=0$. For all simulations the initial conditions are a uniform and static WNM at a temperature of \mbox{$T = 8000$\,K}. 
The boundary conditions are periodic. The velocity field is modified at every time step by the turbulent stirring in Fourier space.

This section presents the parameter study we did on the initial density and the two properties of the stirring velocity field: the spectral weight $\zeta$ and the large scale velocity $v_{\mathrm{S}}$. We performed 90 simulations with 128$^3$ cells, varying the density from its fiducial WNM value 0.2 to \mbox{10 cm$^{-3}$}, the velocity at large scale $v_{\mathrm{S}}$ from 5 to \mbox{20 km s$^{-1}$} and the spectral weight from 0 (rotation free field) to 1 (divergence free). We computed for each simulation the CNM mass fraction $f_{\mathrm{CNM}}$, the mean Mach number \machobs\ (Eq.~\ref{eq:machobs}) and the mean velocity dispersion (Eq.~\ref{eq:sigturb}) at each Myr and averaged them over time after a stationary state has been reached, that is on the time interval [20\,Myr,40\,Myr].

\subsection{Effect of the initial density increase}

\begin{table*}
\caption{Results of the 128$^3$ simulations with natural turbulence \mbox{$\zeta$=0.5} : the two first columns are the initial density and the large scale velocity of the turbulent stirring, the last five are the obtained results: average theoretical and observational Mach numbers, average and standard deviation of the cold massive fraction, and turbulent velocity dispersion of the line-of-sight velocity component. The average and standard deviation were computed between 20 and 40 Myrs.}
\label{tab:setpw05}
\centering
\begin{tabular}{c c|c c c c c}
\hline\hline
$n_0$ & \vs      & $<\mathcal{M}_{\mathrm{theo}}>$ & $<\mathcal{M}_{\mathrm{obs}}>$  & $<f_{\mathrm{CNM}}>$ & $\sigma(f_{\mathrm{CNM}})$ &$<\sigma_{\mathrm{turb}}>$\\
(\mbox{cm$^{-3}$}) & (\mbox{km s$^{-1}$})  &      &         &                               &    & (km s$^{-1}$)          \\
\hline\hline
0.2 & 20   & 1.8 & 0.8 & 0.0 & 0.0 & 6.1 \\
0.2 & 15   & 1.3 & 0.6 & 0.0 & 0.0 & 5.0 \\
0.2 & 12.5 & 1.1 & 0.5 & 0.0 & 0.0 & 4.3 \\
0.2 & 10   & 0.8 & 0.4 & 0.0 & 0.0 & 3.6 \\
0.2 & 07.5 & 0.6 & 0.3 & 0.0 & 0.0 & 2.8 \\
0.2 & 05   & 0.4 & 0.2 & 0.0 & 0.0 & 1.9 \\
\hline
1.0 & 20   & 1.9 & 0.8 & 0.002 & 0.001& 6.0\\
1.0 & 15   & 1.4 & 0.7 & 0.000 & 0.00 & 5.0\\
1.0 & 12.5 & 1.2 & 0.6 & 0.000 & 0.00 & 4.2\\
1.0 & 10   & 0.9 & 0.5 & 0.000 & 0.00 & 3.6\\
1.0 & 07.5 & 0.7 & 0.4 & 0.000 & 0.00 & 2.8\\
1.0 & 05   & 0.5 & 0.3 & 0.07  & 0.03 & 1.8\\
\hline
1.5 & 20   & 2.0 & 0.9 & 0.02 & 0.02 & 6.0\\
1.5 & 15   & 1.6 & 0.7 & 0.07 & 0.02 & 4.8\\
1.5 & 12.5 & 1.5 & 0.7 & 0.13 & 0.02 & 4.0\\
1.5 & 10   & 1.3 & 0.7 & 0.25 & 0.02 & 3.1\\
1.5 & 07.5 & 1.1 & 0.5 & 0.32 & 0.01 & 2.3\\
1.5 & 05   & 0.7 & 0.4 & 0.47 & 0.02 & 1.5\\
\hline
2.0 & 20   & 2.3 & 1.0 & 0.11 & 0.02 & 5.9\\
2.0 & 15   & 2.0 & 0.9 & 0.22 & 0.02 & 4.4\\
2.0 & 12.5 & 1.9 & 0.9 & 0.30 & 0.02 & 3.7\\
2.0 & 10   & 1.6 & 0.8 & 0.39 & 0.02 & 2.9\\
2.0 & 07.5 & 1.3 & 0.7 & 0.46 & 0.02 & 2.2\\
2.0 & 05   & 0.9 & 0.5 & 0.52 & 0.02 & 1.5\\
\hline
2.5 & 20   & 2.6 & 1.1 & 0.19 & 0.02 & 5.8\\
2.5 & 15   & 2.3 & 1.0 & 0.34 & 0.03 & 4.3\\
2.5 & 12.5 & 2.2 & 1.0 & 0.41 & 0.02 & 3.6\\
2.5 & 10   & 1.9 & 1.0 & 0.50 & 0.01 & 2.8\\
2.5 & 07.5 & 1.5 & 0.8 & 0.54 & 0.02 & 2.1\\
2.5 & 05   & 1.0 & 0.5 & 0.60 & 0.01 & 1.4\\
\hline
3.0 & 20   & 2.9 & 1.2 & 0.27 & 0.03 & 5.7\\
3.0 & 15   & 2.7 & 1.2 & 0.43 & 0.02 & 4.3\\
3.0 & 12.5 & 2.4 & 1.2 & 0.50 & 0.02 & 3.5\\
3.0 & 10   & 2.1 & 1.1 & 0.57 & 0.02 & 2.8\\
3.0 & 07.5 & 1.6 & 0.9 & 0.60 & 0.01 & 2.1\\
3.0 & 05   & 1.1 & 0.6 & 0.65 & 0.01 & 1.4\\
\hline
5.0 & 20   & 3.8 & 1.6 & 0.47 & 0.02 & 5.6\\
5.0 & 15   & 3.6 & 1.7 & 0.63 & 0.02 & 4.0\\
5.0 & 12.5 & 3.3 & 1.7 & 0.68 & 0.01 & 3.4\\
5.0 & 10   & 2.7 & 1.5 & 0.73 & 0.01 & 2.8\\
5.0 & 07.5 & 2.1 & 1.2 & 0.75 & 0.01 & 2.0\\
5.0 & 05   & 1.4 & 0.8 & 0.77 & 0.01 & 1.4\\
\hline
10.0 & 20   & 5.6 & 2.4 & 0.71 & 0.01  & 5.4\\
10.0 & 15   & 4.9 & 2.6 & 0.81 & 0.01  & 4.0\\
10.0 & 12.5 & 4.4 & 2.5 & 0.84 & 0.008 & 3.3\\
10.0 & 10   & 3.6 & 2.2 & 0.87 & 0.005 & 2.7\\
10.0 & 07.5 & 2.8 & 1.7 & 0.88 & 0.004 & 2.0\\
10.0 & 05   & 1.9 & 1.2 & 0.88 & 0.003 & 1.4\\
\hline
\end{tabular}
\end{table*}

\begin{table*}
\caption{Results of the 128$^3$ simulations with initial density \mbox{n=1 cm$^{-3}$} : the first two columns are the turbulence parameters (solenoidal over compressible modes rate and amplitude of the turbulent forcing), the next five are the obtained results: average theoretical and observational Mach numbers, average and standard deviation of the cold massive fraction and turbulent velocity dispersion of the line-of-sight velocity component. The average and standard deviation were computed between 20 and 40 Myrs.}
\label{tab:setn01}
\centering
\begin{tabular}{c c|c c c c c}
\hline\hline
$\zeta$ & \vs      & $<\mathcal{M}_{\mathrm{theo}}>$ & $<\mathcal{M}_{\mathrm{obs}}>$  & $<f_{\mathrm{CNM}}>$ & $\sigma(f_{\mathrm{CNM}})$ &$<\sigma_{\mathrm{turb}}>$\\
           & (\mbox{km s$^{-1}$})  & &    &                               &    & (km s$^{-1}$)          \\
\hline\hline
0.0 & 20   & 1.8 & 0.7 & 0.39 & 0.05 & 4.1\\
0.0 & 15   & 1.1 & 0.5 & 0.45 & 0.03 & 2.9\\
0.0 & 12.5 & 0.8 & 0.4 & 0.42 & 0.03 & 2.3\\
0.0 & 10   & 0.6 & 0.3 & 0.44 & 0.03 & 1.8 \\
0.0 & 07   & 0.4 & 0.2 & 0.49 & 0.04 & 1.1\\
0.0 & 05   & 0.2 & 0.16& 0.50 & 0.07 & 0.7\\
\hline
0.1 & 20   & 1.7 & 0.7 & 0.36 & 0.04 & 4.0\\
0.1 & 15   & 1.1 & 0.5 & 0.37 & 0.04 & 3.0\\
0.1 & 12.5 & 0.8 & 0.4 & 0.36 & 0.02 & 2.4\\
0.1 & 10   & 0.6 & 0.3 & 0.42 & 0.04 & 1.8\\
0.1 & 07   & 0.4 & 0.2 & 0.45 & 0.05 & 1.2\\
0.1 & 05   & 0.2 & 0.14& 0.46 & 0.09 & 0.8\\
\hline
0.2 & 20   & 1.7 & 0.7 & 0.24 & 0.03 & 4.2\\
0.2 & 15   & 1.1 & 0.5 & 0.26 & 0.03 & 3.3\\
0.2 & 12.5 & 1.0 & 0.45& 0.28 & 0.03 & 2.7\\
0.2 & 10   & 0.7 & 0.4 & 0.30 & 0.03 & 2.2\\
0.2 & 07   & 0.4 & 0.2 & 0.38 & 0.02 & 1.5\\
0.2 & 05   & 0.3 & 0.17& 0.42 & 0.05 & 1.1\\
\hline
0.3 & 20   & 1.7 & 0.7 & 0.08 & 0.02 & 4.8\\
0.3 & 15   & 1.2 & 0.6 & 0.11 & 0.02 & 3.8\\
0.3 & 12.5 & 1.0 & 0.5 & 0.13 & 0.03 & 3.3\\
0.3 & 10   & 0.7 & 0.4 & 0.12 & 0.04 & 2.6\\
0.3 & 07   & 0.6 & 0.3 & 0.25 & 0.03 & 2.0\\
0.3 & 05   & 0.4 & 0.2 & 0.22 & 0.05 & 1.3\\
\hline
0.4 & 20   & 1.8 & 0.8 & 0.02  & 0.02 & 5.5\\
0.4 & 15   & 1.3 & 0.6 & 0.02  & 0.02 & 4.5\\
0.4 & 12.5 & 1.1 & 0.5 & 0.002 & 0.001& 3.8\\
0.4 & 10   & 0.8 & 0.4 & 0.001 & 0.001& 3.2\\
0.4 & 07   & 0.6 & 0.3 & 0.03  & 0.01 & 2.2\\
0.4 & 05   & 0.5 & 0.2 & 0.13  & 0.05 & 1.5\\
\hline
0.5 & 20   & 1.9 & 0.8 & 0.002 & 0.001& 6.0\\
0.5 & 15   & 1.4 & 0.7 & 0.000 & 0.00 & 5.0\\
0.5 & 12.5 & 1.2 & 0.6 & 0.000 & 0.00 & 4.2\\
0.5 & 10   & 0.9 & 0.5 & 0.000 & 0.00 & 3.6\\
0.5 & 07   & 0.7 & 0.4 & 0.000 & 0.00 & 2.6\\
0.5 & 05   & 0.5 & 0.3 & 0.07  & 0.03 & 1.8\\
\hline
1.0 & 20   & 2.7 & 1.0 & 0.000 & 0.00& 8.3\\
1.0 & 15   & 2.0 & 0.9 & 0.000 & 0.00& 6.7\\
1.0 & 12.5 & 1.6 & 0.8 & 0.000 & 0.00& 5.8\\
1.0 & 10   & 1.3 & 0.6 & 0.000 & 0.00& 4.8\\
1.0 & 07   & 0.9 & 0.5 & 0.000 & 0.00& 3.6\\
1.0 & 05   & 0.7 & 0.4 & 0.000 & 0.00& 2.7\\
\hline
\end{tabular}
\end{table*}

Table~\ref{tab:setpw05} and Figure \ref{fig:setpw05} present the results for the 48 simulations for which the turbulence has been kept to a natural state, meaning that the spectral weight is fixed to \mbox{$\zeta = 0.5$} and that the Helmholtz projection operator (Eq~\ref{eq:helmoltz}) is equal to half the identity and therefore keeps the rate of compressible over solenoidal modes of the velocity field in its initial state. In this case, we only vary the large scale velocity $v_{\mathrm{S}}$ and the density. Obviously, as the initial temperature is 8000~K for each simulation, increasing the initial density is equivalent to increasing the initial pressure.

We first observe that simulations with initial values of the WNM density between 0.2 and 1.0~cm$^{-3}$ do not transit to cold gas. At larger densities, the efficiency of the CNM formation increases rapidly with the density. For $n_0=0.2$ or $1.0\,$cm$^{-3}$, the cold mass fraction is indeed very close to zero while a small increase of the density ($n_0=1.5$\,\cc) leads to \fcnm$=0.3-0.4$ and densities greater then 3.0\,\cc\ allow more than half of the mass to transit (see Fig.\ref{fig:setpw05} left part). This behavior is also well represented by the differences regarding the structures on the integrated density maps (Figure\,\ref{fig:cartespw05}). The map on the left, that only contains WNM,  is completely smooth while the structures appear clearly when the initial density increases.
 One can notice that the equivalent pressure at $n_0=1.5$\,\cc\ is equal to 12000\,K\,\cc, much larger than the maximum pressure allowed for the WNM (see Sec.~\ref{sec:thermal}). The gas starts in the thermally unstable regime; at these values of density and pressure the gas is indeed located above the stable branch of the WNM and thus in a cooling zone that allows it to evolve rapidly and efficiently towards the cold gas. 

\begin{figure*}[t]
  \centering\resizebox{0.40\hsize}{!}{\includegraphics{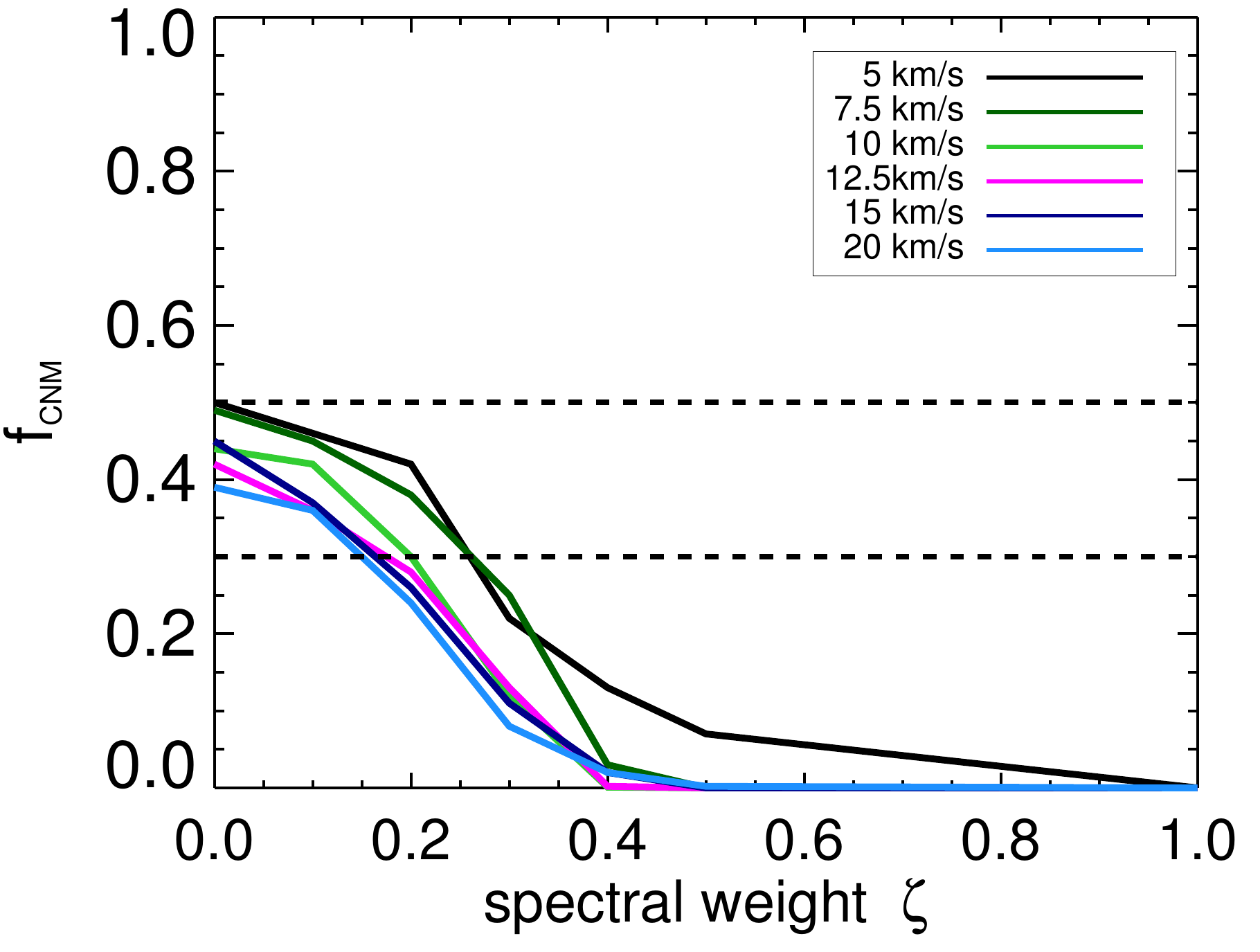}}
  \centering\resizebox{0.40\hsize}{!}{\includegraphics{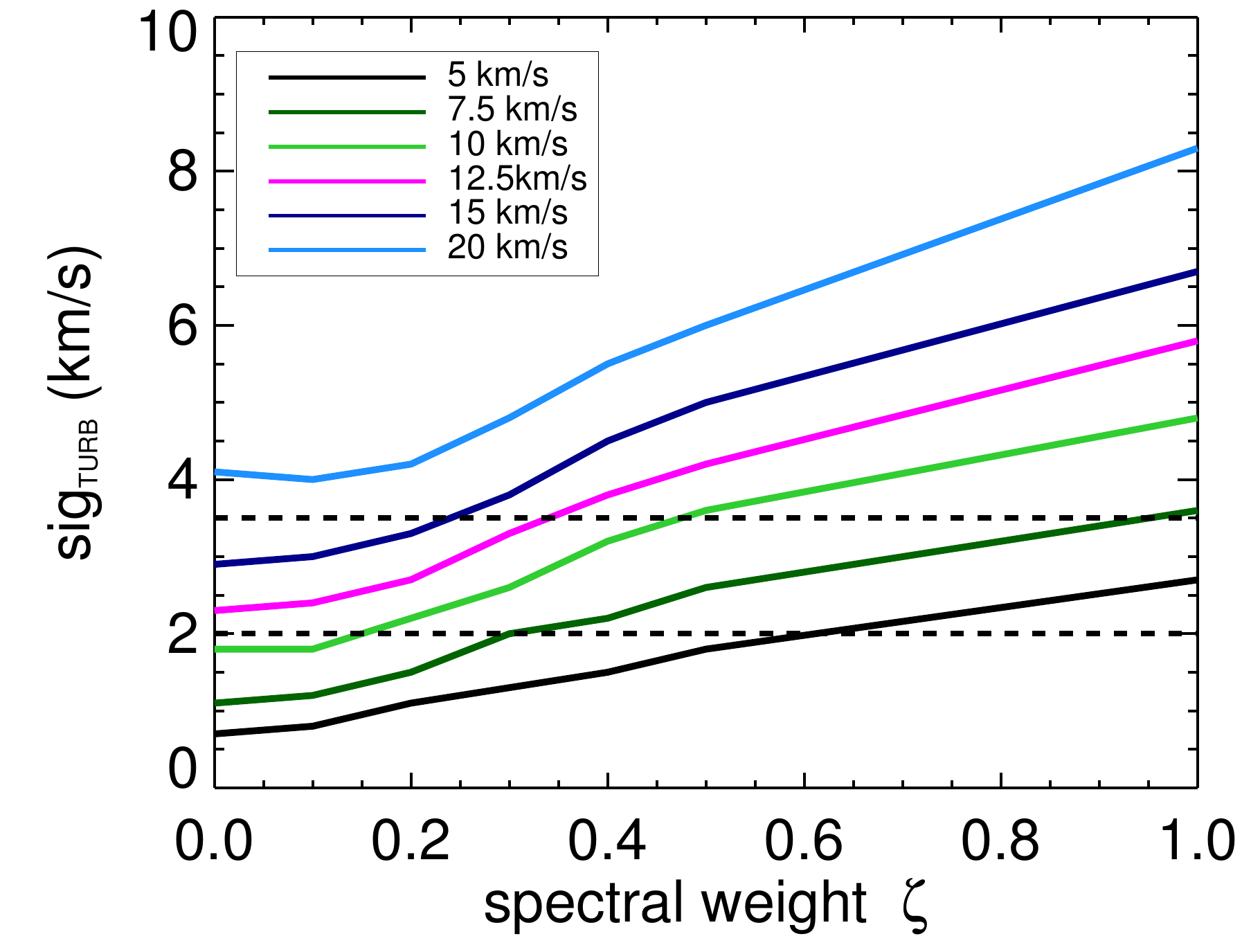}}
  \caption{\label{fig:setn01} Behavior of \fcnm\ ({\it on the left}) and \sigturb\ ({\it on the right}) with the variation of the projection weight $\zeta$, for the different values of the large scale velocity \vs\, (from 5 to 20\,\kms). The dotted lines frame the observational values of \fcnm\ and \sigturb\ that we want to reproduce.}
\end{figure*}

\begin{figure*}[t]
  \resizebox{0.24\hsize}{!}{\includegraphics{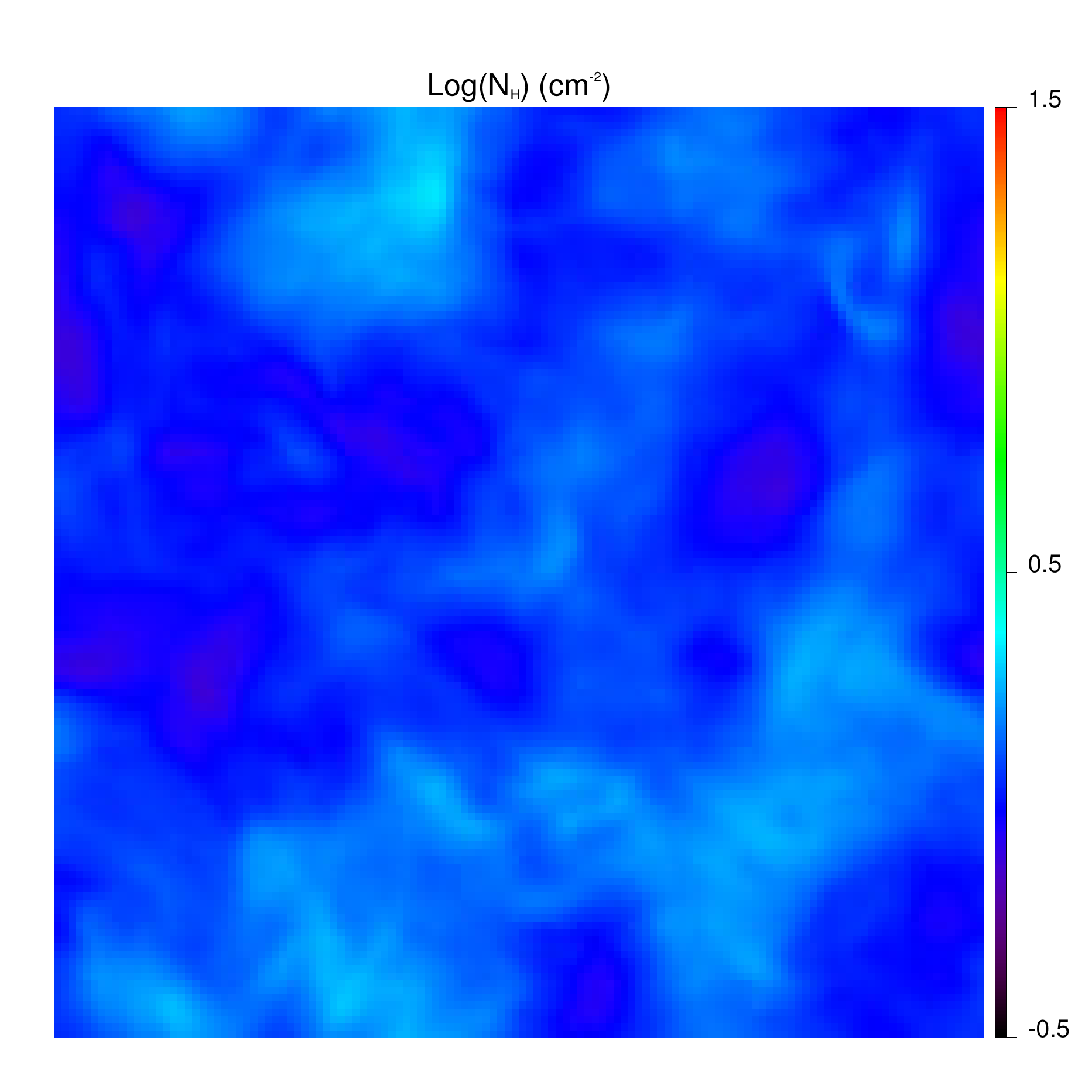}}
  \resizebox{0.24\hsize}{!}{\includegraphics{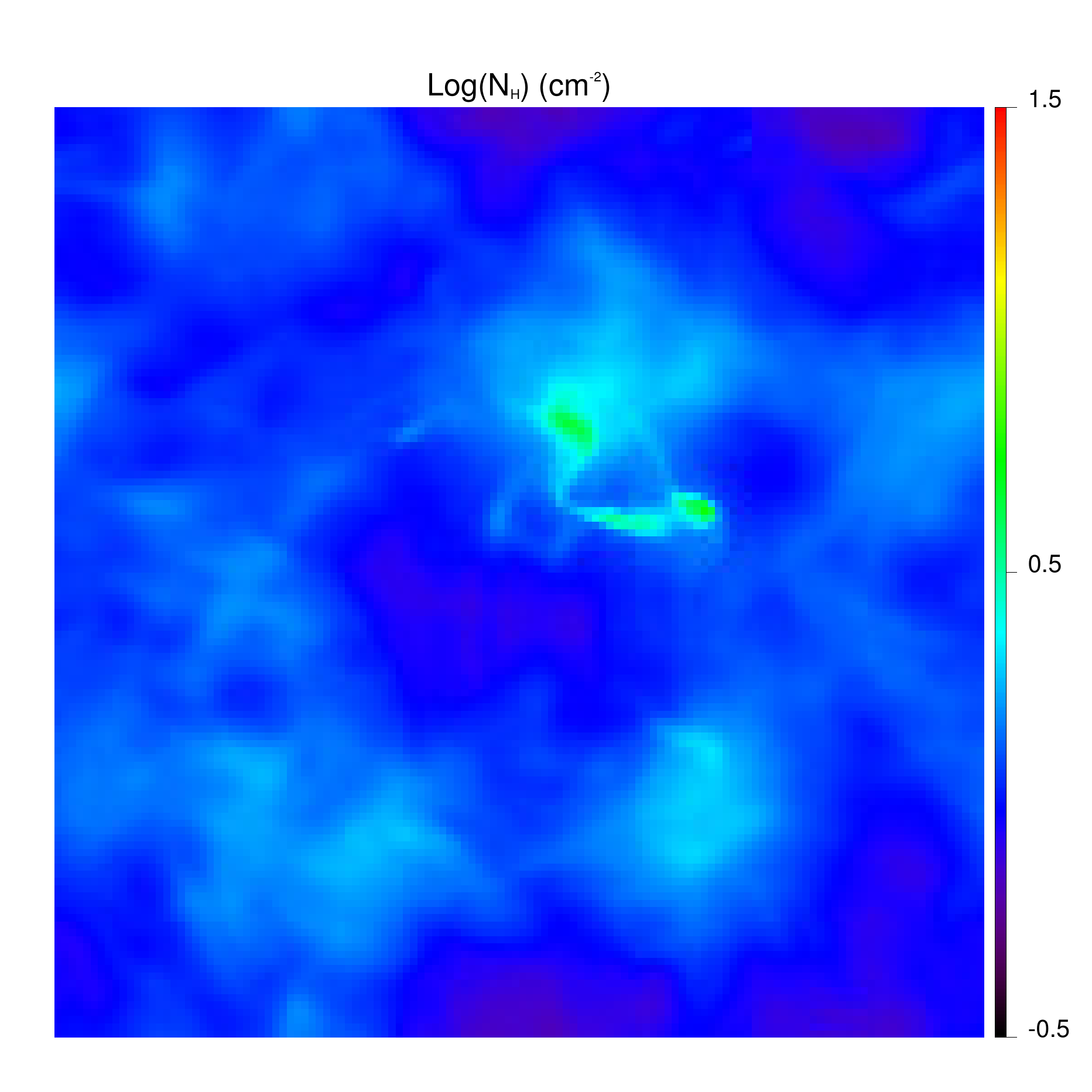}}
  \resizebox{0.24\hsize}{!}{\includegraphics{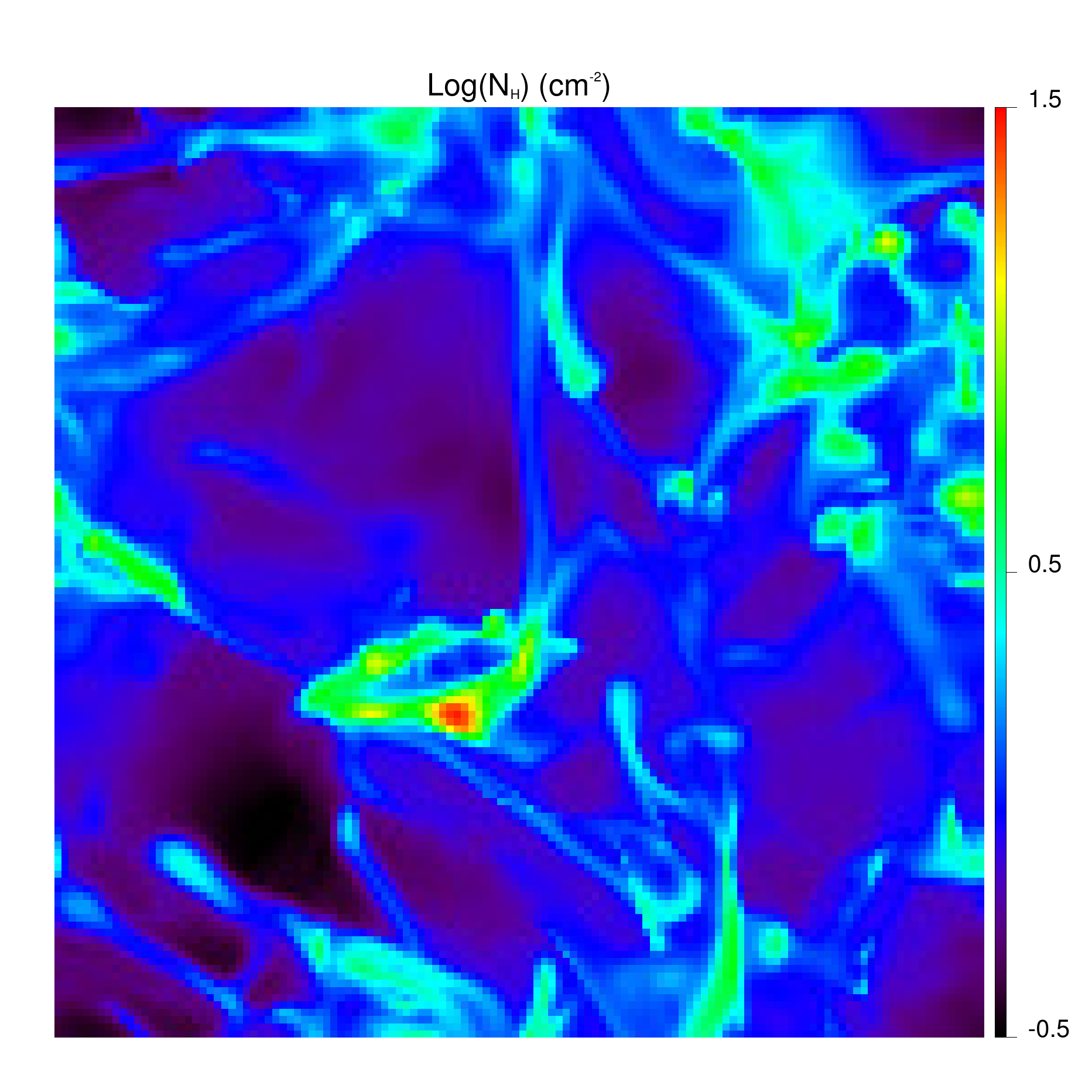}}
  \resizebox{0.24\hsize}{!}{\includegraphics{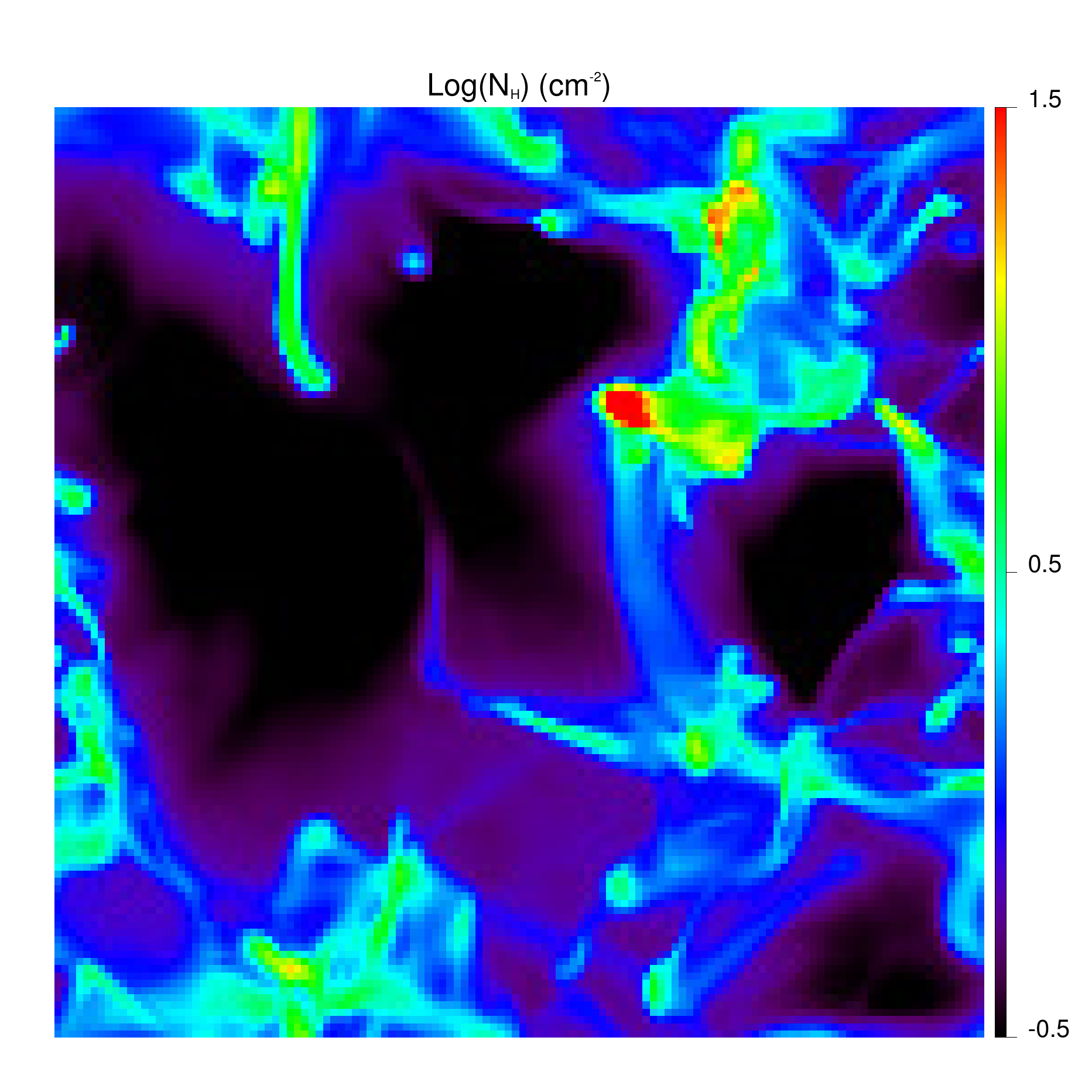}}
  \caption{\label{fig:cartesn01} Maps of the logarithm of the integrated density along the direction $z$. The projection weight decreases from the {\it left} to the {\it right} and takes the following values: 0.5, 0.4, 0.2 and 0.0. The initial density and the large scale velocity are fixed to $n_0=1.0$\,\cc\ and  \vs\,=\,12.5\,\kms. All maps are computed on the same column density range [0.3\,10$^{20}$\,cm$^{-2}$, 32\,10$^{20}$\,cm$^{-2}$].}
\end{figure*}

From Table~\ref{tab:setpw05} and Figure\,\ref{fig:setpw05} one could note that the final turbulent velocity dispersion is lower for higher initial densities; \sigturb\ is thus lower for higher values of the cold mass fraction \fcnm. To understand this behavior we studied the distributions of the three following velocity dispersions: \sigturb, \sigtherm\ and \sigtot\ computed for each thermal phase in a given simulation at high resolution (1024$^3$,$n_0=2.0$\,\cc, $\zeta=0.5$ and \vs=7.5\,\kms). All distributions are plotted on Figure~\ref{fig:histosigmas}. As expected \sigtherm\ depends on the temperature of the gas. We note that it is also the case for the distributions of \sigturb, the one computed in the CNM phase ($T<200$\,K) being especially different from the other two thermal phases. Its peak lies at very small values between 0.1 and 0.2\,\kms\ and presents a wide tail reaching 5\,\kms. 
To better understand the shape of this distribution, we looked at two distinctive lines of sight for which \sigturb(CNM) is equal to 0.14 and 4.7\,km respectively. The density, temperature and velocity ($z$-component) profiles are plotted on Figure~\ref{fig:lofs128}.
The line of sight with a low velocity dispersion (top of Fig.\ref{fig:lofs128}) is composed of four CNM structures with temperatures between 100 and 500\,K. These clumps have homogeneous internal velocities and their relative velocity is small ($\sim\,3$\kms), explaining the low value of \sigturb(CNM) integrated along the line of sight. On the other hand, if the internal velocities of the CNM structures on the second line of sight are also homogeneous, their relative velocity is high and close to 15\,\kms. Thus, the shape of the histograms of \sigturb(CNM) reveals that individual clumps have an internal velocity dispersion close to 0.2\,\kms\ and that the large values are the result of the relative velocities between clumps. Besides, the three graphs on Figure~\ref{fig:histosigmas} show that the shape of the \sigtot\ histograms is dominated by the thermal motions suggesting a subsonic Mach number. Like \citet{heitsch2005,heitsch2006} we conclude that the cold gas structures have subsonic internal motions and that the supersonic Mach number generally observed in the CNM \citep{heiles2003b} can be the result of their relative motions.

A similar explanation is proposed for the increase of the observed Mach number with density (see  Table~\ref{tab:setpw05}).
By increasing the density, the cold mass fraction increases and so the number of cold clumps in the box. 
These have relative velocities inherited from the WNM and therefore supersonic with regard to their low temperature. 

As mentioned by \cite{audit2005}, we note that at a given initial density, the efficiency of the CNM formation decreases when the large scale turbulent velocity increases (see Fig.\ref{fig:setpw05} on the left) and thus when the Mach number increases. \cite{vazquez2012c} argues that supersonic turbulence acts faster than TI and thus dominates the evolution of the cloud. It has been recently confirmed by \cite{walch2011}. As stated in Eq.~\ref{eq:tcooltdyn} and as emphasized by \cite{hennebelle1999} the radiative time needs to be smaller than the dynamical time for the thermal instability to occur. Equivalently, the cooling time must be smaller than the typical time of the stirring. 
\begin{equation}
\frac{k_{\mathrm{B}}T}{(\gamma -1)n\Lambda} < L_{\mathrm{S}}/v_{\mathrm{S}}.
\end{equation}
For fixed box size ($L_S$) and initial temperature, this offers a natural explanation why the transition is inefficient at the lowest densities and the highest amplitudes \vs. In our case $L_S$ is \mbox{20 pc} and, for typical initial conditions of \mbox{$n=1 $ cm$^{-3}$} and \mbox{$T=8000$ K}, the cooling time is $t_{\mathrm{cool}}=1.7\,$Myr implying that \mbox{$v_{\mathrm{s}} < 11.4$km s$^{-1}$}.

\subsection{Effect of the compressive modes}

\citet{federrath2008b,federrath2010} showed on isothermal simulations that the ratio of compressible to solenoidal modes in the forced velocity field changes the structure of the gas. To understand the effect of the nature of the turbulent forcing on the cold gas formation in the thermal instability frame, we performed 42 simulations (128$^3$ pixels) with different values of the projection weight $\zeta$ and a fixed initial density $n_0=1.0$\,\cc, close to the typical WNM density. We voluntarily chose an initial density that we showed does not induce CNM formation in the case of the natural state of turbulent forcing  ($\zeta=0.5$, see section~\ref{subsec:forcing}) in order to see if a different combination of compressive {\it versus} solenoidal modes can trigger the phase transition at low density. Like previously the velocity field amplitude \vs\ was varied from 5 to 20\,\kms. The velocity field is either purely compressive ($\zeta=0$), either purely solenoidal ($\zeta=1$) or a mixing of both components ($0.1\le\zeta\le 0.9$). Table~\ref{tab:setn01} summarizes the values of $\mathcal{M}_{\mathrm{obs}}$, $\mathcal{M}_{\mathrm{theo}}$, $f_{\mathrm{CNM}}$ and $\sigma_{\mathrm{turb}}$ in the stationary regime for the 42 simulations. The behavior of \fcnm\ and \sigturb\ with $\zeta$ are also presented on Figure\,\ref{fig:setn01}.

These results show that a turbulent field with a majority of solenoidal modes or with an energy equilibrium between compressible and vortical modes ($\zeta \ge 0.5$) cannot trigger the transition to CNM with this intermediate value of the initial density (Figure~\ref{fig:setn01}, left). This behavior is obvious on the representations of the integrated density on Figure\,\ref{fig:cartesn01}: the more compressive modes the velocity field has (right side), the more cold gas is formed. For a WNM initial density of $n_0=1\,$cm$^{-3}$, $\zeta$ needs to be smaller than 0.3 to reach 30\% of the mass in the cold phase.  Besides, as previously showed and for the same reason, $f_{\mathrm{CNM}}$ increases at lower values of the large scale velocity at a given value of the spectral weight $\zeta$ (Fig.\ref{fig:setn01} left part).

One can notice that \sigturb\ continually increases with the projection weight $\zeta$ (right part of Fig.\ref{fig:setn01}), even when no cold gas is created in the simulation. The turbulent motions estimated with \sigturb\ are thus affected by their repartition in the compressive and solenoidal modes. \sigturb\ increases with the energy fraction injected in the solenoidal modes for the same total amount of energy injected. The solenoidal modes are efficient at mixing and uniformizing the density more efficiently than the compressive modes, which create higher density contrasts correlated with the velocity field \citep{federrath2010}. The \sigturb\ computation gives more weight to the high density regions and thus to the compression regions where the velocity field is close from uniform, explaining why \sigturb\ decreases when the compressive modes become dominant.

\section{High resolution simulations}
\label{sec:1024}

\begin{figure*}[ht]
  \resizebox{0.5\hsize}{!}{\includegraphics{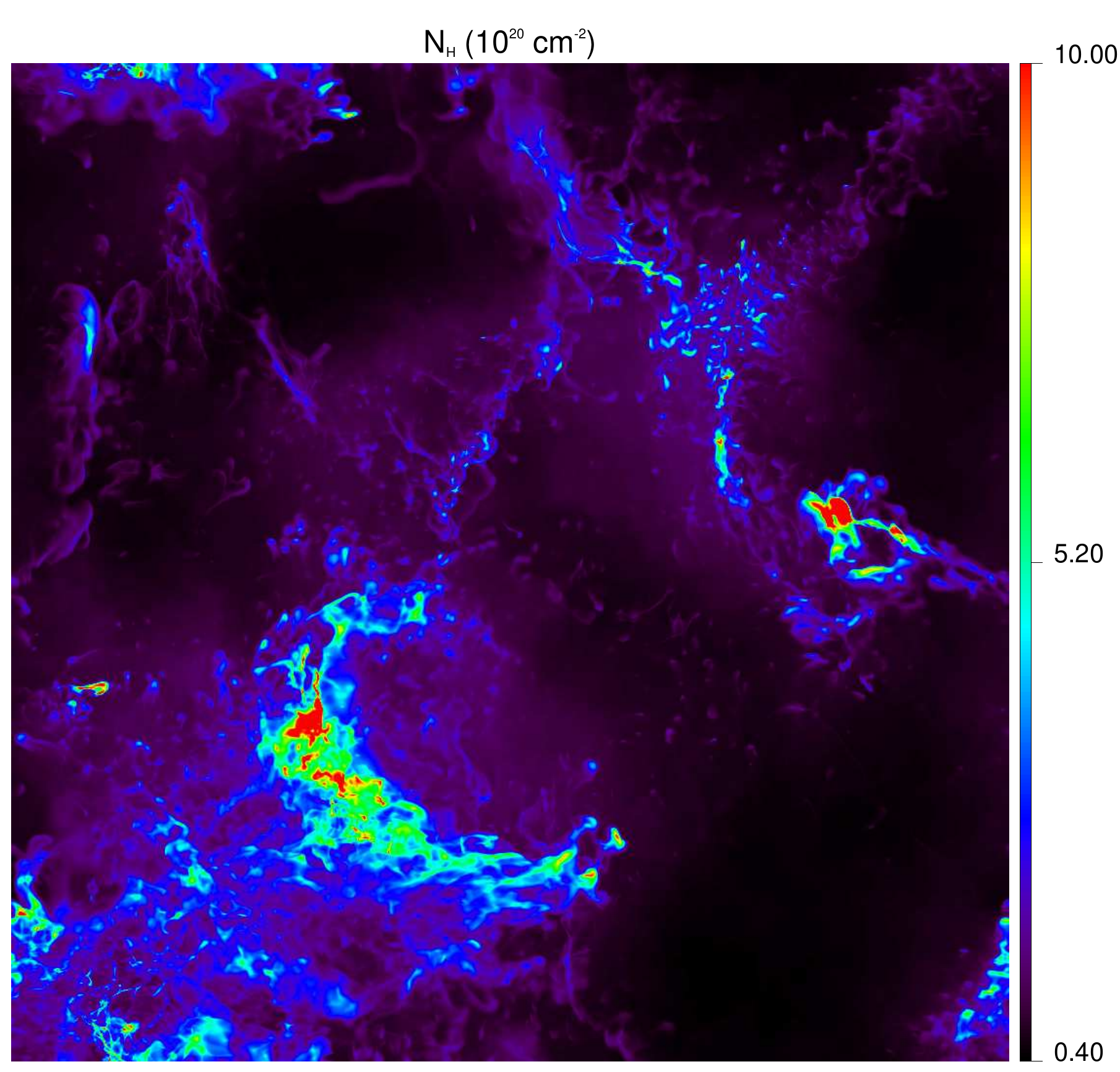}}
  \resizebox{0.5\hsize}{!}{\includegraphics{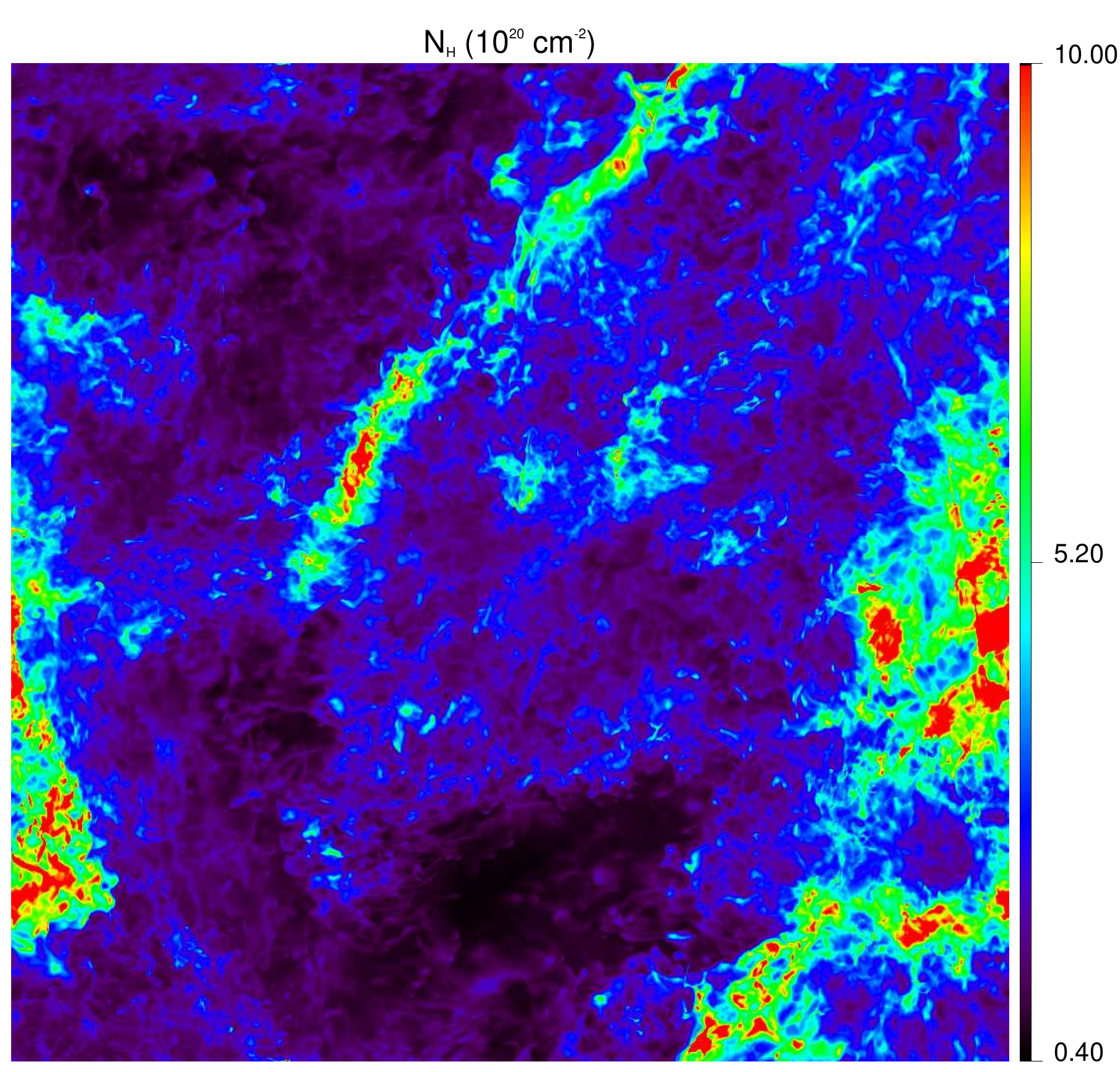}}
  \caption{\label{fig:NH01_02} Integrated density (along the $z$-axis) maps for 1024N01 {\it on the left} and 1024N02 {\it on the right}, in $10^{20}\,{\rm cm}^{-2}$.}
\end{figure*}

All the results presented hereafter are concerning the two \mbox{$1024^3$ cells} simulations after \mbox{16 Myr}. We selected two different sets of initial conditions leading to the formation of 30 to 50\% of cold gas, a subsonic Mach number and a turbulent velocity dispersion between 2 and \mbox{3 km s$^{-1}$}. The first one (hereafter 1024N01) has an initial density of \mbox{1 cm$^{-3}$}, a majority of compressible modes with \mbox{$\zeta$ = 0.2}, and a large scale velocity \vs\ of \mbox{12.5 km s$^{-1}$}. The second one (hereafter 1024N02) has an initial density of \mbox{2 cm$^{-3}$}, high enough to allow us to study the natural state of the turbulent velocity field with \mbox{$\zeta$ = 0.5}, and a large scale velocity \vs\ of \mbox{7.5 km s$^{-1}$}. 

Both simulations are shown on Figure~\ref{fig:NH01_02} that represents the integrated density along the $z$-axis. They reveal structures in clumps and filaments comparable to observations. Besides, 1024N02 creates much more structures than 1024N01, suggesting that it is more efficient at triggering the transition, which we will discuss later.

\subsection{Mach number}

\begin{table}
\caption{\label{tab:mach1024} Values of the theoretical Mach number computed at first on all pixels and then only on the pixels with a temperature greater than 200\,K, of the observational Mach number and of the turbulent velocity dispersion for both simulations at high resolution.}
\centering
\begin{tabular}{c|c c c c}
\hline\hline
Simulation & \machtheo\ & \machtheo(T$>$200) & \machobs\ & \sigturb\ (\kms) \\ 
\hline\hline
1024N01 & 0.85 & 0.80 & 0.47 & 2.77\\
1024N02 & 1.20 & 1.20 & 0.68 & 2.20\\
\hline
\end{tabular}
\end{table}

Table~\ref{tab:mach1024} summarizes the values of \machtheo, computed at first on all the pixels, and then only on the pixels with a temperature greater than 200\,K, \machobs\ and \sigturb. The theoretical Mach numbers show that 1024N01 is subsonic while 1024N02 is transsonic. Besides, removing the cold pixels of the box ($T<200$\,K) in the computation of \machtheo\ does not change the results, suggesting that the turbulent motions are dominated by the WNM motions, as we will discuss it while studying the CNM structures. The observational Mach number is in both cases clearly subsonic, in agreement with observations. Lastly, the computed values of \sigturb\ also show that the observed properties of turbulence are well reproduced in both simulations although it is slightly lower than the estimated value in a WNM cloud of 40\,pc ($\sim3$\,\kms). This is due to the cold structures present in the simulations for which the velocity dispersion is much lower.

\subsection{Thermal distribution of the gas}

\begin{figure}
  \centering\resizebox{0.8\hsize}{!}{\includegraphics{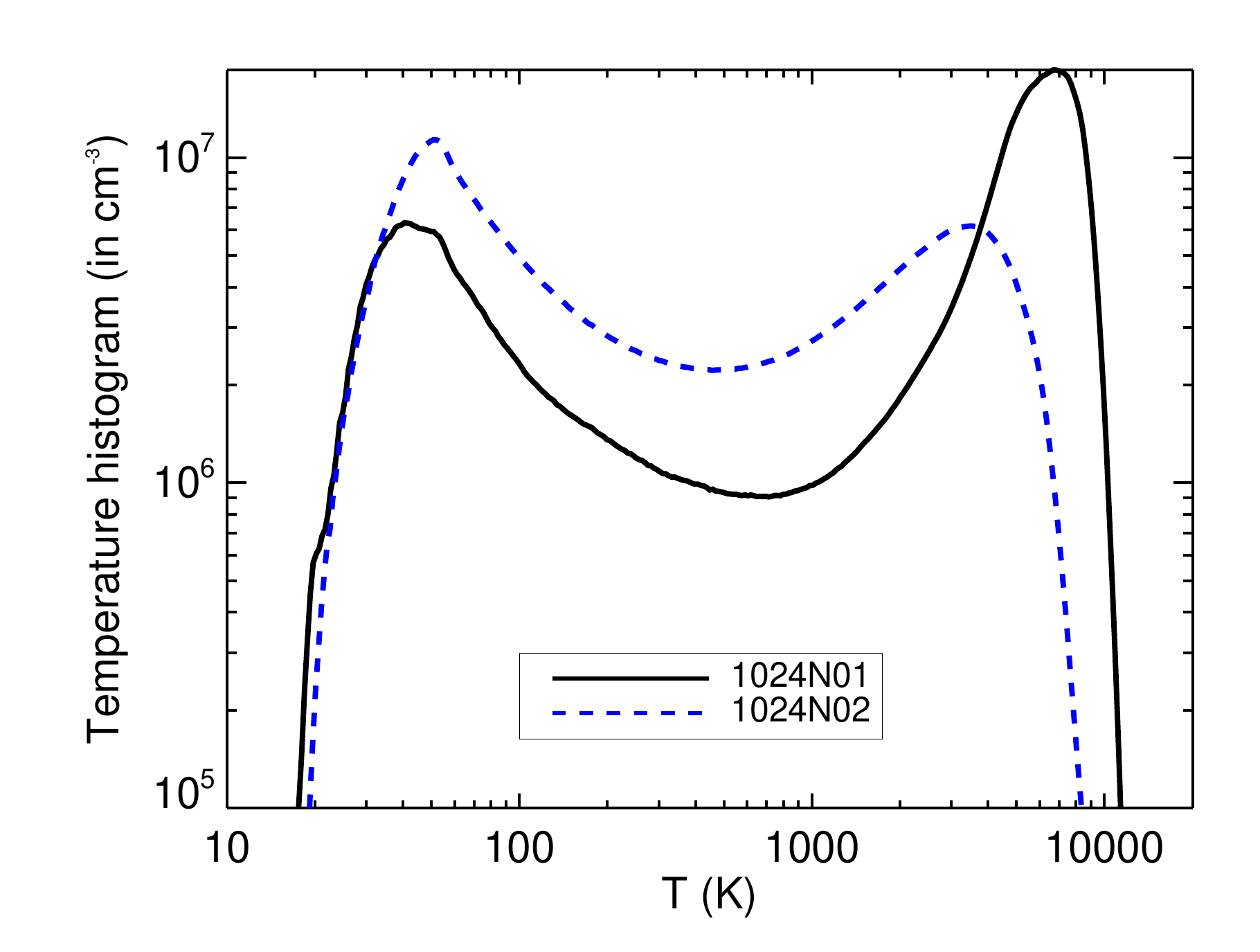}}
  \caption{\label{fig:histoT} Temperature histograms weighted by the density: 1024N01 (solid black line) and 1024N02 (dashed blue line). For the 1024N02 simulation, the histogram is normalized by 2 to be comparable to 1024N01}
\end{figure}

\begin{figure}
  \resizebox{\hsize}{!}{\includegraphics{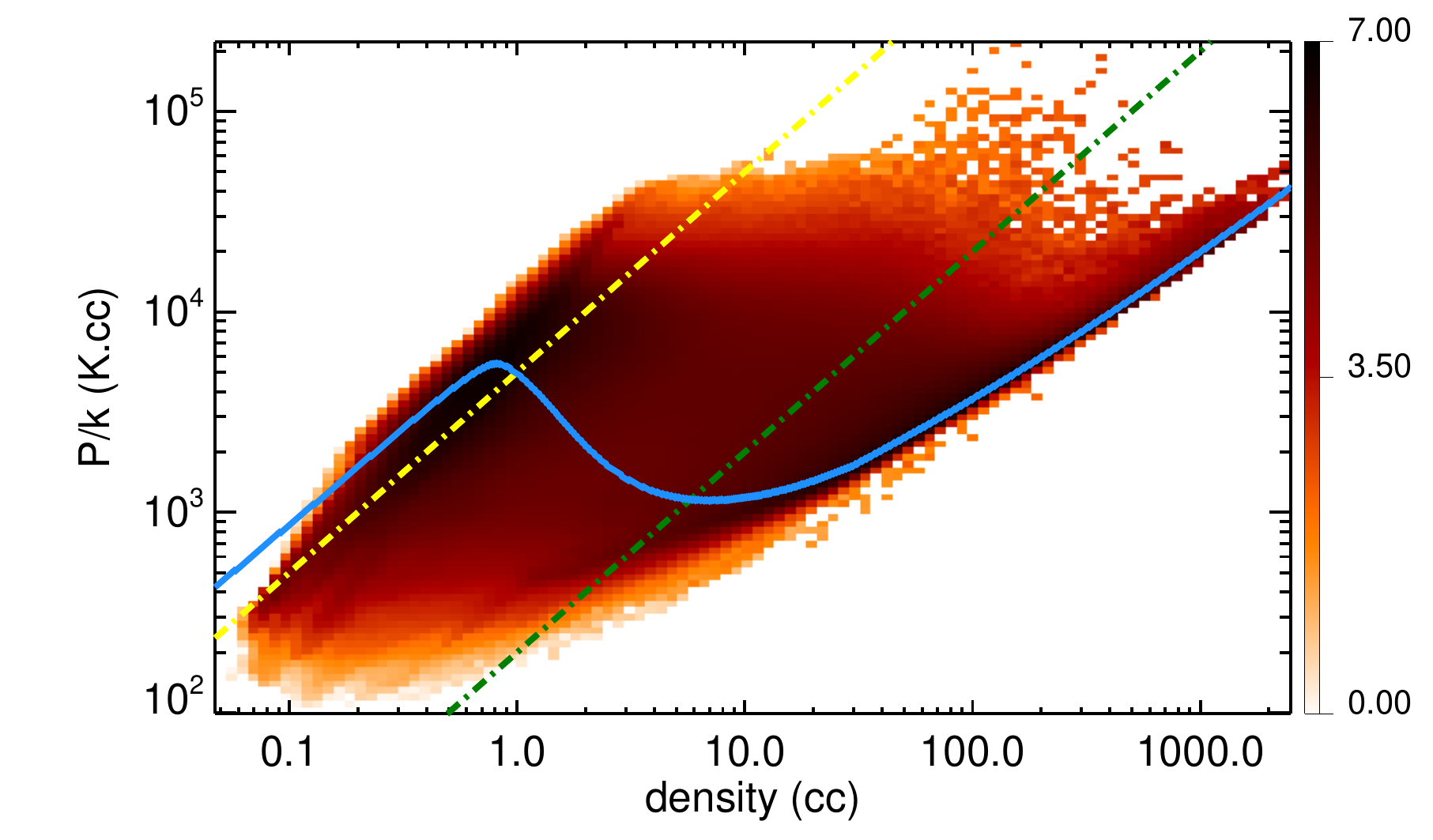}}
  \resizebox{\hsize}{!}{\includegraphics{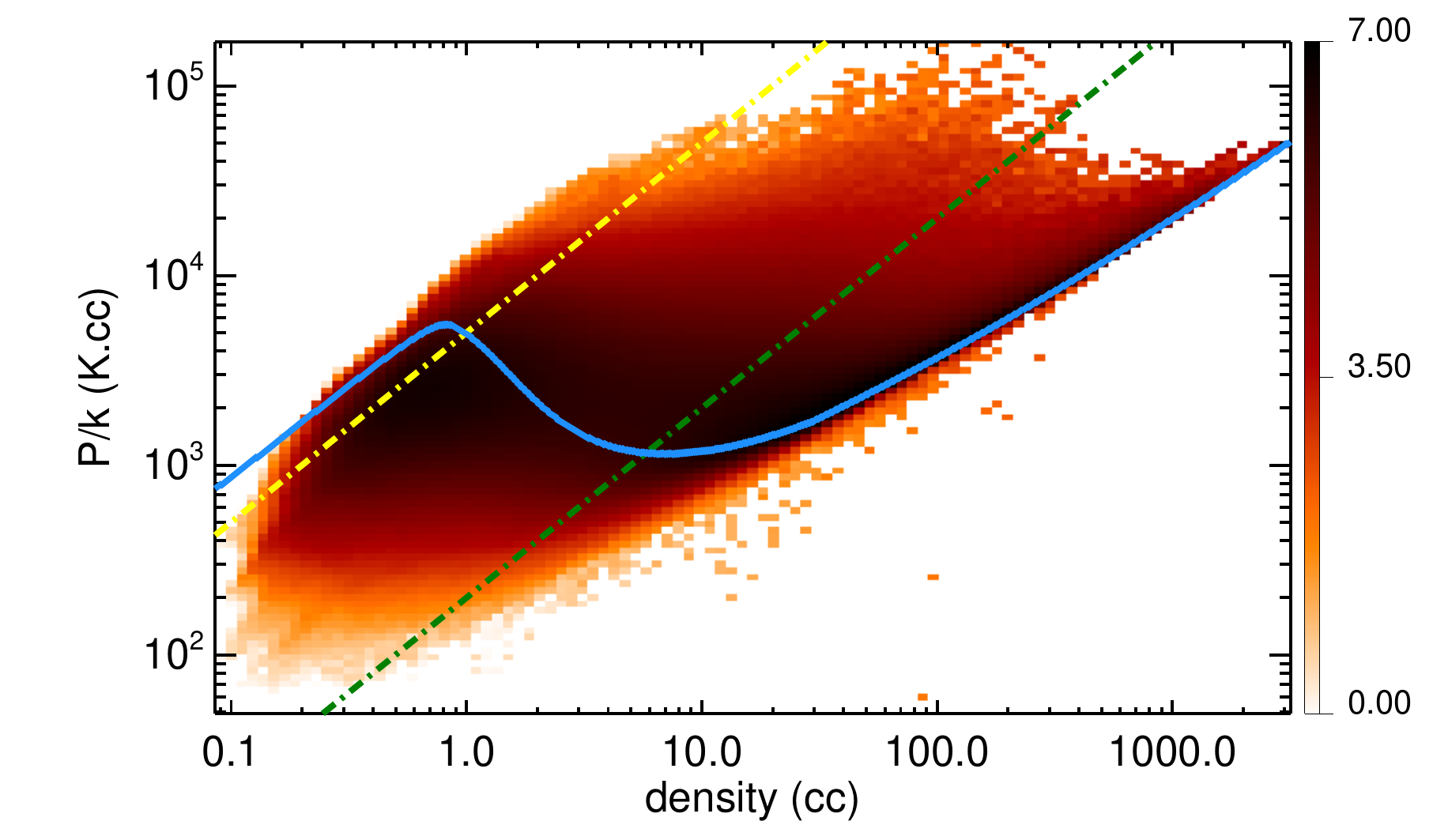}}
  \caption{\label{fig:diagpn} Distributions of mass in a pressure-density diagram with on the top the 1024N01 simulation and on the bottom the 1024N02 simulation. The unit of the 2D histogram is the logarithm of the density (in \cc). The solid blue curve shows the thermal equilibrium (cooling equal to heating) with the processes implemented in HERACLES, the dashed-dotted lines are the 200 K (green) and 5000 K (yellow) isothermal curves.}
\end{figure}

As expected with the thermal instability at low Mach number, the distributions in the $P,n,T$ space present the evidences of a multiphasic medium. The temperature histograms (Fig.~\ref{fig:histoT}) have indeed a clear bimodal shape, with a peak in the warm phase and one in the cold phase. Similarly the massive distributions of the gas in a pressure-density diagram (Fig.~\ref{fig:diagpn}) clearly show two preferential zones: the first one at high densities and the second one at low densities, suggesting again CNM and WNM. 

\begin{figure}
  \centering\resizebox{0.8\hsize}{!}{\includegraphics{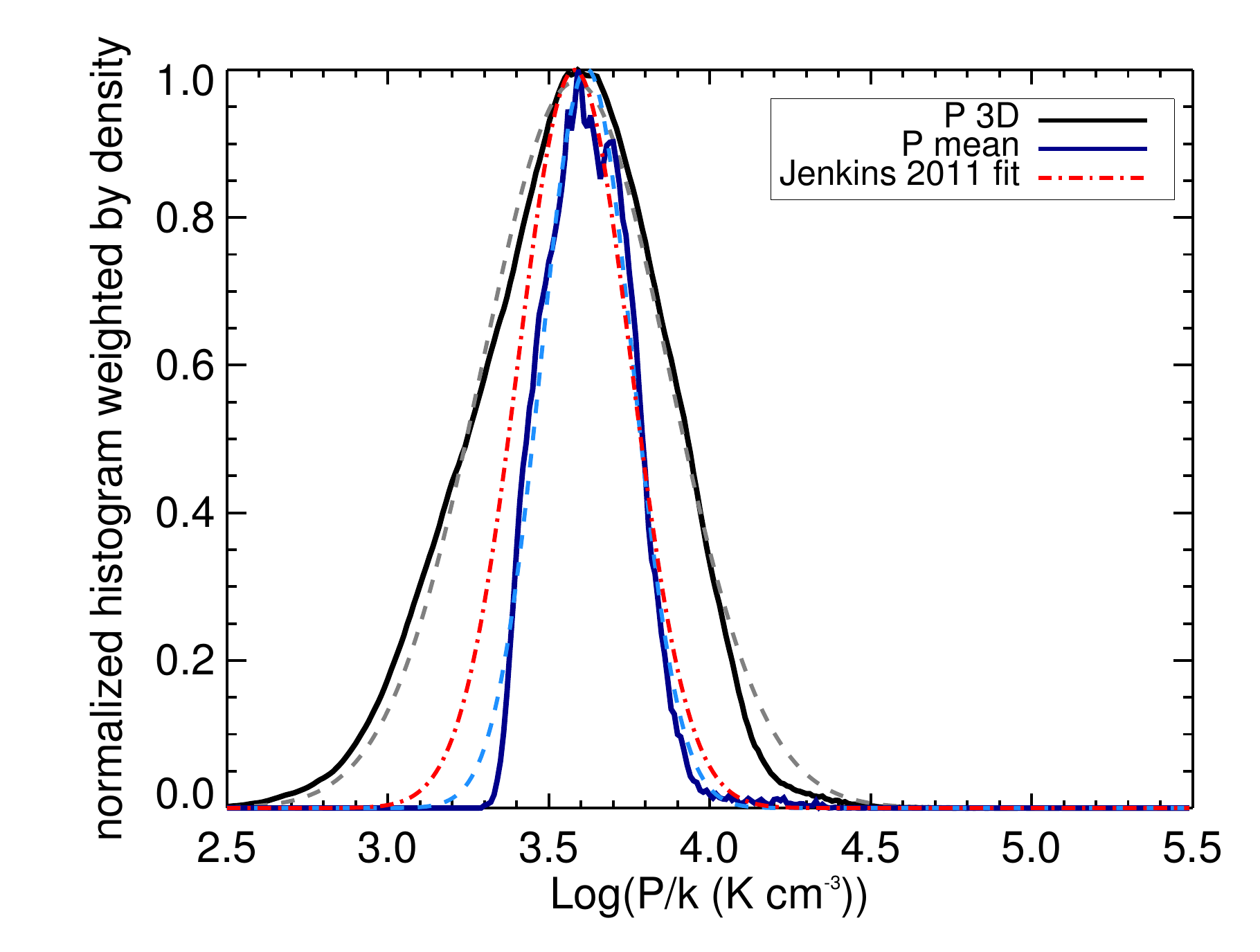}}
  \centering\resizebox{0.8\hsize}{!}{\includegraphics{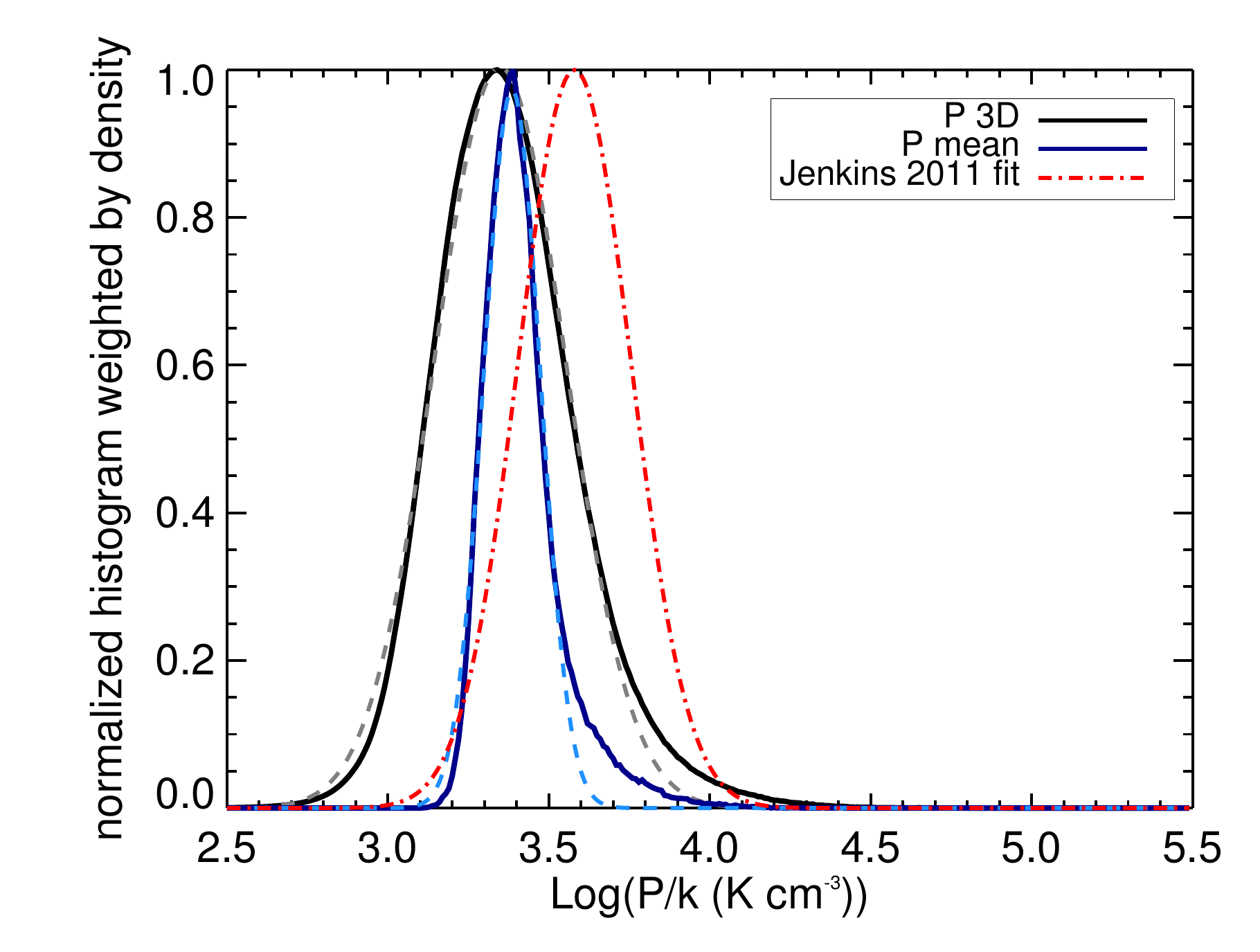}}
  \caption{\label{fig:histoP}  Pressure histograms from 1024N01 (top) and 1024N02 (bottom). The black line is the histogram of the 3 dimensional pressure, the blue one is the histogram of the averaged pressure along the line of sight. The associated colored dashed lines are the lognormal fits. The red dashed-dotted line represent the lognormal fit found by \cite{jenkins2011} on observations.
The lognormal coefficients ([$a_1$, $a_2$] -- see Eq.~\ref{eq:lognormal}) are [3.58,0.175] for \cite{jenkins2011} For the 3D and averaged pressure, they are [3.58, 0.29], [3.58,0.14] for 1024N01 and [3.35, 0.20], [3.38, 0.09] for 1024N02.}
\end{figure}

\begin{figure}
  \centering\resizebox{0.8\hsize}{!}{\includegraphics{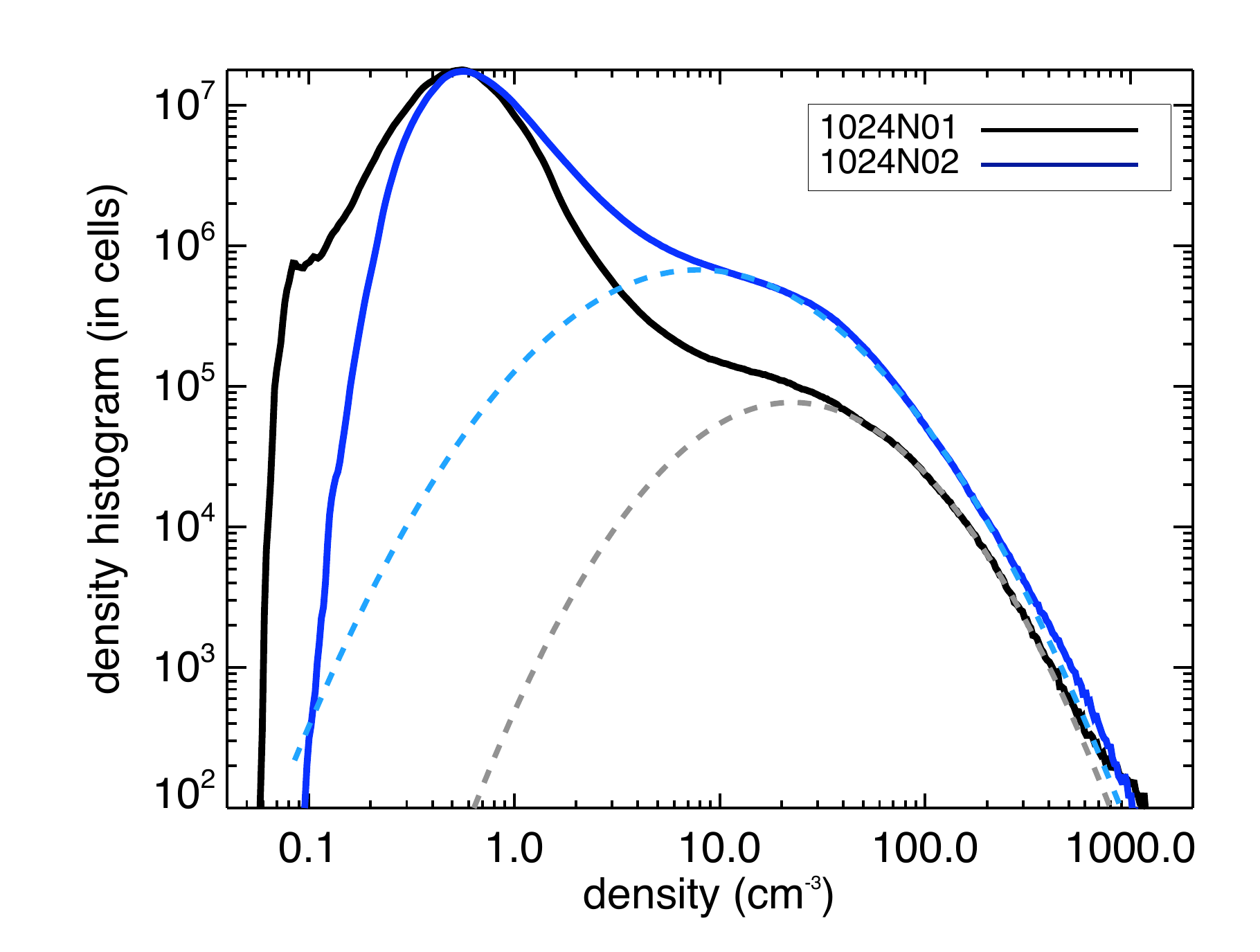}}
  \caption{Density pdfs for both simulations: 1024N01 (solid black line) and 1024N02 (dashed blue line).}
  \label{fig:fitnpdf} 
\end{figure}

However, the properties of the warm gas are slightly different in both simulations. If the cold gas peaks at similar temperatures in both cases (40\,K for 1024N01 and 50\,K for 1024N02), it is not true for the warm gas: while the WNM of 1024N01 has a mean temperature of 7000\,K, close from the prediction of the two-phases model \citep{field1969}, the warm gas in 1024N02 peaks around 3500\,K, two times smaller. This effect can also be seen in the pressure-density diagrams (Fig.~\ref{fig:diagpn}), where the pressure of 1024N02 is smaller than the one of 1024N01 at low densities. Therefore, while the gas at low density of 1024N01 follows the stable branch of the WNM, the gas at low density of 1024N02 is shifted below the equilibrium curve where the heating is dominant. This pressure difference is confirmed by the pressure histograms presented on Figure~\ref{fig:histoP}, where the mean pressure of 1024N02 (bottom) is lower than the mean pressure of 1024N01 (top). On the other hand, the density distributions presented on Figure~\ref{fig:fitnpdf} peaks around $0.5-0.6$\,\cc\ in both cases, similar to the mean density estimated from observations properties. 

Two factors can be considered to explain the presence of such a warm gas out of equilibrium. First, raising the initial density induces the decrease of the cooling time in the WNM (eq.~\ref{eq:tcool}): $t_{\rm cool}$ goes from 1.75\,Myr for $n_0=1.0$\,\cc\ to 1.15\,Myr for $n_0=2.0$\,\cc. 
Therefore, the warm gas at low density located below the equilibrium in 1024N02 is cooled down faster than in 1024N01 and does not have time to be heated enough to reach the equilibrium curve. 
Secondly, the thermal equilibrium curve considered in the pressure-density diagrams is established for a static gas and does not take into account the energy contribution of turbulence. 
Depending on the properties of the turbulence, WNM gas can stabilized at a temperature significantly lower than the prediction of the two-phases model. 

The mass fractions of the different thermal phases have been calculated using two different temperature thresholds and one density threshold. 
All results are summarized in Table~\ref{tab:stat1024}. We first note that the amount of CNM created by the simulations is 30\% for 1024N01 and 50\% for 1024N02. These quantities are in perfect agreement with the predictions deduced from the 128$^3$ simulations and with
the observational constraint of \citet{heiles2003b} who emphasized that 40\% of the mass of the \hi\ lie in the cold phase. Besides, we estimated in Section~\ref{sec:obs} the cold mass fraction around 44\% and the CNM volume filling factor at 1\% from observations properties of \hi. The CNM volume filling factor is here also well reproduced with 1\% for 1024N01 and 4\% for 1024N02. 

Regarding the warm gas, about 20\% of its mass is lying between 2000 and 5000\,K in both simulations. In 1024N02, where the WNM-CNM transition is highly efficient, the unstable regime holds the major part of the mass (46\% between 200 and 5000\,K) and half of it has a temperature lower than 2000\,K; in 1024N01, where the WNM-CNM transition is less efficient, only 10\% are lying in the unstable regime under 2000\,K and a significant amount of the gas lies in the WNM range beyond 5000\,K (40\%). 

If we use a density threshold to separate the phases, we note that both simulations hold the same amount of mass of unstable gas ($\sim 30$\%) and that the difference between the two simulations only lies in the CNM and WNM phases : the amount of gas that does not transfer to CNM in 1024N01 stays in the WNM phase. 
Besides, in the 1024N01 case, the three thermal phases are distributed like 1/3-1/3-1/3, as observed by \citet{heiles2003b}, while in the 1024N02 case, the mass fraction of WNM is lower of a factor 2 and more than 50\% of the mass is in the CNM. \\

\begin{figure*}
  \resizebox{0.5\hsize}{!}{\includegraphics{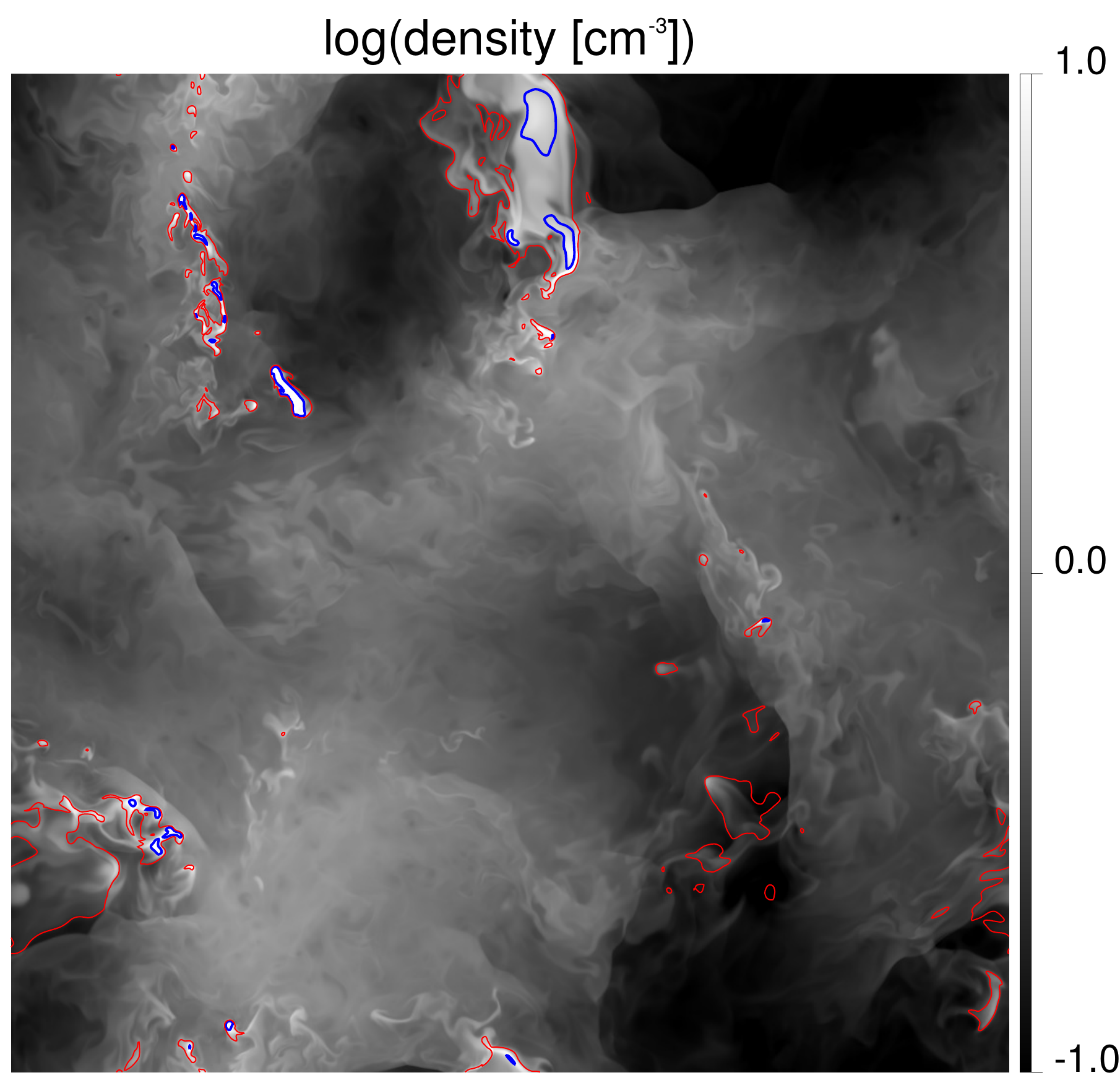}}
  \resizebox{0.5\hsize}{!}{\includegraphics{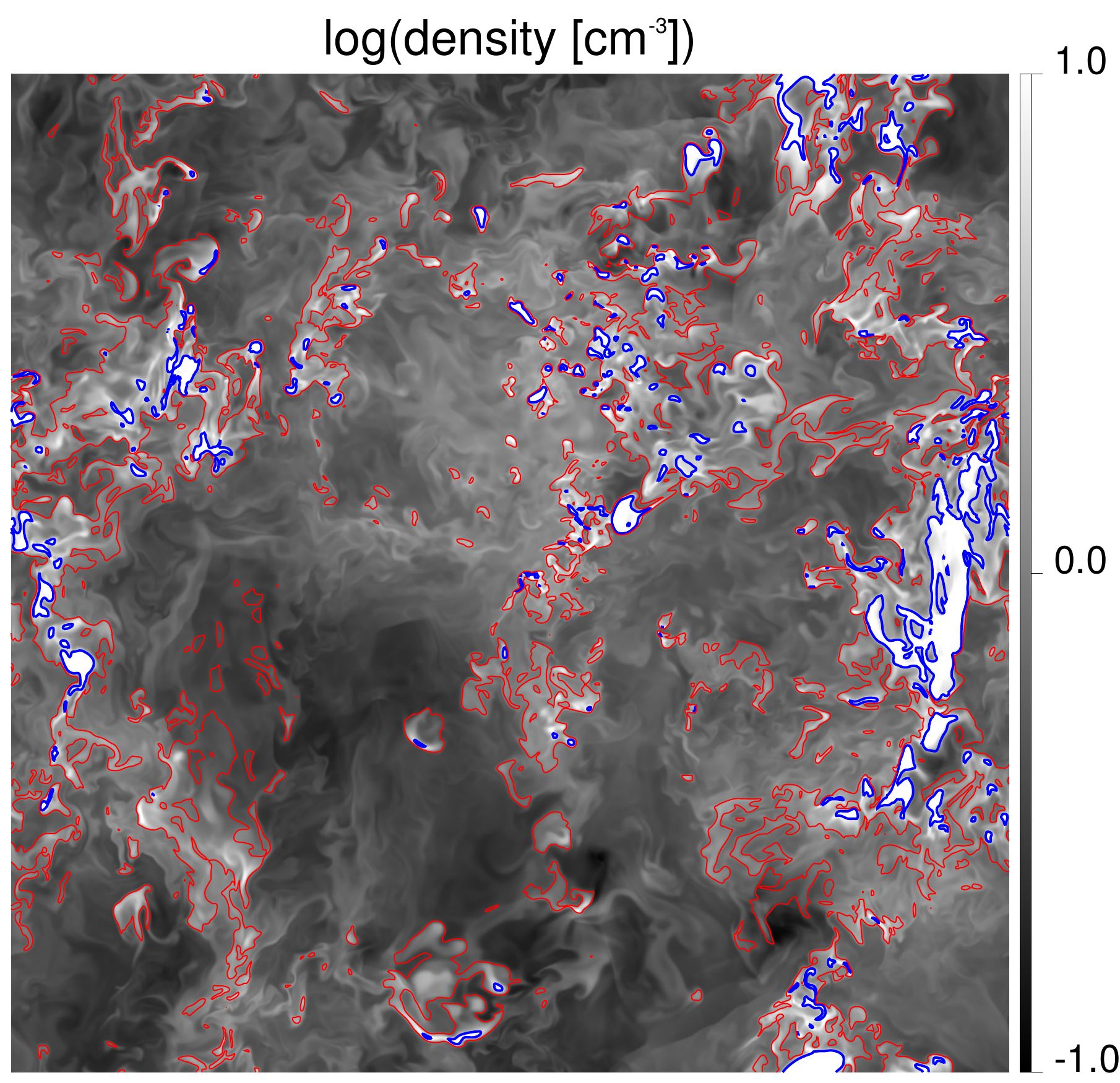}}
  \caption{Logarithme of a density (cm$^{-3}$) snapshots from the $1024^3$ simulations, with contours from the temperature at 200\,K (blue lines) and 2000\,K (red lines). Left: 1024N01, right: 1024N02}
  \label{fig:densT} 
\end{figure*}

2D density slices with isocontours placed at 200\,K (blue) and 2000\,K (red) are presented on Figure \ref{fig:densT}. The highest temperature threshold has been consciously chosen lower than the usual limit at 5000\,K in order to see the distribution of the thermally unstable gas. The gas  lying between 2000 and 5000\,K is indeed widely distributed and does not show any preferential location. 
On the other hand, the thermally unstable gas below 2000\,K lies preferentially around the cold structures as sheets around the denser regions, in agreement with \citet{gazol2010}.

\begin{table*}
\caption{Mass fractions for the three thermal phases : cold, warm, unstable range, Mach number and turbulent velocity dispersion at \mbox{16 Myr} for two different criteria in temperature and a density criterion. }
\label{tab:stat1024}
\centering
\begin{tabular}{c|c c c c c}
\hline\hline
simulation & $f_{\mathrm{CNM}}$ & $f_{\mathrm{UNSTABLE}}$ & $f_{\mathrm{WNM}}$ & CNM volume & WNM volume \\
           &  T$<$200K      & 200K$<$T$<$5000K    & T$>$5000K      & filling factor & filling factor \\
\hline\hline
1024N01 & 0.30 & 0.29 & 0.41 & 1\% & 66\% \\
1024N02 & 0.51 & 0.46 & 0.03 & 4\% & 14\% \\
\hline\hline
&&&&&\\
&  T$<$200K      & 200K$<$T$<$2000K    & T$>$2000K      &  \\
\hline\hline
1024N01 & 0.30 & 0.10 & 0.59 & 1\% & 95\% \\
1024N02 & 0.51 & 0.25 & 0.24 & 4\% & 73\% \\
\hline\hline
&&&&&\\
&  $n>7.0$\,\cc      & 0.8\,\cc$<n<$7.0\,\cc    & $n<0.8$\,\cc      &  & \\
\hline\hline
1024N01 & 0.33 & 0.32 & 0.35 & 1\% & 75\%\\
1024N02 & 0.54 & 0.31 & 0.15 & 4\% & 60\%\\
\hline
\end{tabular}
\end{table*}

\subsection{Pressure range}

\cite{jenkins2011} used observations in absorption to study the distribution of the thermal pressure in the diffuse, cold neutral medium. They found that the pressure distribution, weighted by the mass density, is well fitted by a lognormal.
The pressure histograms weighted by the density of our simulations are also well represented by lognormal distributions.
Figure~\ref{fig:histoP} show the 3D pressure histograms weighted by the density (black lines), the averaged pressure weighted by the column density (blue lines), the fitted lognormal distributions (dashed lines) and the lognormal distribution found by \cite{jenkins2011} (red dotted line). The averaged pressure is computed on each line of sight and is itself weighted by the density of each cell. By doing this, we try to get closer to real observations in which the 3-dimensional pressure is not available. 
Following the definition of \cite{jenkins2011},
\begin{equation}
\label{eq:lognormal}
\frac{{\rm d}n}{{\rm d}\log{P/k}} = a_0 \times \exp\big(-\frac{(\log{P/k} - a_1)^2}{2\times a_2^2}\big),
\end{equation}
the values found for our lognormal fits are given in the caption of Fig.~\ref{fig:histoP}. The simulation 1024N01 peaks at the same value than their results, meaning 3800\,K\,\cc. The pressure peak of 1024N02 stays close although slightly lower at 3.35\,dex (2200\,K\,\cc). These two values lie exactly in the pressure range allowing a two-phase medium according to \citet{field1969}. { As expected, we note that the width of the distributions are larger for the 3D pressure histograms than the averaged pressure histograms in both cases (0.29 and 0.20 compared to 0.14 and 0.09 respectively). The average along the line of sight indeed smoothes the most extreme values. Besides, the distributions of 1024N01 are larger than 1024N02, which can be explained by the lower amplitude of the turbulent stirring used in 1024N02. However, the width obtained in these simulations are close to the result of \cite{jenkins2011} and thus are in agreement with the observations. }

\subsection{Density pdfs}

The density distribution is of great importance in turbulence studies. Many simulations of molecular clouds have shown that the density distribution of an isothermal gas follows a lognormal law with a widening that depends on the Mach number \citep{vazquez1994,passot1998,federrath2008b}. The density distribution of a two-phases gas is however more complex and often presents a bimodal shape \citep{hennebelle2008a,audit2005,audit2010}. \citet{piontek2005}, \cite{gazol2005} and \cite{walch2011} showed that density pdfs are broadened and loose the bimodal shape when the Mach number increases, as most of the gas move to the thermally unstable range. The density distributions presented on Figure~\ref{fig:fitnpdf} point the evidence of a transsonic biphasic gas with two distinct peaks, one at low densities, 0.4-0.5\,\cc, in agreement with the density estimation in the WNM,  and the other one around 10-20\,\cc, in agreement with the density in the CNM.

To compare with studies of molecular clouds, we tried to fit lognormals on cold peaks for each simulation. Regarding the high density part, it is possible to fit the right tail with a lognormal, but for both simulations, the lognormals are overly broad. The left tail of the lognormals are indeed going to too low densities to be a good approximation of the CNM density pdf. To confirm this, we separated the cells of the box where \mbox{$T<200$K} and plotted the pdfs of the CNM. They did not show any characteristic shape, and it was impossible to fit lognormals on them. We also didn't find any evidence of a power law tail as \cite{seifried2011} did. Therefore, if some lognormal fit is possible on the high density part of the pdf, it cannot be used to draw any physical conclusion.

\subsection{Power spectra}

\begin{figure}
  \centering\resizebox{0.9\hsize}{!}{\includegraphics{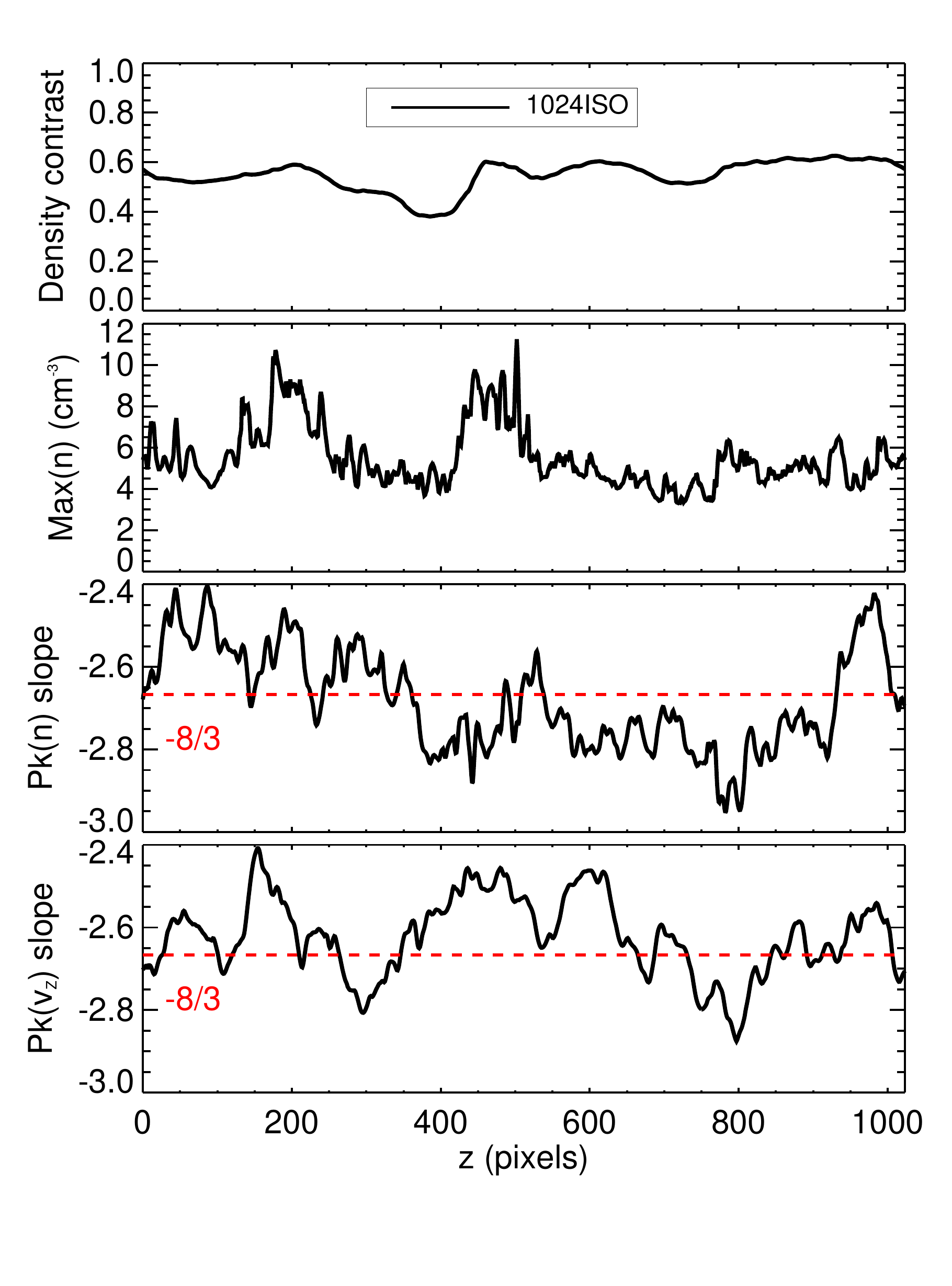}}
  \caption{\label{fig:densps2d_iso}Isothermal simulation. Bottom: 2D power spectra slopes of density (Pk(n)) and velocity (Pk(v$_z$)) cuts along z. For each cut, the plots on the top give the maximum of the density map and the density contrast.} 
  \centering\resizebox{0.9\hsize}{!}{\includegraphics{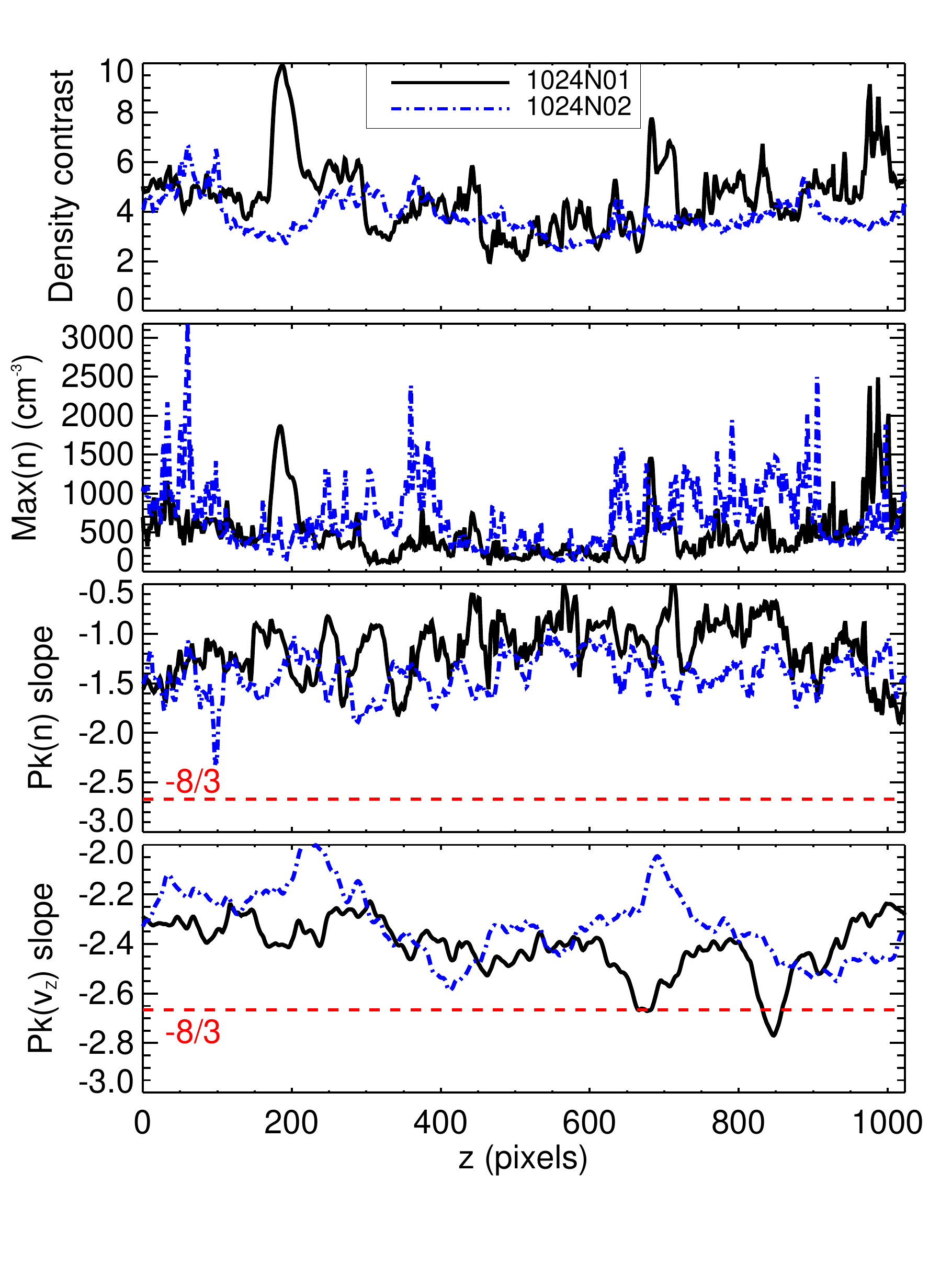}}
  \caption{\label{fig:densps2d_n01n02}Simulations with thermal instability. Bottom: 2D power spectra slopes of density (Pk(n)) and velocity (Pk(v$_z$)) cuts along z. For each cut, the plots on the top give the maximum of the density map and the density contrast.}
\end{figure}

\begin{figure}
  \resizebox{\hsize}{!}{\includegraphics{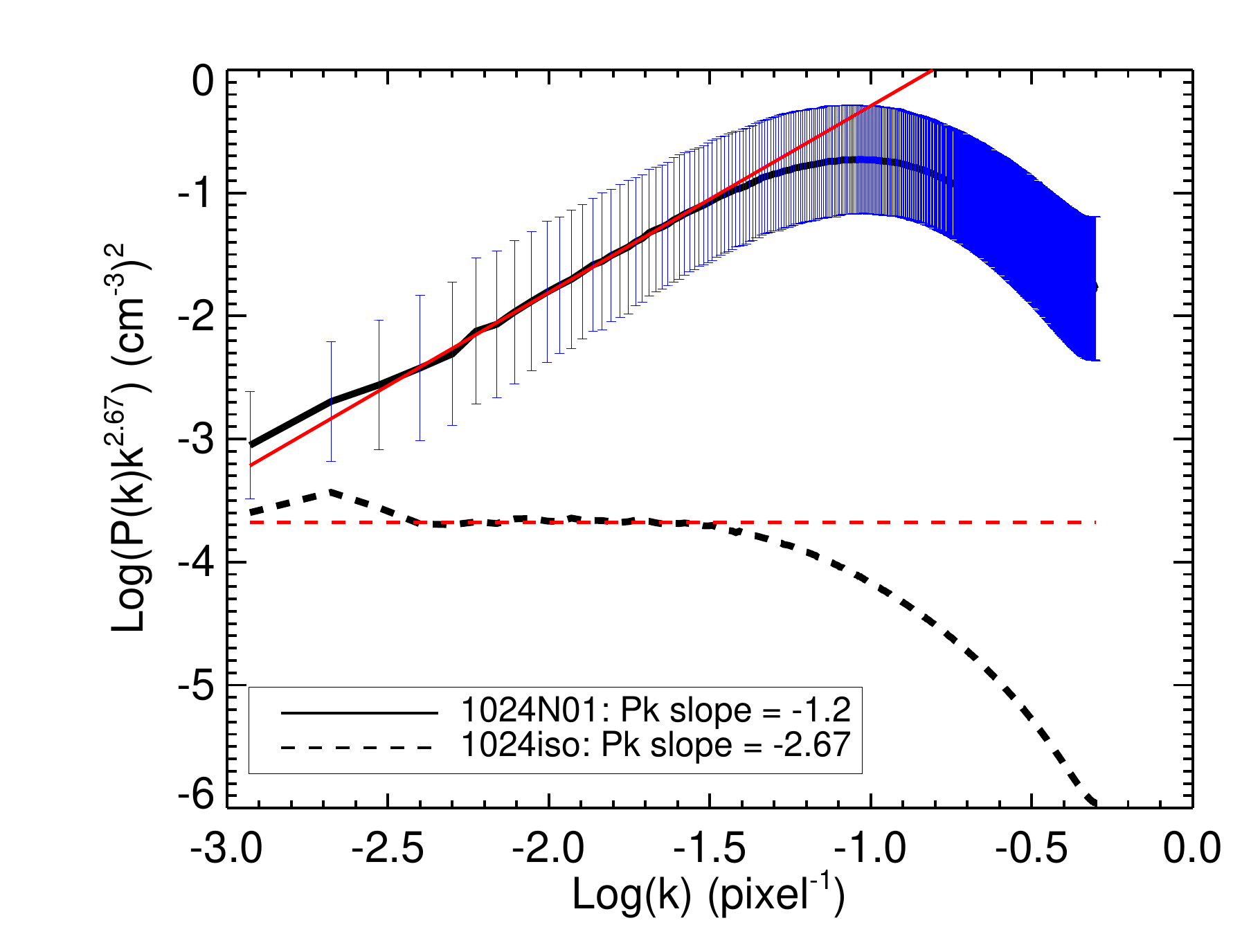}}
  \caption{\label{fig:pkdens} Mean density compensated power spectra $P_k k^{2.67}$ for 1024N01}
\end{figure}
\begin{figure}
  \resizebox{\hsize}{!}{\includegraphics{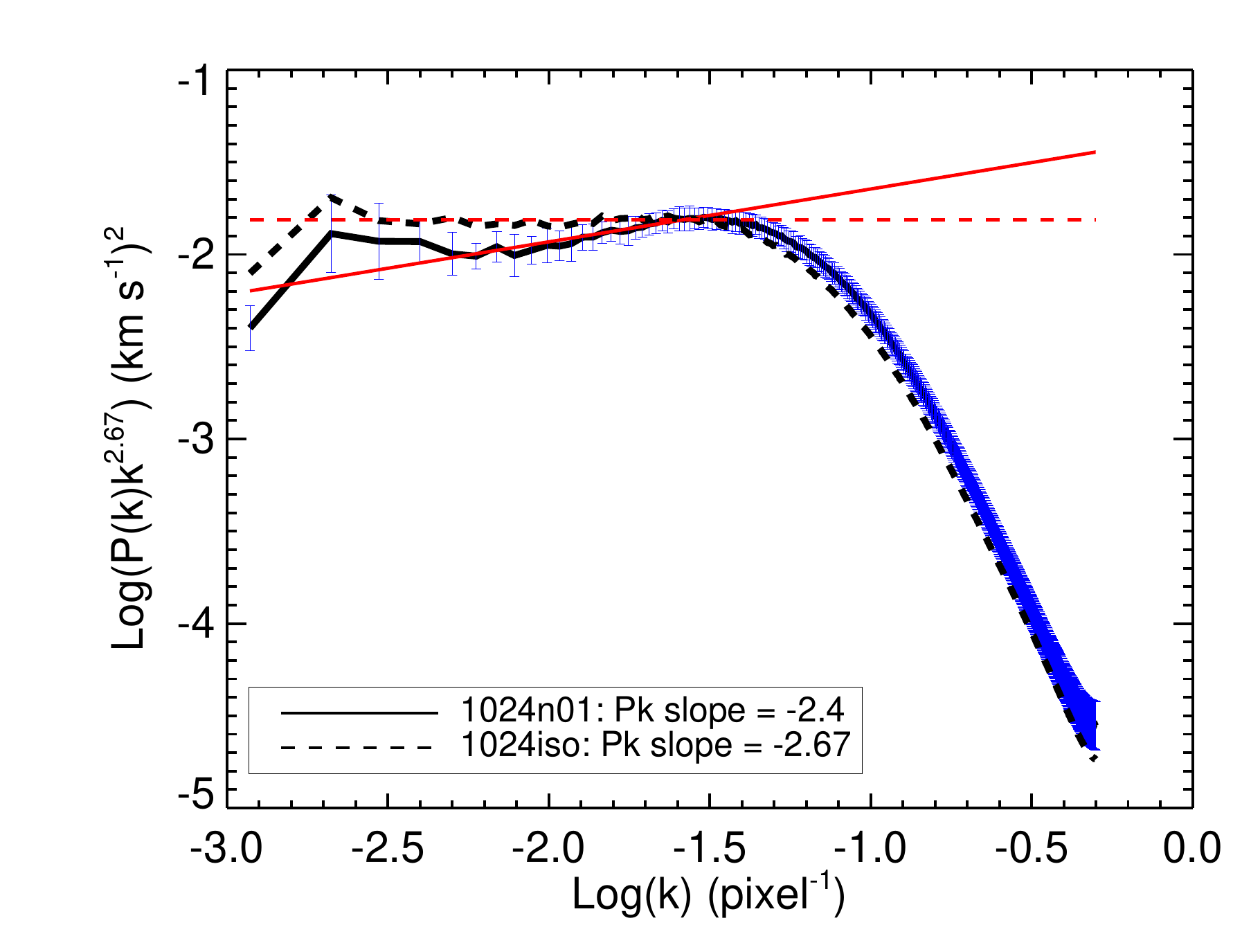}}
  \caption{\label{fig:pkvz} Mean velocity compensated power spectra $P_k k^{2.67}$ for 1024N01}
\end{figure}

Compared with converging flow simulations, the stirring in Fourier provides a way to better study the velocity and density power spectra and therefore characterize the specific properties of the turbulence in a two-phase medium. What is shown here is a study of the statistical properties of the density and velocity fields in 2D slices along the $z$-direction\footnote{the choice of the axis is not important as the simulations are isotropic due to the turbulent stirring in Fourier space.}. 
We choose the $z$-component of the velocity and not its norm to stay closer to what observations have acces to.
For comparison purposes, we also ran a 1024$^3$ cells isothermal simulation, using the same turbulent stirring than 1024N01 (\mbox{$\zeta$ = 0.2} and \mbox{$v_S$  = 12 km s$^{-1}$}), an initial density of \mbox{n = 1 cm$^{-3}$} and a temperature of 8000\,K. 
For each simulation, we calculated for each cut along the $z$-axis the density contrast $<n>/\sigma(n)$, the maximum of the density and computed 2D power spectra of the density and the velocity. We present on Figures \ref{fig:densps2d_iso} and \ref{fig:densps2d_n01n02} the evolution of the contrast, the density maximum, and the slopes of the power spectra with $z$ and on Figures \ref{fig:pkdens} and \ref{fig:pkvz} the compensated mean power spectra $ k^{8/3}P(k)$ of the density and velocity cuts for 1024N01 and for the isothermal simulation (these power spectra for 1024N02 being similar to those of 1024N01, they are not displayed here). The errors bars represent the ($1\sigma$) variations on the 1024 cuts.

As shown by \cite{kim2005}, the density power spectrum of the isothermal and subsonic simulation has a Kolmogorov slope of \mbox{-8/3}. We note that the velocity power spectrum also follows the Kolmogorov law. The estimation of the slope of the power spectrum in the inertial zone of numerically simulated turbulence can only be done on a small scale range due to the turbulent forcing at large scale and the numerical dissipation at small scales. The size of the simulations (1024$^3$) and the use of an isothermal simulation are of great help in the estimation of the inertial range, since the result is known ($P(k)\propto k^{-8/3}$ in 2D). We deduce from the power spectra of the isothermal simulation that the Kolmogorov slope is well reproduced on the decade of scales $0.003<k<0.03$ (or $-2.52<\log{k}<-1.52$). All the values of slopes fitted on the simulations with thermal instability have been computed on this range of scales. 

Regarding the TI simulations, the velocity power spectra are slightly flatter from Kolmogorov with slopes of \mbox{-2.4} for 1024N01 and \mbox{-2.3} for 1024N02. On the other hand, the density power spectra are far from Kolmogorov with slopes of  \mbox{-1.3} for 1024N01 and \mbox{-1.5} for 1024N02. This is due to the much higher density contrast. Its values are indeed around 4 ($<C_{\rm n01}>=4.6\pm 1.4$ and $<C_{\rm n02}>=3.8\pm0.7$) while the density contrast of the isothermal simulation is almost constant from a cut to another and stays close from 0.6 (fig.~\ref{fig:densps2d_iso}).

In supersonic turbulent flows it is generally observed that the slope of the density power spectrum flattens with the increase of the Mach number. This effect has been reported for isothermal \citep{kim2005} and supersonic, thermally bistable \citep{gazol2010} flows. This effect is related to the increase of the width of the log-normal PDF of the density with Mach number \citep{vazquez2012}. It was also noted that while the the density power spectrum flattens with the Mach number, the velocity power spectrum stays close from Kolmogorov \citep{kritsuk2007}, slightly steeper tough. 

We observe a similar behavior on simulations of transsonic, thermally bistable turbulence; the density power spectrum is much flatter while the velocity power spectra stay close to Kolmogorov estimate. \cite{hennebelle2007a} emphasized that the density fluctuations of the \hi, and thus the density contrast, are due to the thermal instability at low Mach numbers while supersonic shocks dominate at high Mach numbers. The TI simulations are indeed much more structured at small scales than the sub-sonic isothermal one. {We also want to point out that the power spectrum of the logarithm of the density ($Pk(log(n))$ ), which traces more directly the velocity power spectra through the equation of continuity, has a slope very close to the Kolmogorov law for both 1024N01 and 1024N02.}

\subsection{CNM structures}

Two observational facts often attributed to the self-similarity induced by the turbulent cascade in the interstellar medium are the mass-scale  \citep[$M\propto L^{\alpha}$ --][]{larson1981,elmegreen1996, romanduval2010} and velocity dispersion-scale \citep[\sigturb=$\sigma_{\rm 1pc}L^{\gamma}$ --][]{larson1981,heyer2004} relations observed in molecular clouds. If CNM clumps are the precursors of the molecular matter, it is interesting to see of they follow similar laws. To identify CNM clumps in the simulations we selected pixels having a density greater than 5\,\cc. Structures were then identified as islands of spatially connected pixels. Following the approach of \cite{hennebelle2007b}, the size of a structure is defined  using the largest eigenvalue $\lambda_1$ of its inertial matrix $\mathcal{I}$: 
\begin{eqnarray}
I_{ii} &=& \int(x_j^2+x_k^2) \mathrm{d}m \\
I_{ij} &=& \int{x_ix_j \mathrm{d}m} \\
I_{ji} &=& I_{ij} \quad \mathrm{with}\quad i,j,k \in [1,2,3]\quad\mathrm{and}\quad i\ne j\ne k
\end{eqnarray} 
The $x_i$, $i\in {1,2,3}$, are the pixels coordinates and d$m$ the infinitesimal mass. One can thus estimate the size of the structure by computing $L = \sqrt{\lambda_1/M}$, where $M$ is its mass. We point out that \cite{audit2010} investigated other choices than the largest eigenvalue, as the geometrical mean of the eigenvalues, and found very similar results. 

\begin{figure}
  \resizebox{\hsize}{!}{\includegraphics{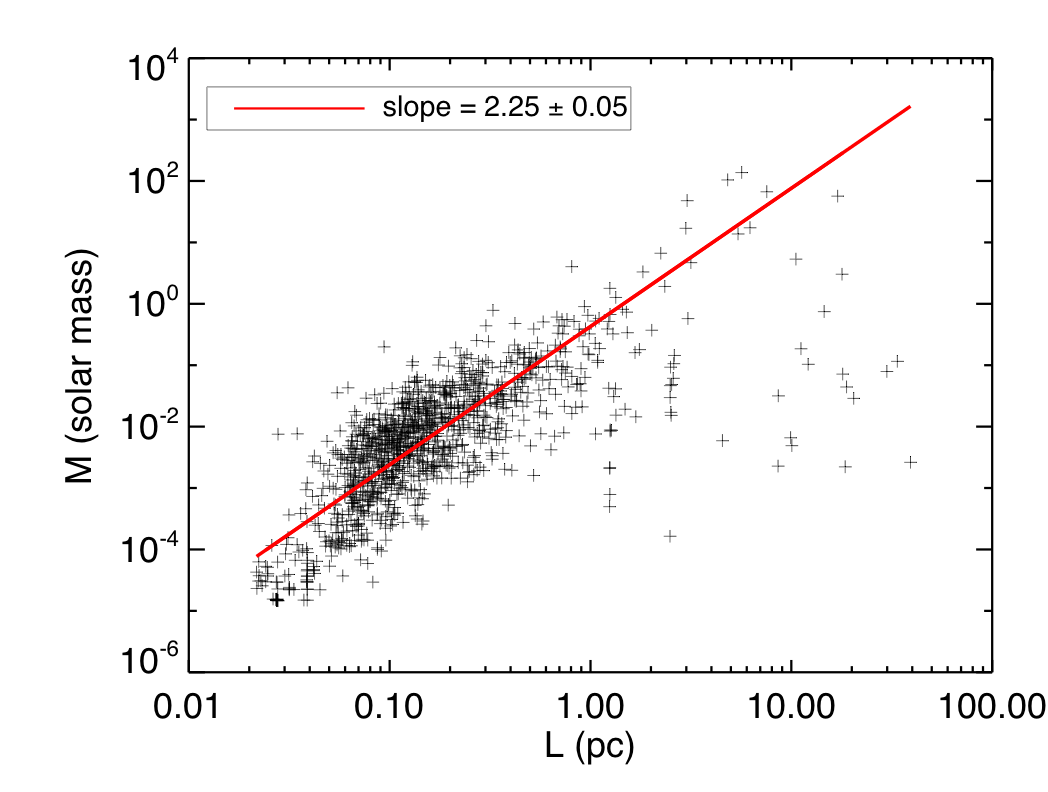}}
  \caption{\label{fig:MvsL} Structures mass versus their size for 1024N01.}
  \resizebox{\hsize}{!}{\includegraphics{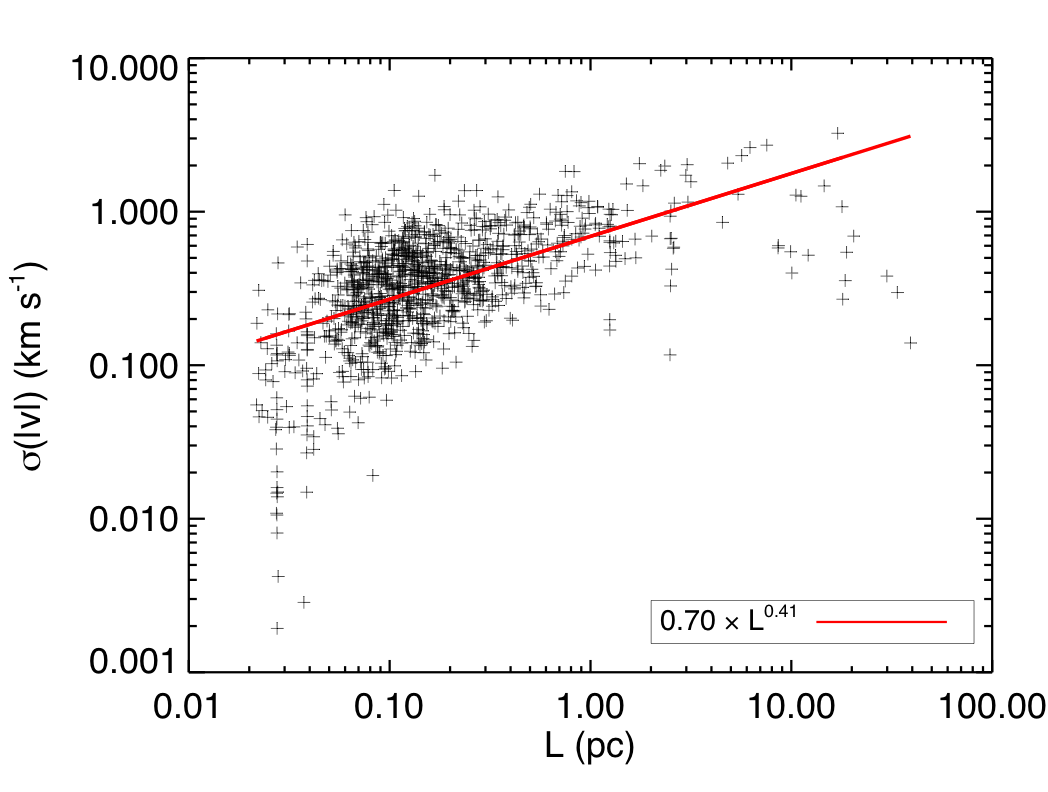}}
  \caption{\label{fig:sigvsL} Distribution of the velocity dispersion inside each clump versus their size. We overplotted in red the fit of the scatter \mbox{$0.70\times L^{0.41}$}}
\end{figure}

\mbox{Figure \ref{fig:MvsL}} shows the relation between the mass and the size of the structures for 1024N01 (this relation for 1024N02 being similar to that of 1024N01, it is not displayed here). We found slopes of \mbox{$2.25 \pm 0.05$} and \mbox{$2.28 \pm 0.03$} for 1024N01 and 1024N02 respectively. This results are in agreement with previous work on isothermal simulations \citep{kritsuk2007,federrath2009a} and on simulations including the thermal instability \citep{audit2010}. They are also in agreement with the results obtained on CO observations by \cite{elmegreen1996} who found a value between 2.2 and 2.5 on a large sample of independent molecular clouds. Recently, \cite{romanduval2010} measured an index of 2.36 on 580 molecular clouds. The structure observed here on \hi\ simulations is similar to molecular clouds.

Similarly we investigated the relation between the velocity dispersion and the size of the CNM clumps. The velocity dispersion is computed as:
\begin{equation}
 \sigma(|\mathbf{v}|)= \frac{\sum\rho(|\mathbf{v}|-|v_0|)^2}{\sum\rho}
\end{equation}
with $v_0$ the mean velocity of each structure weighted by the density $\rho$
\begin{equation}
v_0 = \frac{\sum\rho|v|}{\sum\rho}.
\end{equation}
Figure \ref{fig:sigvsL} represents the values of $\sigma(|\mathbf{v}|)$ versus size for the structures found in simulation 1024N01 (this relation for 1024N02 being similar to that of 1024N01, it is not displayed here). The fits obtained are \mbox{$0.70\times L^{0.41}$} for 1024N01 and \mbox{$0.51\times L^{0.42}$}. These are compatible with the results of \cite{audit2010} who found $0.33 < \gamma < 0.53$. They are also in agreement with the values measured on observations : 0.37 in \hi\ clouds \citep{larson1979} and molecular clouds \citep{larson1981}, and 0.5 obtained on a large sample of molecular clouds by \cite{heyer2004}. All these values are also close to the estimated value of $\gamma$ for subsonic turbulence. On the other hand, the values obtained here for $\sigma_{\rm 1pc}$, 0.70 and 0.51\,\kms, are lower than the results presented by \cite{audit2010} who found values between 1.2 and 3.3\,\kms but with a different forcing scheme. The turbulence in their simulation is indeed produced by WNM converging flows with a Mach number close from 1.5. This type of forcing is very efficient to create cold structures but induces a Mach number in the WNM higher than the one observed in the diffuse medium. 

\begin{figure}
  \resizebox{\hsize}{!}{\includegraphics{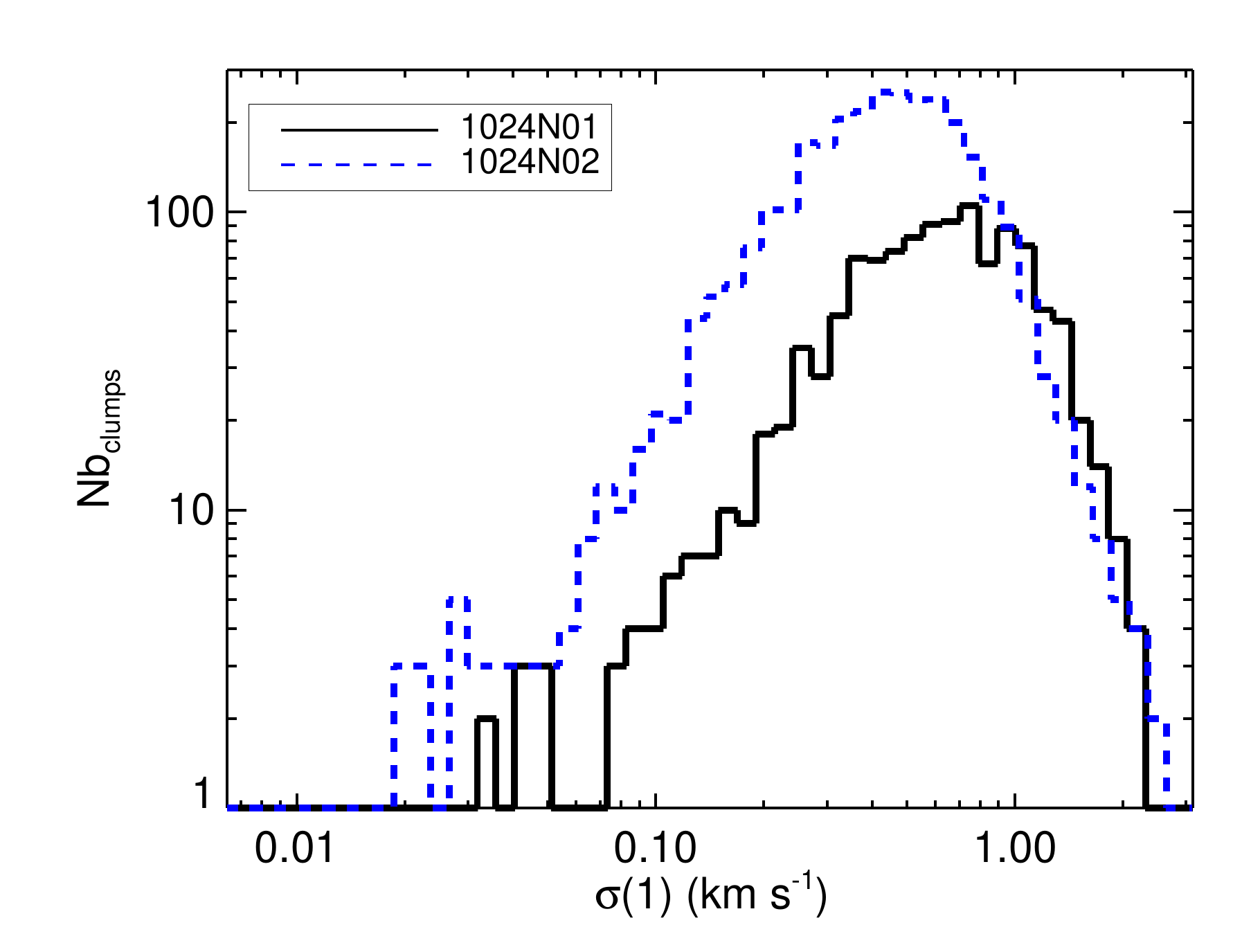}}
  \caption{\label{fig:histosigma1} Histograms of $\sigma(v_Z)/L^{1/3}$ for the 1024N01 simulation in the solid black line, and 1024N02 in the dashed blue line.}
\end{figure}

To compare to the work of \cite{wolfire2003} and to the value 0.92\,\kms\ estimated in \S~\ref{sec:obs}, we computed $\sigma(1) = \sigma(|\mathbf{v}|)/L_{\rm pc}^{1/3}$ for each CNM structure using the the theoretical value $\gamma=1/3$ expected for a subsonic turbulence.
The histograms presented on Figure~\ref{fig:histosigma1} peak around 0.8 and 0.6\,\kms\ for 1024N01 and 1024N02 respectively. These values are in agreement with the velocity dispersion at 1~pc we wanted to reproduce in these simulations and with the work of \cite{wolfire2003}. The fact that $\sigma_{\rm 1pc}$ is well reproduced suggests that the internal dynamics of the CNM clumps is related to the dynamics in the WNM. To confirm this assertion, we computed the mean velocity dispersion $<\sigma(|\mathbf{v}|)>$. We obtained 0.3 and 0.4\,\kms, much lower than the mean thermal velocities inside the clumps: 1.7\,\kms\ for 1024N01 and 1.3\,\kms\ for 1024N02.  The individual cold structures are therefore subsonic. On the other hand, the relative motions between the clumps are greater. We computed the cloud-cloud velocity dispersion as
\begin{equation}
\sigma_{\rm cc}^2=\frac{\sum_0^{Ng} (v_0 - <v_0>)^2}{Ng}
\end{equation}
$Ng$ being the number of clumps.
We obtained $\sigma_{\rm cc}=2.3$\,\kms\ for 1024N01 and $\sigma_{\rm cc}=1.9$\,\kms. These values are very close from the total velocity dispersions in the boxes (2.8\,\kms\ for 1024N01 and 2.2\,\kms\ for 1024N02), suggesting again that the clumps motions are related to the motions in the WNM. {This means that the CNM structures keep the memory of the flow in which they were formed and that little dissipation occurs in cloud-cloud collisions.}
We thus conclude that a significant fraction of the observed 21 cm line broadening is due to the relative motions of the clumps and is not the result of internal supersonic motions. This is in agreement with the work of \cite{heitsch2005,heitsch2006} who simulated the formation of molecular clouds from \hi\ converging flows and reached the same conclusion. 

\section{Summary and discussion}
\label{sec:discussion}

The discussion of the results are divided in two parts. First we discuss the physical conditions that favor the formation of CNM in the local ISM. Second we describe the results on the physical properties of the CNM itself, based on the analysis of the two simulations at high resolution (1024$^3$).

\subsection{Formation of the CNM}

\begin{enumerate}
\item The first main result of the parameter study on 128$^3$ simulations is that the fiducial conditions of the warm neutral medium (\mbox{$n = 0.2$ cm$^{-3}$ and $T = 8000$ K}) do not lead to the formation of cold gas, whatever the turbulence properties in the range we explored. With a slight increase of the density (\mbox{1 cm$^{-3}$}), a majority of compressive modes of the stirring are needed to trigger the transition. A higher increase of the initial WNM density (above \mbox{1.5 cm$^{-3}$}) produces CNM very efficiently. A moderate compression effect is therefore needed to trigger the transition with either a compressive velocity turbulent field or with an increase of the WNM density (or pressure). 
To first order the required increase in pressure (from $P_1$ to $P_2$) for the WNM-CNM transition could happen by the confluence of two flows with a relative velocity $\Delta V$ :
\begin{equation}
\Delta V = \Big[(P_2-P_1)\Big(\frac{1}{\rho_1}-\frac{1}{\rho_2}\Big)\Big]^{1/2}. 
\end{equation}
Considering $n_1=0.5$\,\cc\ (typical density in the WNM), $n_2=2.0$\,\cc\ (initial density of 1024N02) and $T=8000$\,K, one obtains $\Delta V=10.3$\,\kms. The pressure increase needed to place the WNM into suitable conditions to trigger the transition can happen with converging flows with a relative velocity of 10\,\kms, which is completely feasible in cases of supernov\ae\ explosions or stellar winds
and compatible with the idea that turbulence at large scales is produced by converging flows induced by the star formation activty \citep{ferriere2001,elmegreen2004,mckee2007,deavillez2007,kim2010,kim2011}.

\item The second main result is that a sub or transsonic turbulence is required to get CNM gas with properties in agreement with observations (\fcnm$\sim 40\%$ and \sigturb$\sim3$\,\kms). 
In opposition to an isothermal gas, it is rather inefficient to inject supersonic turbulence to produce high density structure in a thermally bistable turbulent flow. This result ties up with the idea of \citet{vazquez2012} and can be explained by Eq.~\ref{eq:tcooltdyn}: when $t_{\rm dyn}$ is lower than $t_{\rm cool}$, turbulent compressions and depressions act quicker than the thermal instability and prevent its development. On the contrary, when $t_{\rm cool} < t_{\rm dyn}$, the gas has time to cool down while turbulence is compressing it and to reach the stable branch of the CNM before its reexpansion. This argument is in agreement with previous studies. \citet{walch2011} observed that, in the conditions of the solar neighborhood, the two phases appear only for low Mach number. \citet{gazol2005} also gathered that the ratio $t_{\rm cool}/t_{\rm dyn}$ increases with the Mach number and that the structures become transient and move away from the cold branch of the thermal equilibrium. At the same time, \citet{heitsch2005} deduced from their simulations  of \hi\ converging flows that the thermal instability is dominant when the density is high and the velocity of the flow low. 

\item The third important result is related to the impact of the distribution of the turbulent energy between the solenoidal and compressive modes. Simulations with moderate initial WNM density 
($n_0=1.0$\,\cc) and a purely compressive forcing field ($\zeta=0$) leads to a good fraction of CNM, close to what is deduced from observations, but in a very long time ($>20\,$Myr). Besides, the Mach number and the turbulent velocity dispersion also need more than 10\,Myr to converge to the observed values, suggesting that the turbulence takes a long time to be fully developed, namely to distribute the energy of the compressive modes towards the solenoidal modes that are responsible for the turbulent cascade. The convergence and the conversion of the WNM in CNM is much faster ($\sim 1\,$Myr) in the case of turbulence with a natural ratio of solenoidal to compressive modes but in this case a slightly higher WNM density ($n_0 > 1.5$\,\cc) is required to trigger the transition.
Given these timescales, we conclude that it is more likely that the formation of CNM structures occurs by an increase of the WNM density rather than purely compressive turbulent energy injection at lower density.\\

\end{enumerate}

\subsection{Structure of the bistable and turbulent \hi}

We summarize here the main results obtained on the \hi\ simulations at high resolution.

\begin{enumerate}
\item We initially showed that these simulations reproduce well the observed physical quantities used as constraints for the parameter study : a cold mass fraction around 40\% with a CNM volume filling factor going from 1 to 4\%, a transsonic Mach number and a turbulent velocity dispersion estimated around 3\,\kms\ for 40\,pc. 
Furthermore, the mass of the gas is distributed in the simulations as it is observed by \cite{heiles2003b} between the thermally unstable regime and the WNM defined by a density criterion. Finally, the pressures are in very good agreement with the observations of \cite{jenkins2011}, for both the mean and the dispersion. It is remarkable that the pressures are stabilized right in the pressure range allowed by the two-phases medium \citep{field1969,wolfire2003}, even when the initial pressures are higher. 

\item We noted that the thermally unstable gas, defined with temperatures included between 200 and 5000\,K behave in two different ways, depending on its temperature. The gas between 200 and 2000\,K is located at the interface of the cold clumps and the warm gas while the gas beyond 2000\,K is dynamically connected to the WNM and is widely distributed in the box. This wide distribution of the unstable gas has also been noticed by \cite{gazol2010} who observed that the boundaries between CNM and WNM are sharper when the Mach number is very low (0.2), due to the unperturbed development of the thermal instability by turbulence. 

\item The density distribution is bimodal, bearing the characteristic of a bistable gas, as noticed by \cite{piontek2004} and \cite{audit2010} in a multiphasic medium. This characteristic can be seen here because the gas is transsonic. Indeed, a high turbulence maintains the gas in the thermally unstable regime \citep{gazol2005,walch2011} and breaks the bimodality of the distribution that tends to be close from a single peak lognormal function. As mentioned by \citet{vazquez2012c}, this is due to the fact that for strong turbulence, the gas is not in the thermal equilibrium locus of the pressure-density phase diagram but it is rather mostly in the thermally unstable range. 
This  might not be representative of the Solar neighborhood where the WNM is thought to be transsonic, implying a bimodal density PDF. In this case we showed that the modelling of the density PDF as a lognormal is not convincing, not even the high density part where we do not observed a power law tail either. 

\item The velocity power spectra of the simulations with thermal instability follow the Kolmogorov law (-8/3 in 2D) with slopes equals to -2.4 and -2.3 for 1024N01 and 1024N02 respectively. On the other hand, the density power spectra are much flatter (-1.3 and -1.4) and the density contrasts high (around 4). These signatures are the direct result of the formation of dense structure by the thermal instability in a low Mach number flow \citep{hennebelle2007a}. It is interesting to point out that the power spectra behavior observed here is similar to the one seen in supersonic isothermal flows where the density power spectrum flattens with the increase of the the Mach number \citep{kim2005} while the velocity power spectrum stays close to Kolmogorov \citep{kritsuk2007} or even steeper. 

\item The mass-scale relation $M\propto L^{\alpha}$ found ($\alpha=2.25$ and 2.28) for clumps with a density higher than 5\,\cc\ are in good agreement with observations \citep{elmegreen1996,romanduval2010} as well as with previous numerical studies, even when different conditions and algorithm are used \citep{kritsuk2007,federrath2009a,audit2010}.  The $\gamma$-index of the velocity dispersion-scale relation ($\sigma\propto L^{\gamma}$) is also well reproduced in both simulations with values of 0.41 and 0.42. These values are similar to the ones obtained on numerical simulations \citep{audit2010}, to the observed in molecular clouds \citep{larson1981,heyer2004}, but also to the value expected for a subsonic turbulence $\gamma=1/3$. These results suggest that the self-similar density structure observed in molecular clouds could be inherited from the \hi\ from which they formed.

\item The internal velocity dispersion of the cold structures (0.3-0.4\,\kms)  indicate that they are clearly subsonic (the thermal velocity dispersion is 1.3-1.7\,\kms). Like for the WNM, the turbulent velocity dispersion of clumps increases with scale with a normalisation at 1\,pc of $\sigma(1)=0.6-0.8$\,\kms, very close to the value found in the WNM. Similarly, the cloud-cloud velocity dispersion is 1.9-2.3\,\kms, close to the total velocity dispersions in the boxes (2.2-2.8\,\kms). This implies that the turbulent motions inside clumps and the relative velocities between them are related to the motions of the WNM from which they were formed. 
Individually, the clouds are subsonic but their relative velocities are supersonic relative to the CNM sound speed \citep[in accordance with][]{koyama2002,heitsch2005}.  This result is in complete agreement with the work of \citet{hennebelle2007a} who noticed that the CNM velocity dispersion estimated from 21\,cm spectra is greater than the mean velocity dispersion of the CNM structures identified in their simulation. This means that a non negligible fraction of the measured velocity dispersion is caused by the relative motions of the clumps along the line of sight, suggesting that the observed line broadening is likely to be due to the relative clumps motions rather than supersonic turbulence.
 
\end{enumerate}

\section{Conclusion}
\label{sec:conclusion}

We have presented a set of 90 low resolution simulations ($128^3$ pixels) in order to study the physical conditions that lead to the production of CNM structures
out of purely WNM gas, given the observational constraints (velocity dispersion, density, pressure) and the model of \citet{wolfire2003} of the
heating and cooling processes at play in the diffuse ISM. 
These simulations show that WNM gas at typical values of density, temperature and turbulent motions do not transit to CNM gas. The typical volume density of the WNM in the disk lies between 0.2 and 0.5 cm$^{-3}$. We showed that WNM with initial density under \mbox{1 cm$^{-3}$} never transit into a cold phase, whatever the turbulent conditions. Stronger turbulent motions clearly does not help in producing cold gas, in fact we showed that it is the opposite. On the other hand a moderate increase of the WNM density (or equivalently of its pressure) is very efficient; all simulations with an initial density higher than \mbox{1.5 cm$^{-3}$} do transit and the amount of cold gas in the simulation increases with the density. When the initial density reaches \mbox{3 cm$^{-3}$}, all simulations achieve to create at least 90\% of CNM. These results were obtained with a natural balance between compressive and solenoidal modes in the injected velocity field. In this case the formation of CNM gas is almost instantaneous as soon as the turbulence is fully developed.
We also explored the case of an intermediate initial density (\mbox{1 cm$^{-3}$}) but with varying the ratio of solenoidal to compressive mode in the turbulent forcing. We showed that, at this density, the transition into CNM only occurs when there is a majority of compressible modes in the turbulent velocity field but this process is much slower (10\,--\,20 Myr). 

We conclude that the ISM turbulence cascade on its own cannot induce a transition from WNM into CNM at the average density of the WNM.
The production of CNM is likely to be due to turbulent motions of moderate amplitude associated to a compressive event of the WNM that could be led by transient phenomena such as outflows, supernova explosion or spiral density waves. This would trigger the phase transition without delay. Once formed, the amount of CNM gas is dynamically stable. Therefore there is no need for a constant energy injection to maintain the high density contrast of the \hi.\\

We also have presented a detailed study of the \hi\ based on two higher resolution ($1024^3$ pixels) simulations. To our knowledge these are the largest simulations to date that include the thermal instability and a pseudo-spectral turbulent stirring. The bimodal temperature and density histograms show the evidence of a bistable medium and the massive fractions of each phase, the Mach number of the WNM and the pressure distributions are all in agreement with observations. With these simulations we confirm a number of results of previous studies based on converging flows \citep{heitsch2006,hennebelle2007a,audit2010} that showed that the structure of the CNM and WNM are tightly interwoven: the two media share the same velocity fields and the CNM cloud-cloud velocity dispersions is close to the WNM sound speed. While the individual CNM structures are clearly subsonic, their relative motions are supersonic with respect to their own temperature. 
In addition, thanks to the pseudo-spectral forcing method, the statistical properties of the turbulent bi-stable flow could be analysed on a larger range of scales than could be done in previous studies. We found that for such sub/transsonic bistable flows, the velocity field keeps the properties of subsonic turbulence with $P(k)\propto k^{-8/3}$ while the density is highly contrasted, reminiscent of what is observed for supersonic flows and in molecular clouds. In the case of the \hi\ simulated here, the joined action of the turbulence and the thermal instability allows the formation of long-lasting cold and dense structures without need for supersonic motions, providing a favorable terrain for the formation of highly contrasted molecular couds.
We now intent to use these high resolution simulations to create synthetic observations and to test and improve some analysis methods used on observations.

\begin{acknowledgements}
Heracles is available on the following link: http://irfu.cea.fr/Projects/Site\_heracles/ \citep{gonzalez2007}.\\
The simulations at low resolution have been performed on the cluster Sunnyvale at CITA, University of Toronto, and the $1024^3$ cells simulations have been performed on the cluster Jade, Cines, Montpellier. 
\end{acknowledgements}

\bibliographystyle{aa}
\bibliography{mypaper}

\end{document}